\pdfoutput=1
\documentclass[a4paper,12pt]{article}

\usepackage{amsfonts}
\usepackage{mathrsfs}
\usepackage{amssymb}
\usepackage{amsmath}
\usepackage{shuffle}
\usepackage{framed} 
\usepackage{subfigure}
\usepackage[medium]{titlesec}
\usepackage{bm}
\usepackage{cite}

\usepackage[normalem]{ulem}
\usepackage{extarrows}
\usepackage{slashed}
\usepackage{isodateo}
\usepackage{graphicx}
\usepackage{array}
\usepackage[dvipsnames]{xcolor}
\usepackage[bookmarksnumbered=true,bookmarksopen=true]{hyperref}
 \hypersetup{colorlinks,%
             linkcolor=NavyBlue, %
             citecolor=PineGreen, %
             urlcolor=PineGreen}
\usepackage[hmargin=.7in,vmargin=1.1in]{geometry}
\usepackage{indentfirst}
\usepackage{booktabs}

\usepackage{bbm}

\newcommand{\FR}[2]{\displaystyle\frac{\,{#1}\,}{#2}}

\newcommand{\n}{\nonumber}

\graphicspath{{fig/}}

\def\bge{\begin{equation}}
\def\ede{\end{equation}}
\def\bga{\begin{aligned}}
\def\eda{\end{aligned}}
\def\bgb{\begin{bmatrix}}
\def\edb{\end{bmatrix}}
\def\bgp{\begin{pmatrix}}
\def\edp{\end{pmatrix}}
\def\bgm{\begin{matrix}}
\def\edm{\end{matrix}}
\def\bgs{\begin{subequations}}
\def\eds{\end{subequations}}
\newcommand{\order}[1]{\mathcal{O}({#1})}
\def\di{{\mathrm{d}}}

\def\mb{\mathbf}

\def\to{\rightarrow}

\def\ii{\mathrm{i}}

\def\ep{\epsilon}

\def\si{\sigma}

\def\aa{\mathsf{a}}
\def\bb{\mathsf{b}}

\newcommand{\ft}[1]{\big[{#1}\big]}

\DeclareFontFamily{U}{bigshuffle}{}
\DeclareFontShape{U}{bigshuffle}{m}{n}{
  <5-8> s*[1.7] shuffle7
  <8->  s*[1.7] shuffle10
}{}
\DeclareSymbolFont{BigShuffle}{U}{bigshuffle}{m}{n}
\DeclareMathSymbol\bigshuffle{\mathop}{BigShuffle}{"001}
\DeclareMathSymbol\bigcshuffle{\mathop}{BigShuffle}{"002}

\usepackage{mdframed}

\newmdenv[backgroundcolor=gray!15,%
          skipabove=5pt,%
          skipbelow=5pt,%
          leftmargin=0pt,%
          rightmargin=0pt,%
          innertopmargin=8pt,%
          innerbottommargin=8pt,%
          innerleftmargin=12pt,%
          innerrightmargin=12pt,%
          splittopskip=0pt,%
          splitbottomskip=0pt,%
          linewidth=0pt,%
          nobreak=true]%
          {keytext}

\newmdenv[backgroundcolor=gray!15,%
          skipabove=0pt,%
          skipbelow=5pt,%
          leftmargin=0pt,%
          rightmargin=0pt,%
          innertopmargin=-5pt,%
          innerbottommargin=7pt,%
          innerleftmargin=2pt,%
          innerrightmargin=2pt,%
          splittopskip=0pt,%
          splitbottomskip=0pt,%
          linewidth=0pt,%
          nobreak=true]%
          {keyeqn}

\makeatletter
\@ifpackageloaded{hyperref}%
  {\newcommand{\mylabel}[2]
    {\protected@write\@auxout{}{\string\newlabel{#1}{{#2}{\thepage}%
      {\@currentlabelname}{\@currentHref}{}}}}}%
  {\newcommand{\mylabel}[2]
    {\protected@write\@auxout{}{\string\newlabel{#1}{{#2}{\thepage}}}}}
\makeatother

\usepackage{titlesec}          
\titleformat{\section}
{\normalfont\fontsize{15}{20}\bfseries}{\thesection}{1em}{}

\newcommand{\wt}[1]{\mkern 2mu \widetilde{\mkern -2mu #1 \mkern -2mu}\mkern 2mu}
\newcommand{\wh}[1]{\mkern 2mu \widehat{\mkern-2mu#1\mkern-2mu}\mkern 2mu}

\newcommand{\fnemail}[1]{\footnote{Email: \href{mailto:#1}{\nolinkurl{#1}}}}

\begin{document}

\title{\Large\textbf{Anatomy of Family Trees in Cosmological Correlators\\[2mm]}}

\author{Bingchu Fan$^{\,a\,}$\fnemail{fbc23@mails.tsinghua.edu.cn}~~~~~ and ~~~~~Zhong-Zhi Xianyu$^{\,a,b\,}$\fnemail{zxianyu@tsinghua.edu.cn}\\[5mm]
\normalsize{\emph{$^a\,$Department of Physics, Tsinghua University, Beijing 100084, China}}\\ 
$^b\,$\normalsize{\emph{Peng Huanwu Center for Fundamental Theory, Hefei, Anhui 230026, China}}
}

\date{}
\maketitle

\vspace{20mm}

\begin{abstract}
\vspace{10mm}
The time-ordered multilayer integrals have long been cited as major challenges in the analytical study of cosmological correlators and wavefunction coefficients. The recently proposed family tree decomposition technique solved these time integrals in terms of canonical objects called family trees, which are multivariate hypergeometric functions with energies as variables and twists as parameters. In this work, we provide a systematic study of the analytical properties of family trees. By exploiting the great flexibility of Mellin representations of family trees, we identify and characterize all their singularities in both variables and parameters and find their exact series representations around all singularities with finite convergent domains. These series automatically generate analytical continuation of arbitrary family trees over many distinct regions in the energy space. As a corollary, we show the factorization of family trees at zero partial-energy singularities to all orders. Our findings offer essential analytical data for further understanding and computing cosmological correlators.

\end{abstract}

\newpage
\tableofcontents

\newpage

\section{Introduction}

Recent years have seen growing interest in studying high-energy physics through primordial curvature and tensor fluctuations. A focus of attention has been the spontaneous on-shell production of heavy particles driven by cosmic inflation. Such a process is kinematically forbidden in the Minkowski vacuum but allowed in a time-dependent background. The produced heavy particles may leave various observable signatures via couplings to long-lived curvature or tensor fluctuations. In this so-called Cosmological Collider (CC) program \cite{Chen:2009we,Chen:2009zp,Arkani-Hamed:2015bza}, many viable particle-physics models have been identified \cite{Chen:2016nrs,Chen:2016uwp,Chen:2016hrz,Lee:2016vti,An:2017hlx,Iyer:2017qzw,Kumar:2017ecc,Tong:2018tqf,Chen:2018sce,Chen:2018xck,Wu:2018lmx,Saito:2018omt,Li:2019ves,Lu:2019tjj,Liu:2019fag,Hook:2019zxa,Hook:2019vcn,Kumar:2018jxz,Kumar:2019ebj,Wang:2019gbi,Wang:2020uic,Li:2020xwr,Wang:2020ioa,Aoki:2020zbj,Bodas:2020yho,Lu:2021gso,Sou:2021juh,Lu:2021wxu,Pinol:2021aun,Cui:2021iie,Tong:2022cdz,Reece:2022soh,Chen:2022vzh,Niu:2022quw,Chen:2023txq,Chakraborty:2023qbp,Tong:2023krn,Jazayeri:2023xcj,Jazayeri:2023kji,Aoki:2023tjm,McCulloch:2024hiz,Craig:2024qgy,Pajer:2024ckd,Wang:2025qww,Bodas:2025wuk}, and some of them have been searched for in cosmological data \cite{Sohn:2024xzd,Cabass:2024wob,Bao:2025onc}. 

The key observables of CC physics are the correlation functions of primordial fluctuations, known as \emph{cosmological correlators}. The dynamics of inflation and the history of subsequent cosmic evolution make these objects both fascinating yet paradoxical in two ways: First, they represent the data we can collect at the largest length scale (10\;Gpc\;$\sim\order{10^{26}}$m) in the universe but carry messages of fundamental particles and interactions at inflation Hubble scale which is probably the tiniest scale we can ever probe (down to $(10^{13}\;\text{GeV})^{-1}\sim \order{10^{-28}}$\;m) \cite{Planck:2018jri,ACT:2025fju}. Second, the large-scale fluctuations appear to us stationary in the sky, but they are inevitably consequences of active time evolution of quantum fields during inflation \cite{Achucarro:2022qrl}. 

It is not a priori clear what the best theoretical approach is to study these cosmological correlators, as the time-dependent dynamics of quantum fields during inflation could be very complex and even chaotic. Fortunately, observational data have shown that primordial curvature fluctuations at the largest scales are weakly coupled \cite{Planck:2019kim}. Therefore, it is natural to consider models of CC physics described by weakly coupled quantum field theory (QFT) in an inflationary background which is close to (the Poincaré patch of) de Sitter spacetime (dS). For such models, we expect the usual perturbative approach based on Feynman diagram expansion to be a good approximation. 

It has long been appreciated that computing general Feynman diagrams is a daunting task for cosmological correlators involving massive fields. From the perspective of standard calculations using the Schwinger-Keldysh (SK) formalism in bulk dS \cite{Chen:2017ryl}, the difficulties are essentially due to an evolving spacetime background and thus a lack of 4-momentum representation. A time-dependent background distorts massive Green functions into various types of special functions and the lack of 4-momentum representation forces us to evaluate time integrals over all interaction vertices, accounting for all possible time orderings. As a result, we get time-ordered multilayer integrals over special functions which are challenging even for tree graphs. 

These difficulties have been noted in the literature, where various approaches have been developed to circumvent direct integration, such as the differential equation method \cite{Arkani-Hamed:2018kmz,Baumann:2019oyu,Pimentel:2022fsc,Jazayeri:2022kjy,Qin:2022fbv,Qin:2023ejc,Aoki:2023wdc,Aoki:2024uyi,Chen:2024glu,Liu:2024str}, the full Mellin-space approach \cite{Sleight:2019mgd,Sleight:2019hfp,Sleight:2020obc,Sleight:2021plv}, the partial Mellin-Barnes (PMB) representation \cite{Qin:2022lva,Qin:2022fbv,Qin:2023bjk,Qin:2023nhv,Qin:2024gtr}, the family-tree decomposition (FTD) \cite{Xianyu:2023ytd,Fan:2024iek}, the dispersive method \cite{Liu:2024xyi,Werth:2024mjg}, the spectral decomposition \cite{Xianyu:2022jwk,Zhang:2025nzd,Bodas:2025wuk}, etc. The field has been developing fast; Many calculations that seemed unimaginable a few years ago have now been completed.

However, with new results accumulating, it gradually becomes clear that these massive correlators are inherently complicated: Unlike gauge-boson amplitudes in flat spacetime, where highly complicated intermediate calculations simplify dramatically in the final result, these massive correlators are generally described by multivariate hypergeometric functions (MHFs) with high complexity \cite{Liu:2024xyi}. The complexity can be quantified using the concept of transcendental weight, which is essentially the layer of iterated integrals and increases with the number of vertices and loops in a correlator. Functions of higher weights cannot be reduced to simpler functions. This poses a challenge: no matter what approach we adopt, we inevitably arrive at results involving these poorly understood, complicated functions. Therefore, to make further progress, we need to understand them.\footnote{For special choices of mass and twist, the MHFs representing cosmological correlators could reduce to more familiar functions such as rational functions or multiple polylogarithms \cite{Fan:2024iek}.}

At the moment, there are two approaches capable of yielding explicit analytical results for arbitrary tree-level massive correlators: One can reduce a tree graph into either the sum of a \emph{massive family tree} and its cuts \cite{Liu:2024str} or a Mellin integral of a set of somewhat simpler \emph{family trees} weighted by Euler $\Gamma$ factors \cite{Xianyu:2023ytd}. In both cases, these family-tree functions are MHFs. They can be viewed as building blocks of correlators and their family-tree structure (reviewed in Sec.\;\ref{sec_CCandFT}) is a vivid manifestation of time orderings in the bulk. Thus, to better understand massive cosmological correlators in an analytical way, we need to know the analytical properties of family trees as much as we can.\footnote{Incidentally, as shown in \cite{Fan:2024iek}, a simple linear combination of family trees reproduces tree-level conformal-scalar amplitudes in a general power-law FRW universe. Therefore, these family trees are also solutions to the differential equations of kinematic flow \cite{Arkani-Hamed:2023kig,Arkani-Hamed:2023bsv}. These differential equations have been shown to be more easily derived starting from FTD \cite{He:2024olr}.} 

As MHFs, family trees are analytic in the (complex) parameter space except for a finite number of singularities. These singularities are exclusively poles or branch points, with divergences bounded by finite complex powers. As has been very familiar in physics, an amplitude or correlator does not diverge without a physical reason; Each singularity of the family tree has a particular physical meaning. Thus, an understanding of all singularities of a family tree provides a firm grasp of what is physically going on in a QFT process in dS. Furthermore, the information about singularities can uniquely fix the family tree function and thus could be particularly useful either for a dispersive bootstrap program of correlators (using singularities to recover the full function) or for a numerical implementation of the function. Therefore, understanding the analytical property (or equivalently, the singularity structure) of family trees is both physically relevant and computationally useful. 

\paragraph{Outline and main results}
In this work, we provide a comprehensive study of the analytical structure of family trees introduced in \cite{Xianyu:2023ytd}, by identifying all singularities of an arbitrary family tree and, more importantly, deriving a hypergeometric series that converges over a finite domain around each singularity. Although the derivations of these results could be technical at times, the main results are simple, which we summarize briefly here:
\begin{description}
  \item[A conceptual preparation (Sec.\;\ref{sec_CCandFT})] After a brief review of cosmological correlators, PMB, and FTD, we introduce the family tree to be studied in this work. A family tree is defined by a partially ordered time integral and endowed with a graphic notation as well as a square-bracket notation. For example:
  \begin{align}
    \parbox{25mm}{\includegraphics[width=25mm]{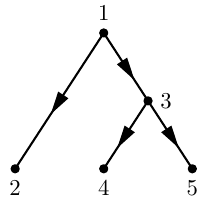}}=\ft{1(2)(3(4)(5))}\equiv (-\ii)^5\int\prod_{i=1}^5\Big[\di\tau_i(-\tau_i)^{q_i-1}e^{\ii\omega_i\tau_i}\Big]\theta_{43}\theta_{53}\theta_{31}\theta_{21}.
  \end{align}
Here $\omega_i$ are called the \emph{energies} and $q_i$ the \emph{twists}. 
The graphic and square-bracket notations should be self-explanatory. See (\ref{eq_FamilyTreeExamples}) for examples. The ancestor site (with outflows only) is also called the \emph{root} of the family tree. In the case of $\ft{1(2)(3(4)(5))}$, the root is Site 1.

  \item[Result 1:] \textbf{All singularities of a family tree (Sec.\;\ref{sec_SingularityProof})}  A family tree has a singularity in the complex space of energies if and only if the sum of energies of a root-bearing subgraph of the family tree goes to either zero or infinity. We call them zero partial-energy singularities and infinite partial-energy singularities, respectively. In the case of $\ft{1(2)(3(4)(5))}$, there are 20 singularities, listed in (\ref{eq_5treeSing}) and graphically in Fig.\;\ref{fig_singularity}.
  
  \item[Result 2:] \textbf{Large partial-energy series (Sec.\;\ref{sec_LargeEnergy})} The series representation at an infinite partial-energy singularity is given by (\ref{eq_InfPESeries}). This generalizes previously known special cases, such as the large single-energy series (\ref{eq_LargeMaximalSeries}) and the large total-energy series (\ref{eq_LargeTEseries}) \cite{Xianyu:2023ytd,Fan:2024iek}. Remarkably, a family tree is represented by a single MHF multiplied by a complex power at each of these singular points and the singularity is entirely from the complex power.\footnote{Although we sometimes use terms such as ``singular points'' or ``branch points'' borrowed from complex analysis of one variable, for general energy space of $N$ complex dimensions, these objects are actually complex codimension-1 surfaces. They become points only when projected to a transverse complex plane.}  
    
  \item[Result 3:] \textbf{Small partial-energy series (Sec.\;\ref{sec_SmallEnergy})} The series representation at a zero partial-energy singularity is given by (\ref{eq_SmallPESeries}), of which the small root-energy series (\ref{eq_ZeroRootSeries}) and the small total-energy series (\ref{eq_TEseries}) are special cases. All these results are new. In contrast to the infinite partial energy's case, a family tree at a small partial-energy singularity separates into a singular piece and a regular piece. The singular piece is universal and is given by a single MHF multiplied by a singular power function, while the regular piece depends on energy ordering in a more complicated way and is normally a sum of several MHFs multiplied by power functions that are regular in those limits. 
  
  \item[Result 4:] \textbf{Factorization at small partial-energy singularities (Sec.\;\ref{sec_Factorization})} The singular part of a family tree at one of its small partial-energy limit factorizes into products of disjoint subgraphs, and the singularity comes entirely from the small total-energy singularity of a subgraph. See (\ref{eq_PEFactThm}). The factorization is not only a leading-order result but also holds to all orders in the small partial energy.   
  
  \item[Result 5:] \textbf{Regularized family trees are entire functions of twists (Sec.\;\ref{sec_twists})} Finally, we also investigate the singularity structure of family trees in complex twist space, and show that all family trees are meromorphic in the twist space where singularities are all simple poles coming from Euler $\Gamma$ factors. After regularizing the family trees by removing these $\Gamma$ factors, the regularized family trees are entire functions of twists when the arguments (energies) are away from their own singularities. 
  
  \item[Technical aspects] A major technical tool we use in this work is the Mellin integral representation of family trees (not to be confused with the PMB representation of massive correlators). This includes 1) choosing appropriate representations for nested time integral and 2) carrying out Mellin integrals by a pole collecting algorithm. Every step can be done with great flexibility and we exploit this flexibility to get all desired results. We introduce this method in Sec.\;\ref{sec_MBRofTI}, and show its use with many examples in the subsequent sections. Our pole collecting algorithm is more systematically summarized in App.\;\ref{app_MB} where we also show that it is equivalent to the conic hull method. 

\end{description}

The results of this paper enable a more thorough understanding of analytical and even numerical aspects of cosmological correlators and form the starting point of many future projects, on which we briefly comment in Sec.\;\ref{sec_Outlook}. Also, for readers' convenience, we collect special notations and conventions in App.\;\ref{app_Notations}, the special functions and their useful properties in App.\;\ref{app_Functions}. In App.\;\ref{app_CPN} we discuss a technical point of how to compactify the energy space to meaningfully talk about infinite energy limits of family trees. Finally, in App.\;\ref{app_Examples} we tabulate a few fully worked out examples which may be of some use in future studies.

\paragraph{Notations and conventions} Throughout the paper, we use a time variable $\tau\in(-\infty,0)$ which coincides with the conformal time in dS, although our results for family trees apply to more general spatially flat FRW spacetime as well. We often use a shorthand for summation such as $q_{ij\cdots}\equiv q_i+q_j+\cdots$ and $n_{12\bar 3}=n_1+n_2-n_3$. Euler $\Gamma$ products and fractions are often written in a condensed manner, such as $\Gamma[a,b,\cdots]\equiv\Gamma(a)\Gamma(b)\cdots$ and $\Gamma[\begin{smallmatrix}a,b,\cdots \\ c,d,\cdots\end{smallmatrix}]\equiv \Gamma[a,b,\cdots]/\Gamma[c,d,\cdots ]$. Other special notations will be introduced when they first appear and are collected in App.\;\ref{app_Notations}.

\section{Cosmological Correlators and Family Trees}
\label{sec_CCandFT}
In this section, we review the general structure of the nested time integral in cosmological correlators. We also give a brief introduction to the family tree decomposition method developed in previous works \cite{Xianyu:2023ytd,Fan:2024iek}. Along the line, we will gain insights into how to find hypergeometric series representations of the family trees in wider regions.

\subsection{General integrals for tree correlators}

We consider a general tree-level correlator in dS background, with an arbitrary number of external lines of massless ($m=0$) or conformal ($m=\sqrt2 $) scalars, and an arbitrary number of massive ($m>0$) internal lines. With the diagrammatic rules in SK formalism \cite{Chen:2017ryl}, a tree-graph contribution to such a correlator can be computed by the following integral:
\begin{align}
    \mathcal{I}=\sum_{\aa_1,\cdots,\aa_N = \pm} \int_{-\infty}^0 \prod_{j=1}^N \Big[\di\tau_j\, \ii \aa_j (-\tau_j)^{p_j} e^{\ii\aa_j E_j \tau_j}\Big] \prod_{i=1}^I D^{(\wt\nu)}_{\aa_{i_1} \aa_{i_2}} (K_i;\tau_{i_1},\tau_{i_2}) \ .
    \label{eq_GeneralCorrelator}
\end{align}
Here we have $V$ vertices and $I$ internal lines. (For tree graphs, $V=I+1$.) For each vertex $j$, we associate with four variables: 1) A conformal time variable $\tau_j\in(-\infty,0)$; 2) An SK index $\aa_j=\pm$, which is to be summed over in the final expression; 3) A \emph{vertex energy} $E_j$ as appearing in the exponential factor $e^{\ii\aa_jE_j\tau_j}$, which is defined to be the sum of the magnitudes of the external 3-momenta connected to the vertex; 4) A power function of the conformal time $(-\tau_j)^{p_j}$  where the exponent $p_j$ takes care of various types of time dependences from the coupling and the mode functions.\footnote{The parameter $p_j$ is called twist in \cite{Liu:2024str}, but in this work, we reserve the term twist to another parameter $q_j$ to be introduced below.} For instance, the mode functions for massless scalar $\varphi$ and conformal scalar $\phi_\text{c}$ have the following time dependences: 
\begin{align}
    \varphi(k,\tau)\propto(1+\ii k \tau) e^{- \ii k \tau},\ \phi_\text{c}\propto \tau e^{-\ii k \tau} \  .
\end{align} 
So, by choosing the values of $p_j$ and taking linear combinations, it is easy to cover both cases with the expression (\ref{eq_GeneralCorrelator}). More generally, $p_j$ can take arbitrary complex values, with the only requirement that the integrals for the correlator $\mathcal{I}$ are convergent in the infrared limit $\tau_j\to 0$ for all $j$. For instance, a non-(half-)integer value of $p_j$ can describe the effect of internal particle decay or slow-roll corrections, while the imaginary part of $p_j$ can describe background resonances. 

On the other hand, for each internal line, we associate with a massive propagator $D_{\aa\bb}(K;\tau_1,\tau_2)$ with two more parameters: 1) The mass parameter $\wt\nu\equiv\sqrt{m^2-9/4}$. For the applications of CC physics, the main interest lies in the case of principal series where $\wt\nu>0$ ($m>3/2$) although our result applies easily to more general $m>0$ for which $\wt\nu$ can take complex values. 2) A \emph{line energy} $K_i$, which is defined to be the magnitude of the momentum flowing in the line. The internal propagator takes a more complicated form, involving both Hankel functions $\text{H}^{(1,2)}_{\nu}(z)$ and time-ordering Heaviside $\theta$ functions:
\begin{align}
    D_{-+}^{(\wt{\nu})}(k,\tau_1,\tau_2)&=\frac{\pi}{4} e^{-\pi \wt{\nu}} (\tau_1 \tau_2)^{\frac{3}{2}} \text{H}_{\ii \wt{\nu}}^{(1)}(-k \tau_1)\text{H}_{-\ii \wt{\nu}}^{(2)}(-k \tau_2) \ , \notag \\
    D_{+-}^{(\wt{\nu})}(k,\tau_1,\tau_2)&=D_{-+}^{(\wt{\nu})}(k,\tau_1,\tau_2)^* \ , \notag \\
    D_{\pm\pm}^{(\wt{\nu})}(k,\tau_1,\tau_2)&=D_{\mp\pm}^{(\wt{\nu})}(k,\tau_1,\tau_2)\theta(\tau_1-\tau_2)+D_{\pm\mp}^{(\wt{\nu})}(k,\tau_1,\tau_2)\theta(\tau_2-\tau_1) \ .
\label{eq_SK_Propagators}
\end{align}

The integral (\ref{eq_GeneralCorrelator}) is the starting point for the computation of general cosmological correlators. As mentioned, the main complications here come from the two sources: the special function (Hankel function) in the internal modes and the time orderings. The former complication can be resolved by the PMB representation \cite{Qin:2022lva,Qin:2022fbv}, while the latter is overcome by FTD. 

The main idea of PMB representation is to reduce the $\tau$-dependence of special functions to simple power functions, with the price of generating a new Mellin-Barnes (MB) integral. Specifically, we make use of the MB representation for Hankel functions:
\begin{align}
    \text{H}_\nu^{(j)}(a z)=\int_{-\ii \infty}^{+\ii \infty} \frac{\di s}{2\pi \ii} \frac{(a z/2)^{-2s}}{\pi} e^{(-1)^{j+1}(2s-\nu-1)\pi\ii/2} \Gamma\left[\frac{s-\nu}{2},\frac{s+\nu}{2}\right] \quad (j=1,2) \ .
\end{align}
Typically, MB integrands are meromorphic for massive correlators that are well convergent in the late-time limit. Therefore, one can finish the MB integrals by using the residue theorem and summing the residues at relevant poles. Since every step is convergent, we can postpone the MB integrals to the end. So, from now on, we will only focus on the MB integrand.

In the MB integrand, the Hankel function disappears, and the power of $\tau_j$ at Vertex $j$ is shifted by $-2s_j$ where $s_j$ is the sum of Mellin variables for all massive modes connected to Vertex $j$. The only remaining complication is the time integral, which is the main focus of this work. To deal with this time integral, we can neglect all $\tau$-independent coefficients which may depend on $K_i,E_j,s_j$. Then, the most general time integral involves the power functions, the exponential functions, and the time ordering functions, and can be put into the following form:   
\begin{align} 
    \label{eq_NestedIntegral}
    \mb{T}^{(q_1\cdots q_N)}_{\mathscr{N}}(\omega_1,\cdots,\omega_N)\equiv (-\ii)^N \int_{-\infty}^0\prod_{j=1}^N\Big[\di\tau_j\,(-\tau_j)^{q_j-1}e^{\ii \omega_j \tau_j}\Big]\prod_{(j_1,j_2)\in\mathscr{N}}\theta(\tau_{j_1}-\tau_{j_2}) \ .
\end{align}
Here we use $\mathscr{N}$ to denote the set of all ordered and adjacent pairs of time variables. That is, we have the pair $(i,j)\in\mathscr{N}$ if and only if there is a time-ordering function $\theta(\tau_i-\tau_j)$ in the integral. The power of time is written as $q_j-1$ so that the dimension of the entire integral is $-\sum\limits_j q_j$, and we call $q_j$ the \emph{twist} parameters at Site $j$. The energy variables $\omega_j$ come from the external-energy factors $e^{\ii\aa_jE_j\tau_j}$ in the original integral (\ref{eq_GeneralCorrelator}). The overall phase factor $(-\ii)^N$ descends from the Feynman rule for vertices. 

Incidentally, the nested time integrals of the form (\ref{eq_NestedIntegral}) also arise from correlators of conformal scalars in power-law FRW spacetime (namely, the scale factor is a power function of the conformal time $\tau$, $a(\tau)\propto |\tau|^P$), as shown in previous works \cite{Arkani-Hamed:2017fdk,Fan:2024iek}. The conformal scalar in power-law FRW universe has a simpler mode function, involving exponential and power functions only. Therefore, no special function appears in the integrands for the conformal-scalar correlators. As a result, the most general time integral for conformal-scalar correlators in power-law FRW universe is still (\ref{eq_NestedIntegral}), although the energy $\omega_j$ in this case can depend on both vertex energies $E_j$ and line energies $K_i$.

\subsection{Family-tree decomposition}
\label{sec_FTD}
To finish the nested time integral (\ref{eq_NestedIntegral}), we use the FTD developed in previous works \cite{Xianyu:2023ytd,Fan:2024iek}. The core idea is to use the identity $\theta(\tau_1-\tau_2)+\theta(\tau_2-\tau_1)=1$ to reverse the direction of some time orderings, so that the integral (\ref{eq_NestedIntegral}) breaks into a sum of products of partially ordered integrals.  

The general procedure is as follows. We start by choosing an arbitrary site as the earliest site, and then adjust the directions of all lines, so that the time flows monotonically out of the earliest site in all resulting nested integrals. Each of these nested integrals is thus a \textit{family tree}. Like in previous works, we define a family tree to be a nested integral satisfying a \textit{partial order} condition: each site is connected to at most one in-going line. Such structure is very similar to a maternal family tree: One site can have many daughters but only one mother, with an eldest ancestor as the earliest site chosen at the beginning. We will adopt such terminology and call the earlier site mother and later site daughter. Occasionally, we also call a daughterless site a \emph{leaf} and the motherless site the \emph{root}. (After reversing a line, it may happen that a nested subgraph does not contain the root site of the whole graph. In this case, one should designate a site within the subgraph as the ``local'' root.)

It is usually convenient to use a graphic representation of these time ordering structure. We use an arrowed line pointing from the mother site to the daughter site, and a dashed line in the factorized term. Also, a useful way of bookkeeping family trees is the parentheses notation introduced in \cite{Fan:2024iek}. In short, for an $N$-site tree without branches, namely a chain, we put the sites in a time-ordered sequence, enclosed by a pair of square brackets: $\ft{12\cdots N}$. For tree with branches in which a site has multiple daughters, we enclose the sub-tree of each branch (the sub-family led by each of her daughter) by an extra pair of parentheses. Some illustrating examples of this notation are 
\begin{align}
  \label{eq_FamilyTreeExamples}
  \ft{123}
  =&~\parbox{25mm}{\vspace{-2mm}\includegraphics[width=25mm]{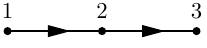}}
  =(-\ii)^3\int\prod_{i=1}^3\Big[\di\tau_i(-\tau_i)^{q_i-1}e^{\ii\omega_i\tau_i}\Big]\theta_{32}\theta_{21},\\
  \ft{1(2)(3)(4)}
  =&~\parbox{25mm}{\vspace{0.8mm}\includegraphics[width=25mm]{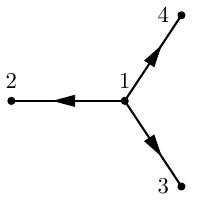}}
  =(-\ii)^4\int\prod_{i=1}^4\Big[\di\tau_i(-\tau_i)^{q_i-1}e^{\ii\omega_i\tau_i}\Big]\theta_{41}\theta_{31}\theta_{21},\\
  \ft{1(2)\big(3(4)(5)\big)}
  =&~\parbox{25mm}{\includegraphics[width=25mm]{fd_FTex_5site}}
  =(-\ii)^5\int\prod_{i=1}^5\Big[\di\tau_i(-\tau_i)^{q_i-1}e^{\ii\omega_i\tau_i}\Big]\theta_{43}\theta_{53}\theta_{31}\theta_{21}.
\end{align}
Here and below, we use $\theta_{ij}\equiv \theta(\tau_i-\tau_j)$ for short. Occasionally, we want to be general and do not specify a particular partial order. In such cases, we write $[\mathscr{P}(\wh{1}2\cdots N)]$, meaning that the $N$ sites are organized into a partial order structure $\mathscr{P}$, with the hatted site, namely $\wh{1}$ in $[\mathscr{P}(\wh{1}2\cdots N)]$, being the earliest site.

\begin{figure}[t]
\centering
\includegraphics[width=0.75\textwidth]{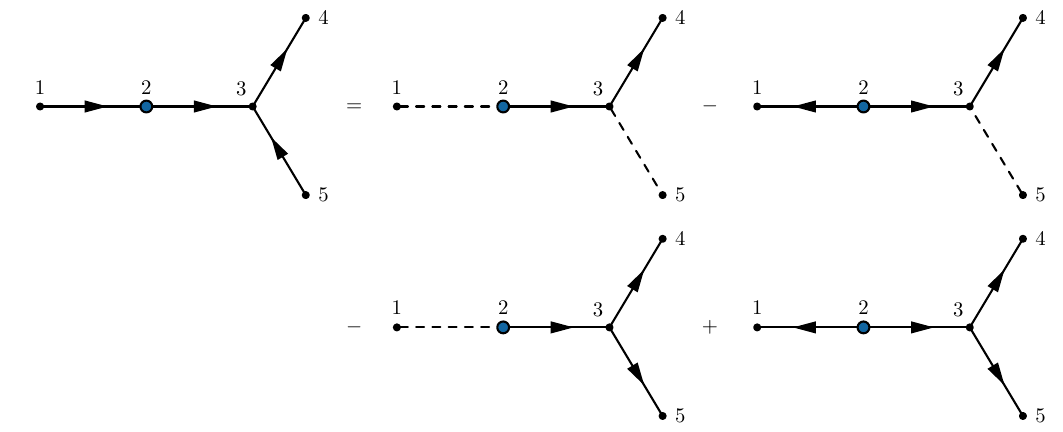}
\caption{A FTD for a five-site graph. Here we choose Site 2 as the earliest site. }
\label{fig_FTD}
\end{figure}

We provide an explicit example of the FTD to show the use of our symbolic and graphic notations:
\begin{align}
 & (-\ii)^5\int\prod_{i=1}^5\Big[\di\tau_i(-\tau_i)^{q_i-1}e^{\ii\omega_i\tau_i}\Big]\theta_{21}\theta_{32}\theta_{43}\theta_{35}\n\\
 =&~\ft{1}\ft{234}\ft{5}-\ft{2(1)(34)}\ft{5}-\ft{1}\ft{23(4)(5)}+\ft{2(1)(3(4)(5))}.
\end{align}
In graphic notations, this example is shown in Fig.\;\ref{fig_FTD}.
 
The main reason for performing FTD is that each family tree is expressible as a single hypergeometric series. As shown in previous works \cite{Xianyu:2023ytd,Fan:2024iek}, a family tree can be expressed as a series in $1/\omega_1$ where $1$ is the earliest site of the tree:
\begin{align}
  \ft{\mathscr{P}(\wh{1}2\cdots N)}=\frac{(-\ii)^N}{(\ii \omega_1)^{\wt{q}_1}} \sum_{n_2,\cdots,n_N=0}^{\infty}\Gamma(\wt{q}_1+\wh{n}_1)\prod_{j=2}^{N} \frac{\left(-\omega_j/\omega_1\right)^{n_j}}{(\wt{q}_j+\wt{n}_j)n_j!} \ .
  \label{eq_LargeMaximal_1}
\end{align}
Here we use tilde notations like $\wt{q}_j$ for the sum of $q_j$ and the $q$'s of all descendants of site $j$. This notation can be used for all ``site parameters'' such as $\wt{n}_j$ and $\wt{\omega}_j$. We also use the hat notation, such as $\wh{n}_j\equiv\wt{n}_j-n_j$. Notice that we never introduce the summation variable $n_1$ for the earliest site (Site $1$), so $\wh{n}_1$ is meaningful but $\wt{n}_1$ is not. Mathematically, the series (\ref{eq_LargeMaximal_1}) defines a class of MHF of Horn type \cite{horn1931hypergeometrische}. These functions have been identified in the literature but many of their properties remain unexplored.

For later convenience, we also introduce a symbol $\mathcal{D}(j)$ to denote the set of all descendants of Site $j$. Using this notation, we can write, for instance, 
\begin{align}
\label{eq_wtqjDef}
  &\wt q_j\equiv q_j+\sum_{i\in\mathcal{D}(j)}q_i,
  &&\wh n_j\equiv \sum_{i\in\mathcal{D}(j)}n_i.
\end{align}
Also, we will use $d_j$ to denote the number of descendants of Site $j$, namely, the number of elements of $\mathcal{D}(j)$.

In \cite{Fan:2024iek}, we also presented a series in the inverse total energy $1/\wt{\omega}_1=1/\omega_{12\cdots N}$ without a proof: 
\begin{align}
   \ft{\mathscr{P}(\wh{1}2\cdots N)}=\frac{(-\ii)^N}{(\ii \wt{\omega}_1)^{\wt{q}_1}} \sum_{n_2,\cdots,n_N=0}^{\infty}\Gamma(\wt{q}_1+n_{2\cdots N})\prod_{j=2}^{N} \left(\frac{\wt{\omega}_j}{\wt{\omega}_1}\right)^{n_j} \Gamma \left[
      \begin{matrix}
          \wt{q}_j+\wh{n}_j \\
          \wt{q}_j+\wt{n}_j+1
      \end{matrix}
      \right] \ .
      \label{eq_LargeTotal_1}
\end{align}
Once again, this defines a MHF, which looks different from but is identical to (\ref{eq_LargeMaximal_1}). Thus, equating (\ref{eq_LargeTotal_1}) with (\ref{eq_LargeMaximal_1}) generates an infinite number of transformation-of-variable identities for these hypergeometric functions. As mentioned, a family tree is always expressed as a single hypergeometric series in the above two expansions, showing that the FTD is particularly meaningful when a site energy or the total energy becomes large. In contrast, in other regions, a family tree may be expanded into the sum of several distinct hypergeometric series, which we will show later.  

\subsection{Mellin-Barnes representation of the time integrals}
\label{sec_MBRofTI}

A main goal of this paper is to study the analytical properties of general family trees. For a given family tree, we aim at identifying the location of all singularities (poles and branch points) and deriving series expansion around each of the singular points, with (\ref{eq_LargeMaximal_1}) and (\ref{eq_LargeTotal_1}) being two simple examples.  
To achieve this goal, we present a general framework to express family trees as MB integrals in many different ways. Note that this is the use of MB representation for a second time within the time integral (\ref{eq_NestedIntegral}), which should not be confused with the PMB representation introduced for the original correlator (\ref{eq_GeneralCorrelator}). 

The basic strategy is to perform the time integral layer by layer following the partial order of the graph, in either chronological or antichronological order. The time integral of a given layer yields certain special functions (due to an indefinite integral limit), which can be reduced to power functions by MB representation. In this way, the time integrals can be carried out throughout all layers, yielding a multi-fold MB integral, and this MB integral can again be finished by collecting residues of appropriate poles. 

The MB representations for various special functions used in this section are collected in App.\;\ref{app_Functions}.

\paragraph{Early-to-late approach} As mentioned, the nested time integrals of a family tree can be performed either following or against the partial order. The early-to-late approach was adopted in \cite{Xianyu:2023ytd}. Later, we will introduce another (late-to-early) approach and use it in the subsequent sections. The comparison of the two gives us useful insights about the structure of nested time integrals. So we first review the early-to-late approach before introducing the new one. 

Let us consider a general family tree $\ft{\mathscr{P}(\wh 12\cdots N)}$. Suppose that we start from the root site (Site 1) and try to integrate out $\tau_1$. Due to the time ordering, the integral range of $\tau_1$ is from $-\infty$ to $\text{min}\{\tau_{i_1},\cdots,\tau_{i_n}\}$ where $\tau_{i_1},\cdots,\tau_{i_n}$ denote the time variables of all daughters of Site 1. Therefore, the integral range of $\tau_1$ is dependent on the relative orderings of the daughters which are undetermined. This introduces unnecessary complication to the problem. To avoid it, a more proper treatment is to consider the \emph{reverse} of a family tree, defined by flipping the directions of all $\theta$-functions, $\theta_{ij}\to\theta_{ji}$ and denoted by $\mathcal{R}\ft{\mathscr{P}(\wh 12\cdots N)}$. Then, for the reversed tree, we start from a leaf, i.e., a daughterless site, say $\tau_i$, whose integral region is from $-\infty$ to the time of her mother $\tau_j$. (The terminology here is that the roles of mothers and daughters always follow the original family tree even in a reversed tree.) Since Site $i$'s mother is unique, this integral is straightforward:
\begin{align}
    \int_{-\infty}^{\tau_j} \di\tau_i\, (-\tau_i)^{q_i-1} e^{\ii \omega_i \tau_i}
    &=(-\tau_j)^{q_i}\text{E}_{1-q_i}(-\ii\omega_i \tau_j) \ .
    \label{eq_time_int_1layer_reversed}
\end{align}
The exponential integral $\text{E}_p(z)$ makes the integral over $\tau_j$ hard. As mentioned above, the solution is to take the MB representation. There are two ways to do this. First, we can resolve $\text{E}_p(z)$ completely into linear superposition of power functions:
\begin{align}
    \text{E}_{1-q}(-\ii\omega\tau)=\int_{-\ii\infty}^{+\ii\infty}\frac{\di s}{2\pi\ii} \frac{(-\tau)^{-s}}{(\ii\omega)^{s}} \Gamma \left[
    \begin{matrix}
        s,s-q \\
        s-q+1
    \end{matrix}
    \right] \ .~~~\text{(completely resolved rep)}
    \label{eq_rep_collect_none_reversed}
\end{align}
Notice that the factor $\Gamma(s-q)/\Gamma(s-q+1)=1/(s-q)$ has a single pole at $s=q$. Here we write it as a $\Gamma$ ratio as a contour prescription. That is, we demand that the MB contour goes around $s=q$ from right. In other words, we treat the pole at $s=q$ as a ``left pole'' similar to those from $\Gamma(s)$. 

Second, we can leave an exponential factor out when taking the MB representation: 
\begin{align}
    \text{E}_{1-q}(-\ii\omega\tau)=e^{\ii\omega\tau}\int_{-\ii\infty}^{+\ii\infty}\frac{\di s}{2\pi\ii} \frac{(-\tau)^{-s}}{(\ii\omega)^{s}} \Gamma \left[
    \begin{matrix}
        s,1-s,s-q \\
        1-q
    \end{matrix}
    \right] \ .~~~\text{(partially resolved rep)}
    \label{eq_rep_collect_all_reversed}
\end{align}
After taking the representation (\ref{eq_rep_collect_all_reversed}), the prefactor $e^{\ii\omega\tau}$ can be combined with a similar exponential factor in the mother site. As a result, the energy on the mother site is shifted by $\omega$, so we say it \emph{collects} the energy $\omega$ along the line. If we use this representation for the integral at every site, we will collect all energies at these sites, ultimately arriving at a total-energy expansion. It is also possible to collect energies in a more flexible manner, which we will discuss later.

From (\ref{eq_time_int_1layer_reversed}), we see that a factor of $(-\tau)^{-q_i}$ is always transmitted to the mother site (Site $j$) whatever representation for $\text{E}_{1-q}(-\ii\omega\tau)$ we choose. In our terminology, the twists are always collected throughout all sites.  

With the above points clarified, it is straightforward to transform the time integral for the reversed tree into a MB integral: We integrate the time variables of all leaves layer by layer. At each site, we have an integral of the form (\ref{eq_time_int_1layer_reversed}), and we use either (\ref{eq_rep_collect_none_reversed}) or (\ref{eq_rep_collect_all_reversed}) to resolve the exponential integral into powers (and exponential factors). Since all twists are collected, and since we have a MB kernel $(-\tau_j)^{-s_j}$ from each site, when we integrate $\tau_j$, the corresponding power factor is $(-\tau_j)^{\wt q_j-\wh s_j}$. (As a reminder, $\wt{q}_j,\wt{s}_j$ represent the sum of $q_j,s_j$ and their descendants, while $\wh{s}_j\equiv\wt{s}_j-s_j$.) On the other hand, the exponential factor at each site depends our choices of (\ref{eq_rep_collect_none_reversed}) and (\ref{eq_rep_collect_all_reversed}) for all her descendant sites. After finishing all the time integrals, we arrive at a multi-fold MB integral, 
\begin{align}
    \mathcal{R}\ft{\mathscr{P}(\wh{1}2\cdots N)}=\frac{(-\ii)^N}{(\ii \Omega_1)^{\wt{q}_1}}\int_{-\ii\infty}^{+\ii\infty}\frac{\di s_2}{2\pi\ii}\cdots \frac{\di s_N}{2\pi\ii}  \Gamma(\wt{q}_1-\wh{s}_1) \prod_{j=2}^{N} \left(\frac{\Omega_1}{\Omega_j}\right)^{s_j} \mathcal{I}_j\left(s_j,\wt{q}_j-\wh{s}_j\right) \ .
\label{eq_RevTreeMB}
\end{align}
In this equation, $\Omega_j$ represent the sum of energies collected to site $j$. For instance, if we choose to use (\ref{eq_rep_collect_none_reversed}) to represent $\text{E}_p(z)$ at all sites, and thus no energies are collected, we have $\Omega_j=\omega_j$. Alternatively, if we choose to use (\ref{eq_rep_collect_all_reversed}) and collect all energies, then we have $\Omega_j=\wt\omega_j$. Clearly, there are many more choices between these two extreme cases. 

In (\ref{eq_RevTreeMB}), the factor $\Gamma(\wt q_1-\wh s_1)$ arises from integrating out the time of the root site (Site 1). This factor remains invariant whatever representations we choose for other sites. This factor plays a role when we study the singularity structure of the family tree below. We call it the \emph{root $\Gamma$} in the following.

On the other hand, the factor $\mathcal{I}_j(s,q)$ ($j=2,\cdots,N$) depends on our choices of MB representations but is always a fraction of $\Gamma$ products. From 
(\ref{eq_rep_collect_none_reversed}) and (\ref{eq_rep_collect_all_reversed}), we see that the integrand can be chosen from either of the following: 
\begin{align}
    \Gamma \left[
    \begin{matrix}
        s,s-q \\
        s-q+1
    \end{matrix}
    \right] \ , \quad
    \Gamma \left[
    \begin{matrix}
        s,1-s,s-q \\
        1-q
    \end{matrix}
    \right] \ .
 \label{eq_RevGammaProducts}
\end{align}
Taking into account the shift $q_j\to \wt q_j-\wh s_j$ as discussed above, we can easily write down the expression for $\mathcal{I}_{j}$. For instance, if we choose (\ref{eq_rep_collect_none_reversed}) and collect no energies, the integrand $\mathcal{I}_{j}$ is:
\begin{align}
    \mathcal{I}_j(s_j,\wt{q}_j-\wh{s}_j)=\Gamma \left[
    \begin{matrix}
        s_j,s_j-(\wt{q}_j-\wh{s}_j) \\
        s_j-(\wt{q}_j-\wh{s}_j)+1
    \end{matrix}
    \right]
    =\Gamma \left[
    \begin{matrix}
        s_j,\wt{s}_j-\wt{q}_j \\
        \wt{s}_j-\wt{q}_j+1
    \end{matrix}
    \right] \ .
\end{align}

To conclude the discussion of this approach, we emphasize that integrating from an earlier site to a later one requires a reversed family tree to proceed, as is done in previous works.

\paragraph{Late-to-early approach}
Now we introduce a second approach in which we perform the time integrals in the antichronological order, starting from leaves and working towards the root. This makes use of a new MB representation of special functions which is equivalent to previous ones. However, we don't need to reverse the family tree in this case, and thus the procedure is somewhat simplified. 

Since we are integrating from the leaves, a time variable in this case is always bounded by her mother's time, and the required integral now is:
\begin{align}
    \int_{\tau_j}^0 \di\tau_i\, (-\tau_i)^{q_i-1} e^{\ii \omega_i \tau_i}
    &=(-\tau_j)^{q_i} \ _1\mathcal{F}_1 \left[
    \begin{matrix}
        q_i \\
        q_i+1
    \end{matrix}
    \ \middle| \ +\ii\omega_i \tau_j
    \right] \ .
    \label{eq_time_int_1layer}
\end{align}
Here ${}_1\mathcal{F}_1$ is the dressed version of Kummer confluent hypergeometric; See App.\;\ref{app_Functions} for the definition. Clearly, (\ref{eq_time_int_1layer}) is the same as (\ref{eq_time_int_1layer_reversed}) up to a time integral from $-\infty$ to 0 which yields a $\Gamma$ function. This connection can be made more transparent with MB representation. Indeed, we can make use of the following MB representation for ${}_1\mathcal{F}_1$:
\begin{align}
    \ _1\mathcal{F}_1 \left[
    \begin{matrix}
        q \\
        q+1
    \end{matrix}
    \ \middle| \ +\ii\omega \tau
    \right]=\int_{-\ii\infty}^{+\ii\infty}\frac{\di s}{2\pi\ii}\frac{(-\tau)^{-s}}{(\ii\omega)^{s}} \Gamma\left[
    \begin{matrix}
        q-s,s \\
        q-s+1
    \end{matrix}
    \right] \ .~~~\text{(completely resolved rep)}
    \label{eq_rep_collect_none}
\end{align}

\begin{figure}[t]
\centering
\includegraphics[width=0.95\textwidth]{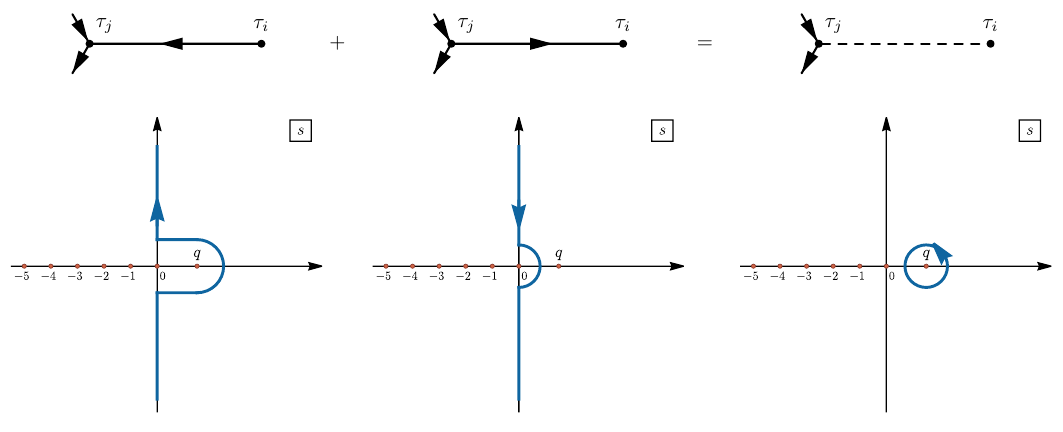}
\caption{The relation between the time-ordered integral and the contour prescription in the MB representation. The left two graphs show the time integral (\ref{eq_time_int_1layer_reversed}) and the corresponding MB representation in (\ref{eq_rep_collect_none_reversed}); The middle graphs show the time integral (\ref{eq_time_int_1layer}) and the corresponding MB representation in (\ref{eq_rep_collect_none}); The right graphs shows the non-time-ordered integral, which is given by the residue of the single pole at $s=q$ in the MB representation. }
\label{fig_contour}
\end{figure}

Comparing this with (\ref{eq_rep_collect_none_reversed}), we see that they are almost the same (up to a minus sign). The only difference is that the single pole at $s=q$ is treated differently. Strictly speaking, the integral path of (\ref{eq_rep_collect_none_reversed}) lies to the right of the single pole (the pole is treated as left pole) while the path of (\ref{eq_rep_collect_none}) lies to the left of the single pole (the pole is treated as right pole). In fact, the residue of that single pole gives exactly a non-time-ordered integral, where we integrate from $-\infty$ to $0$. This is a useful insight: the single pole in this representation corresponds to a factorized term with no time ordering. We show this relation in Fig.\;\ref{fig_contour}.

Similar to the previous approach, we can also choose to MB-represent ${}_1\mathcal{F}_1$ up to an exponential factor: 
\begin{align}
    \ _1\mathcal{F}_1 \left[
    \begin{matrix}
        q \\
        q+1
    \end{matrix}
    \ \middle| \ +\ii\omega \tau
    \right]=e^{\ii\omega \tau} \int_{-\ii\infty}^{+\ii\infty}\frac{\di s}{2\pi\ii}\frac{(-\tau)^{-s}}{(\ii\omega)^{s}}(-1)^{s} \Gamma\left[
    \begin{matrix}
        s,1-s,q \\
        q-s+1
    \end{matrix}
    \right] .~~\text{(partially resolved rep)}
    \label{eq_rep_collect_all}
\end{align}
Again, for both representations we transmit a $(-\tau)^{-s}$ to the next site and the $q_i$ at every site is shifted to $\wt{q}_i-\wh{s}_i$. On the other hand, we can choose to collect or not to collect energies at any site. Thus, following the same considerations as before, we can finish all the time integrals and arrive at a similar MB integral:
\begin{align}
\label{eq_FTMB}
    \ft{\mathscr{P}(\wh{1}2\cdots N)}=\FR{(-\ii)^N}{(\ii \Omega_1)^{\wt{q}_1}}\int_{-\ii\infty}^{+\ii\infty}\frac{\di s_2}{2\pi\ii}\cdots \frac{\di s_N}{2\pi\ii}  \Gamma(\wt{q}_1-\wh{s}_1) \prod_{j=2}^{N} \left(\FR{\Omega_1}{\Omega_j}\right)^{s_j} \mathcal{I}_j\left(s_j,\wt{q}_j-\wh{s}_j\right) \ .
\end{align}
Again, the energy variable $\Omega_j$ depends on how we choose to collect the energy at each descendant site of $j$. Also, we have the root $\Gamma$, $\Gamma(\wt q_1-\wh s_1)$, which is robust. The factor $\mathcal{I}_j(s,q)$ can be found from representations (\ref{eq_rep_collect_none}) and (\ref{eq_rep_collect_all}), and can be either of the following two:
\begin{align}
{\color{RoyalBlue}
    \Gamma\left[
    \begin{matrix}
        q-s,s \\
        q-s+1
    \end{matrix}
    \right]} \ , \quad
{\color{BrickRed}    (-1)^{s} \Gamma\left[
    \begin{matrix}
        s,1-s,q \\
        q-s+1
    \end{matrix}
    \right]} \ .
    \label{eq_GammaProducts}
\end{align}
Since both representations will be used below, we use a color code for easier keeping track of our choices, in which the completely resolved representation is put in blue and the partially resolved one in red.

\paragraph{Multi-fold Mellin-Barnes integral} 
The upshot of the discussion above is that we can turn the time integral (\ref{eq_NestedIntegral}) for any family tree into a MB integral (\ref{eq_FTMB}). The remaining problem is to finish the MB integral. Since the MB integral is essentially a pole-collecting procedure over $\Gamma$ products, we anticipate that a possible final result will be a multivariate hypergeometric series in the complex space of energies. 

At this point, a property of the MB integrand in (\ref{eq_FTMB}) turns out important for us. That is, all MB variables $s_j$ $(j=2,\cdots, N)$ in the MB integrand are \emph{balanced}, in the sense that the sum of all arguments of $\Gamma$ functions from the numerator minus that from the denominator is independent of any $s$ variable. Although this property is trivial to prove from the explicit expressions in (\ref{eq_FTMB}) and (\ref{eq_GammaProducts}), it ensures that the MB integrand does not get exponentially large when any $s$ goes to infinity. (One can see this point by applying Stirling formula for all $\Gamma$ factors with large arguments.) In other words, the MB integrand is polynomially bounded at infinity. 

As a result, the convergence of the resulting series is largely determined by the power factors $(\Omega_1/\Omega_j)^{s_j}$, and thus the domain of convergence is typically finite in the energy space. This shows that we need to develop different series expansions in different regions, and therefore, the procedure of pole collection is unavoidably intertwined with the choice of parameter region. This is a typical problem of multi-fold MB integral with balanced variables, and the solution is known in the literature which makes use of a geometric conic hull construction or its variations \cite{Ananthanarayan:2020fhl,Banik:2023rrz}. We will show that it is easy to carry out the MB integral following a simple algebraic procedure in the following sections. We show the equivalence between our algebraic procedure and the conic hull method in App.\;\ref{app_MB}.

The essence of our algebraic procedure is the following: 
\begin{enumerate}
  \item Since the resulting series expression has a finite convergence domain controlled by small energy ratios, we first pick up an order of the magnitude of all energy variables (either site energy $|\omega_j|$ or tilded energy $|\wt\omega_j|$ or capitalized energy $|\Omega_j|$). For our purposes a total order of energy is sufficient though not always necessary. 
  \item Once we have picked up an energy order, all energy ratios in the MB integrand, such as $|\Omega_1/\Omega_j|$ in (\ref{eq_FTMB}), will be definitely $>1$ or $<1$. Meanwhile, all poles of $s_j$ are from $\Gamma$ factors in (\ref{eq_FTMB}), which are either ``left poles'' (from factors like $\Gamma(\cdots+s_j)$) or ``right poles'' (from factors like $\Gamma(\cdots-s_j)$). Then, for $(\Omega_1/\Omega_j)^{s_j}$ to yield a convergent series, we should pick up left poles of $s_j$ for $|\Omega_1/\Omega_j|>1$ and right poles for $|\Omega_1/\Omega_j|<1$. 
  \item It would be simple if each of the relevant $\Gamma$ factors (i.e., the $\Gamma$ factors that we want to pick up poles from) has only one MB variable in the argument, such as $\Gamma(\cdots\pm s_j)$ where ``$\cdots$'' does not contain any $s$. In this case, we only need to pick up poles of $s_j$ layer by layer. This is the case of infinite energy limits to be considered in Sec.\;\ref{sec_LargeEnergy}.
  \item However, we do encounter cases where relevant $\Gamma$ factors contain multiple MB variables such as the root $\Gamma$. In such cases, we need to be more careful to enumerate all correct combinations of $\Gamma$ factors to avoid overcounting or choosing wrong poles. Still, we can perform the integral in a layer-by-layer fashion. We will show the details of this procedure with many examples in Sec.\;\ref{sec_SmallEnergy}.
\end{enumerate} 
Following the above procedures, we will obtain series representations of family trees in various distinct and finite domains. To end this section, we make an important remark that the resulting series may not always be convergent for entire parameter region specified by the total order. Therefore, when we use conditions like ``$|\Omega_1/\Omega_j|<1$'' to find a series, what we are sure is that the series will be convergent when $|\Omega_1/\Omega_j|<\ep$ for some $\ep$ where $\ep\in(0,1]$ and may be dependent on other energy ratios. So the convergence domain is finite, but determining its exact boundary requires a more sophisticated study. We leave this question to a future work. In the following, we will sometimes refer to a series to be \emph{formally convergent} in $|\Omega_1/\Omega_j|<1$ in the sense that it is convergent in a finite subdomain of $|\Omega_1/\Omega_j|<1$.

\section{Analytical Structure of Family Trees in Energy Space}
\label{sec_SingularityProof}

In this section, we describe the analytical structure of an arbitrary family tree. Specifically, we will show that a general family tree, when viewed as a function of complex energies $\omega_i$, is analytic everywhere except for a finite set of complex codimension-1 singularities. All these singularities are poles or branch points of finite degrees, and thus the family tree can be expanded as a Laurent series at each of these singularities with finite convergence domain. 

All singularities of a family tree come in two types. Singularities of the first type appear when some of the energy variables are sent to infinity, and we call them \textit{infinite-energy singularities}. The second type appears when the sum of several energies goes to zero, and we call them \textit{zero-energy singularities}. The zero-energy limits are in the interior of the energy space $\mathbb{C}^N$ and are trivial. To study the infinite-energy limits, we need to properly compactify the energy space. It turns out that an appropriate choice is to compactify $\mathbb{C}^N$ into $\mathbb{CP}^{N}$. We address this issue in App.\;\ref{app_CPN}.

The strategy to identify all singularities of a family tree is to convert the time integral representation (\ref{eq_NestedIntegral}) into an energy integral via Fourier transform, so that the integrand involves only powers and fractions of energy variables. Then, a standard Landau analysis applies, with which we can conveniently show that all singularities are of endpoint type.\footnote{Alternatively, we can stick to the time integral representation. However, the essential singularities from exponential functions such as $e^{\ii\omega_i\tau_i}$ make the original Landau analysis not directly applicable. Nevertheless, the singularity structure is still determined by these exponential functions in the early-time limit. Therefore, we can use a slightly modified Landau analysis adapted to these singularities, as detailed in \cite{Liu:2024xyi}.}

\paragraph{Energy integrals}
To apply the Landau analysis, we go to the Fourier space and rewrite an arbitrary family tree as an energy integral. As shown in \cite{Fan:2024iek}, this is achieved in two steps, which we briefly review here.

The first step is to perform the ``family chain'' decomposition for the family tree, by repeatedly using a simple identity of Heaviside $\theta$ function: $\theta_{21}\theta_{31}=\theta_{32}\theta_{21}+\theta_{23}\theta_{31}$. In practice, this amounts to recursively taking the shuffle products of all subfamilies within a family tree until all branchings disappear. Then every term in the result is a family tree without branchings, and is called a \emph{family chain}. In other words, a family chain decomposition of a family tree is simply the sum of all distinct chains whose total order is consistent with the family tree's partial order. This procedure can be compactly denoted as: 
\begin{align}
    \ft{\mathscr{P}(1\cdots N)}=\ft{\bigshuffle\mathscr{P}(1\cdots N)} \ .
\end{align}
More details can be found in \cite{Fan:2024iek}. Here we show this decomposition with an example: 
\begin{align}
    \ft{1(2)(3(4)(5))}
    =&~\ft{1(2)\shuffle(3(4)\shuffle (5))}\n\\
    =&~\ft{12345}+\ft{13245}+\ft{13425}+\ft{13452} \notag \\
    &\quad +\ft{12354}+\ft{13254}+\ft{13524}+\ft{13542} \ .
\end{align} 
With our diagrammatic notation, this decomposition can be expressed as:
\bge
\parbox{0.85\textwidth}{
\includegraphics[width=0.85\textwidth]{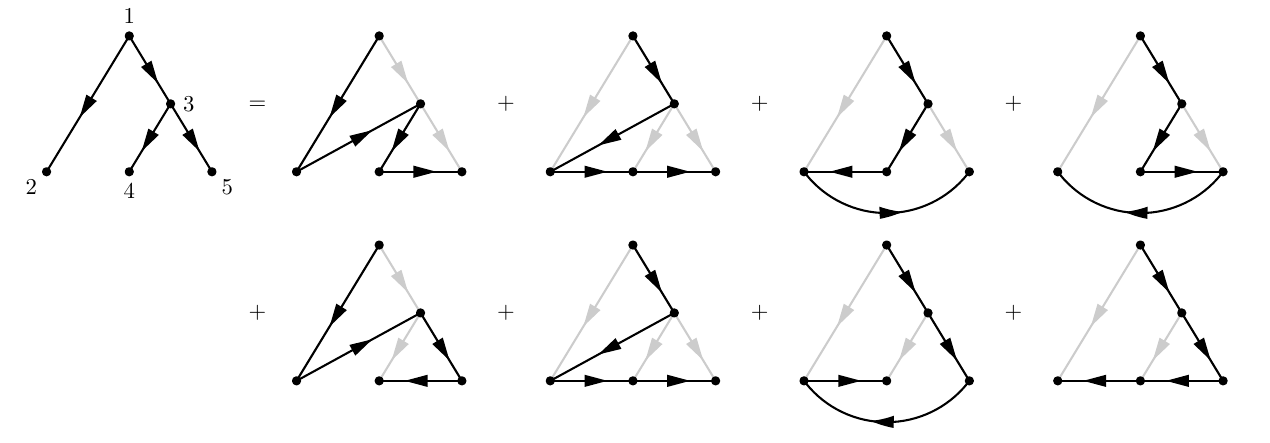}}
\ede

The second step is to perform the Fourier transform of the time $\tau$ to the dual energy variable $\ep$. In particular, the power functions $(-\tau)^{q-1}$ in the time integral (\ref{eq_NestedIntegral}) can be transformed as:
\begin{align}
    (-\tau)^{q-1}=\frac{\ii^{1-q}}{\Gamma(1-q)} \int_0^{+\infty} \di \epsilon\, e^{\ii \epsilon \tau} \epsilon^{-q} \ .
\end{align}
Now, with the above Fourier transform to all power factors, the total order of a family chain makes it trivial to finish all time integrals, after which we get the following energy integral representation for an $N$-site family chain:
\begin{align}
\label{eq_ChainEnergyInt}
    \ii^N\ft{12\cdots N}= \int_0^\infty \prod_{i=1}^{N}\left(\frac{\di \epsilon_i (\ii\epsilon_i)^{-q_i}}{\Gamma(1-q_i)}\right)\FR{1}{(\ep_1+\omega_1)(\ep_{12}+\omega_{12}) \cdots (\ep_{12\cdots N}+\omega_{12\cdots N})} \ .
\end{align}

\paragraph{Landau analysis} We can perform a Landau-like analysis to the energy integral (\ref{eq_ChainEnergyInt}). For nonexperts, the Landau analysis is a set of lemmas to identify all candidates of singularities for a contour integral over a complex function, provided that the complex function in question involves only poles. Typically, the location of poles in the $\ep$-space depends on external parameters $\omega_i$. More specifically, a pole, as a singularity of the integral, means that a generally convergent integral fails to converge at this particular point. In essence, Landau analysis says that this divergence can happen only when the poles of the integrand hits the integration contour in a way that cannot be avoided by any contour deformation. In practice, there are two possibilities: Either a pole in $\ep$-space approaches an endpoint of the integration contour, or two poles simultaneously pinch the contour from two sides. The two possibilities are respectively called endpoint singularities and pinched singularities. Generalizations of this picture to higher dimensions are also known. See \cite{Eden:1966dnq} for a more detailed introduction. 

In our cases, the integrand of (\ref{eq_ChainEnergyInt}) has both poles and branch points and does not meet the requirement of the Landau analysis. Nevertheless, all branch points are inert in the sense that they sit at $\ep_i=0$ and $\ep_i=\infty$ and do not move with ``external'' variables $\omega_i$. Thus, the Landau analysis is still applicable for finding out all singularities, and the effects of these inert branch points are to give nontrivial monodromy to the resulting singularities.

Clearly, there are $N$ poles in the integrand  of (\ref{eq_ChainEnergyInt}), at $\ep_1=-\omega_1$, $\ep_{12}=-\omega_{12},\cdots, \ep_{1\cdots N}=-\omega_{1\cdots N}$. So, we get endpoint singularities of the family chain when any successive sum of energies from the root site goes to zero or infinity, namely, when any of $\{\omega_{1},\omega_{12},\cdots,\omega_{1\cdots N}\}$ goes to zero or infinity. According to our classification, when any one of these $N$ variables goes to zero, we call it a zero-energy singularity, and when it goes to infinity, we call it an infinite-energy singularity. 

Now, we can read off all singularities of a family tree: They are simply given by singularities of all family chains after the chain decomposition. We call the sum of all energies of a connected subgraph a partial energy. Then, the above analysis leads to the following result:
\begin{keytext}
\textbf{Singularities of a family tree:} \emph{all possible singularities of a family tree come from root-bearing partial energies going to zero or infinity}.
\end{keytext}
 In particular, since the family tree is a subgraph of itself, the zero and infinity total-energy singularities are included.

\begin{figure}[t]
\centering
\includegraphics[width=0.95\textwidth]{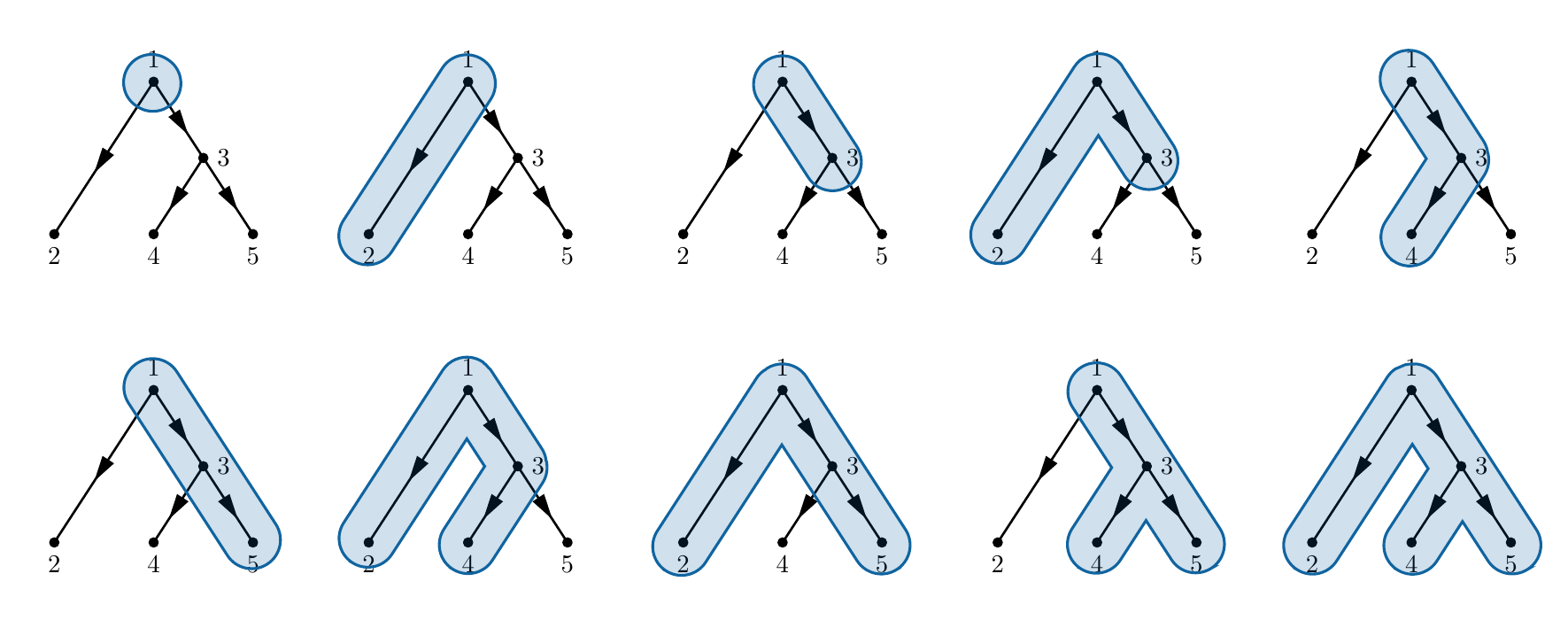}
\caption{All subgraphs of the family tree $\ft{1(2)(3(4)(5))}$ which give rise to singularities when the total-energy of the subgraph goes to zero or infinity. }
\label{fig_singularity}
\end{figure}

Once again, we use $\ft{1(2)(3(4)(5))}$ as an example. This family tree has singularities when any of the following variables goes to zero or infinity: 
\begin{align}
\label{eq_5treeSing}
  &\omega_1, &&\omega_{12}, &&\omega_{13}, &&\omega_{123}, &&\omega_{134}, &&\omega_{135}, &&\omega_{1234}, &&\omega_{1235}, &&\omega_{1345}, &&\omega_{12345}=0\text{ or }\infty.
\end{align}
The subgraphs corresponding to all these partial energies are shown in Fig.\;\ref{fig_singularity}.

\section{Series Representations at Infinite-Energy Singularities}
\label{sec_LargeEnergy}

In this section, we study the series representations for all infinite-energy singularities. Some special cases of this class have been reviewed in Sec.\;\ref{sec_FTD}, including the infinite individual-energy and total-energy limits, previously called maximal energy and total-energy expansions, respectively. Below, we will provide new and simplified derivations for both of them. Furthermore, we will also derive series representations in the infinite limit of an arbitrary root-bearing connected partial energy, which are new results. In all cases, the series representations are rather simple in the sense that there is only one hypergeometric series involved at a time, which justifies the family tree decomposition in the first place. 

\subsection{Large single-energy series}

For a generic family tree $\ft{\mathscr{P}(\wh 12\cdots N)}$ with Site 1 the root, the large single-energy series, called the maximal energy representation in \cite{Fan:2024iek}, has a finite convergent domain around $\omega_1 \to \infty$. With all preparations in Sec.\;\ref{sec_MBRofTI}, we can directly get this series from the MB representation (\ref{eq_FTMB}). For all sites, we use the representation (\ref{eq_rep_collect_none}) and do not collect any energies. Then, the general MB integral becomes: 
\begin{align} 
    \ii^N\ft{\mathscr{P}(\wh{1}2\cdots N)}=\int_{-\ii\infty}^{+\ii\infty}\frac{\di s_2}{2\pi\ii}\cdots \frac{\di s_N}{2\pi\ii}\frac{\Gamma(\wt{q}_1-\wh{s}_1)}{(\ii \omega_1)^{\wt{q}_1}}  \prod_{j=2}^{N} {\color{RoyalBlue}\Gamma \left[
    \begin{matrix}
        \wt{q}_j-\wt{s}_j,s_j \\
        \wt{q}_j-\wt{s}_j+1
    \end{matrix}
    \right] }
    \Big(\FR{\omega_1}{\omega_j}\Big)^{s_j} \ .
\end{align}
We are considering the infinite $\omega_1$ limit, around which the energy ratio $|\omega_1/\omega_j|>1$ for all $j=2,\cdots,N$. This shows that we should choose all ``left poles'' from $\Gamma(s_j)$ factors to finish the MB integral. Collecting all residues at these poles ($s_j=-n_j$), we arrive at a series: 
\begin{align}
    \ft{\mathscr{P}(\wh{1}2\cdots N)}=\FR{(-\ii)^N}{(\ii \omega_1)^{\wt{q}_1}} \sum_{n_2,\cdots,n_N=0}^{\infty}\Gamma(\wt{q}_1+\wh{n}_1)\prod_{j=2}^{N} \frac{\left(-\omega_j/\omega_1\right)^{n_j}}{(\wt{q}_j+\wt{n}_j)n_j!} \ .
    \label{eq_LargeMaximalSeries}
\end{align}
This is exactly the maximal energy representation we obtained in previous works. The formal convergent region for this series is clearly $|\omega_j/\omega_1|<1$ for all $j=2,\cdots,N$, while the true convergent region is a finite subset of the formal convergent region. In particular, the convergence of this series does not require any particular ordering among $\omega_j$ for $j\neq 1$. We leave a more precise characterization of the real convergent region to a future work, which may be done by a Stirling expansion. Here we only mention that the point of infinite energy limit ($\omega_1\to\infty$ and all other $\omega_j$ held finite and fixed) lies in the interior of the real convergent region.

\subsection{Large total-energy series}
\label{sec_LargeTotalSeries}
The large total-energy series (previously called the total-energy representation) converges at $\wt\omega_1=\omega_{1\cdots N}\to \infty$ and corresponds to collecting all energies on all sites. Thus, in (\ref{eq_FTMB}), we have $\Omega_j=\wt\omega_j$ for all $j=1,\cdots,N$. Also substituting in the partially resolved representation (\ref{eq_rep_collect_all}), we arrive at the following explicit form of the MB integral: 
\begin{align}
    \ii^N\ft{\mathscr{P}(\wh{1}2\cdots N)}=\int_{-\ii\infty}^{+\ii\infty}\frac{\di s_2}{2\pi\ii}\cdots \frac{\di s_N}{2\pi\ii}\FR{\Gamma(\wt{q}_1-\wh{s}_1)}{(\ii \wt{\omega}_1)^{\wt{q}_1}}  \prod_{j=2}^{N} \left(-\frac{\wt{\omega}_1}{\wt{\omega}_j}\right)^{s_j} {\color{BrickRed}\Gamma \left[
    \begin{matrix}
        s_j, 1-s_j, \wt{q}_j-\wh{s}_j \\
        \wt{q}_j-\wt{s}_j+1
    \end{matrix}
    \right]} \ .    
\label{eq_MBITotalEnergy}
\end{align}
Similarly, since we are considering the infinite total-energy limit $\wt\omega_1\to \infty$, we can fix the formally convergent region to be $|\wt{\omega}_1|>|\wt{\omega}_j|$ for all $j=2,\cdots,N$. This is clearly true for any point in the interior of the physical region where all $\omega_j>0$. Then, since all power factors $|\wt\omega_1/\wt\omega_j|>1$, we can get convergent series by picking up all poles from all $\Gamma(s_j)$. With $s_j=-n_j$, we get:
\begin{align}
\label{eq_LargeTEseries}
    \ft{\mathscr{P}(\wh{1}2\cdots N)}=\FR{(-\ii)^N}{(\ii \wt{\omega}_1)^{\wt{q}_1}} \sum_{n_2,\cdots,n_N=0}^{\infty}\Gamma(\wt{q}_1+\wh{n}_{1})\prod_{j=2}^{N}  \FR{(\wt\omega_j/\wt\omega_1)^{n_j}}{(\wt q_j+\wh n_j)_{n_j+1}}\ .
\end{align}
Again, we recovered the total-energy representation given in the previous works \cite{Fan:2024iek} without a proof. The computation here provides a simple proof.

\subsection{Large partial-energy series}
\label{sec_LargePartialSeries}
Now we proceed to a more general case and expand the family tree at large partial energy. While it is possible to consider an arbitrary partial energy (namely, the total energy of an arbitrary subgraph), we can limit ourselves to  root-bearing partial energies (as shown in Fig.\;\ref{fig_singularity}) without loss of generality, since any partial energy can be converted to a root-bearing partial energy by a further family-tree decomposition. So, we only consider the singular partial energies here, with the previous two subsections being special cases of the results here. 

To begin with, let us choose an arbitrary root-bearing subgraph $\mathcal{C}$ of a family tree, and consider the limit of its total energy being large, while all energies outside of this subgraph are held finite and fixed. Then, we should use the partially resolved representation (\ref{eq_rep_collect_all}) for all but the root sites within the subgraph $\mathcal{C}$, and use the totally resolved representation (\ref{eq_rep_collect_none}) for all sites outside of $\mathcal{C}$. Then, in the MB integral (\ref{eq_FTMB}), we have $\Omega_j=\omega_j$ for all $j\notin\mathcal{C}$, but $\Omega_j$ for all $j\in\mathcal{C}$ are nontrivial: it corresponds to the sum of all energies of Site $j$ and her descendants \emph{within} the subgraph $\mathcal{C}$.

Suppose the subgraph $\mathcal{C}$ has $M$ sites with $1\leq M\leq N$. Then, without loss of generality, we always have the freedom to relabel the family tree such that the first $M$ labels $(1,\cdots,M)$ are given to the $M$ sites of $\mathcal{C}$ and the root site is always Site 1. This relabeling is not necessary here but will be convenient later when we consider zero partial-energy limits. 

With the above points clarified, we see that the MB integral (\ref{eq_FTMB}) now takes the following form:
\begin{align}
    \ii^N\ft{\mathscr{P}(\wh{1}2\cdots N)}=\int_{-\ii\infty}^{+\ii\infty}\frac{\di s_2}{2\pi\ii}\cdots \frac{\di s_N}{2\pi\ii}\frac{\Gamma(\wt{q}_1-\wh{s}_1)}{(\ii \Omega_1)^{\wt{q}_1}} & \prod_{j=2}^M {\color{BrickRed} (-1)^{s_j}\Gamma \left[
    \begin{matrix}
        s_j, 1-s_j, \wt{q}_j-\wh{s}_j \\
        \wt{q}_j-\wt{s}_j+1
    \end{matrix}
    \right] }\Big(\FR{\Omega_1}{\Omega_j}\Big)^{s_j}  \notag \\
    \times & \prod_{k=M+1}^N {\color{RoyalBlue}\Gamma \left[
    \begin{matrix}
        \wt{q}_k-\wt{s}_k,s_k \\
        \wt{q}_k-\wt{s}_k+1
    \end{matrix}
    \right]} \Big(\FR{\Omega_1}{\omega_k}\Big)^{s_k} \ .    
\end{align}

Similar to the previous two subsections, to find a series convergent in the limit $\Omega_1\to \infty$, we choose the energy region $|\Omega_1|>|\Omega_j|$ for all $j\in\mathcal{C}$ and $|\Omega_1|>|\omega_j|$ for all $j\notin\mathcal{C}$. It is again trivial to see that we only need to pick up all poles from $\Gamma(s_j)$ for all $j=2,\cdots,N$. After doing so, we arrive at the following series: 
\begin{keyeqn}
\begin{align}
\label{eq_InfPESeries}
    \ft{\mathscr{P}(\wh{1}2\cdots N)}=\FR{(-\ii)^N}{(\ii \Omega_1)^{\wt{q}_1}} \sum_{n_2,\cdots,n_N=0}^{\infty}\Gamma(\wt{q}_1+\wh n_{1}) & \prod_{j=2}^M\bigg[   \FR{(\Omega_j/\Omega_1)^{n_j}}{(\wt q_j+\wh n_j)_{n_j+1}} \bigg]
 \prod_{k=M+1}^N \bigg[ \frac{(-\omega_k/\Omega_1)^{n_k}}{(\wt{q}_k+\wt{n}_k)n_k!} \bigg]\ .
\end{align}
\end{keyeqn}
This series expansion can be viewed as a mixture or generalization of the previous two subsections. Indeed, we can recover the results in the previous two subsections by taking $\mathcal{C}=\{1\}$ and $\mathcal{C}=\{1,\cdots,N\}$, respectively.

\section{Series Representations at Zero-Energy Singularities}
\label{sec_SmallEnergy}
In this section, we turn to zero-energy singularities defined in Sec.\;\ref{sec_SingularityProof} and derive series representations around all these points. As we shall see, these series representations are much more complicated than the series at large partial energies. So, we will start from simpler examples to build intuition before considering the most general case. 

We begin with the small total-energy series in Sec.\;\ref{sec_SmallTE}, as the total energy is a (simple) special case of partial energies. In this part, we first consider a family tree without branching (a family chain) and then move on to trees with branchings. These examples will give us sufficient insight to work out the general small total-energy series which is summarized in (\ref{eq_TEseries}). Then, in Sec.\;\ref{sec_SmallPE}, we consider general small partial-energy limits, again starting from a simple example and finally arriving at the most general small partial-energy series in (\ref{eq_SmallPESeries}).

\subsection{Small total-energy series}
\label{sec_SmallTE}

As a starting point, we consider the series expansion at small total energy. We collect energies at all sites and arrive at the MB integral (\ref{eq_MBITotalEnergy}) which we rewrite here again for convenience:
\begin{align}
    \label{eq_TotalIntMB}
    \ii^N\ft{\mathscr{P}(\wh{1}2\cdots N)}=\int_{-\ii\infty}^{+\ii\infty}\frac{\di s_2}{2\pi\ii}\cdots \frac{\di s_N}{2\pi\ii}\frac{\Gamma(\wt{q}_1-\wh{s}_1)}{(\ii \wt{\omega}_1)^{\wt{q}_1}}  \prod_{j=2}^{N} \left(-\frac{\wt{\omega}_1}{\wt{\omega}_j}\right)^{s_j} {\color{BrickRed}\Gamma \left[
    \begin{matrix}
        s_j, 1-s_j, \wt{q}_j-\wh{s}_j \\
        \wt{q}_j-\wt{s}_j+1
    \end{matrix}
    \right]  } \ .  
\end{align}
However, contrary to the situation of Sec.\;\ref{sec_LargeTotalSeries}, here we are considering $\wt\omega_1\to 0$ limit, which suggests us to collect right poles. 

It is at this point that we encounter two new complications. 

First, the sum of the arguments of two $\Gamma$ functions in the numerator $1-s_j,\wt{q}_j-\wh{s}_j$ is exactly the argument of the $\Gamma$ function in the denominator $\wt{q}_j-\wt{s}_j+1$. In other words, the three $\Gamma$ factors combine into a Euler B function: $\Gamma(1-s_j)\Gamma(\wt q_j-\wh s_j)/\Gamma(\wt q_j-\wt s_j+1)=\text{B}(1-s_j,\wt q_j-\wh s_j)$. This means that, when executing the pole collecting algorithm, we should never simultaneously pick up the poles from the two upstairs $\Gamma$ factors: These are fake poles whose residues are zero due to the $\Gamma$ downstairs. 

Second, there are $\Gamma$ factors whose arguments involve sums of several Mellin variables, which is the case for the root $\Gamma$: $\Gamma(\wt q_1-\wh s_1)=\Gamma(q_{1\cdots N}-s_{2\cdots N})$. Were we to collect the poles of the root $\Gamma$ to integrate out one Mellin variable (say, $s_2$), we should set $s_2=q_{1\cdots N}-s_{3\cdots N}+n_2$ with $n_2\in\mathbb{N}$. This has two effects: 1) The energy ratios are recombined; 2) Some left poles in the original MB integrand flip to right poles and vice versa. Due to these effects, additional caution must be exercised. Below we use a simple yet nontrivial example to show these points.

\paragraph{Four-site chain}
Consider the family chain $\ft{1234}$. This is the simplest nontrivial case with all general structures we want to show. The MB integral is now 
\begin{align}
\label{eq_4chainMBI}
    &\ft{1234}=\frac{1}{(\ii\omega_{1234})^{q_{1234}}}\int_{s_2,s_3,s_4}\left(-\frac{\omega_{1234}}{\omega_{234}}\right)^{s_2} \left(-\frac{\omega_{1234}}{\omega_{34}}\right)^{s_3} \left(-\frac{\omega_{1234}}{\omega_{4}}\right)^{s_4} \notag \\
    &\times \Gamma(q_{1234}- s_{234})
    {\color{BrickRed}\Gamma\left[
    \begin{matrix}
        s_2,1-s_2,q_{234}-s_{34} \\
        q_{234}-s_{234}+1
    \end{matrix}
    \right]
    \Gamma\left[
    \begin{matrix}
        s_3,1-s_3,q_{34}-s_4 \\
        q_{34}-s_{34}+1
    \end{matrix}
    \right]
    \Gamma\left[
    \begin{matrix}
        s_4,1-s_4,q_4 \\
        q_4-s_4+1
    \end{matrix}
    \right]} \ .
\end{align}
We want to consider energy regions covering the limit $\omega_{1234}\to 0$ where all three energy ratios in (\ref{eq_4chainMBI}) are small. So we could collect right poles, which all come from upstairs $\Gamma(\cdots-s)$ factors. 

Now suppose that we want to integrate over $s_2$ first. Then, we have two choices: Collecting poles of either $\Gamma(1-s_2)$ or $\Gamma(q_{1234}-s_{234})$.

Suppose that we choose $\Gamma(1-s_2)$ and thus set $s_2=n_2+1$ with $n_2\in\mathbb{N}$. This choice yields a series of the form $(-\omega_{1234}/\omega_{234})^{n_2+1}$ which is formally convergent when $|\omega_{1234}/\omega_{234}|<1$. Happily, picking up these poles has no consequences to other Mellin variables. So, next, we can consider integrating out $s_3$. Then, again, there are two possibilities: We collect poles from either $\Gamma(q_{1234}-s_{234})$ or $\Gamma(1-s_3)$.

The procedure goes in the same way if we choose poles of $\Gamma(1-s_3)$ to integrate out $s_3$. Then, again, there are two choices for integrating out $s_4$, using poles of either $\Gamma(q_{1234}-s_{234})$ or $\Gamma(1-s_4)$. To recap, it is easiest when we pick up poles from $\Gamma(1-s_j)$ for all $j=2,3,4$. In this case we simply set $s_j=-n_j$ and we are done. 

Now let us return to the first step where we integrate out $s_2$, and now consider the other choice, i.e., picking up poles from $\Gamma(q_{1234}-s_{234})$. This amounts to setting $s_2=q_{1234}-s_{34}+n_2$ with $n_2\in\mathbb{N}$. After doing so, the three power factors of energy ratios in (\ref{eq_4chainMBI}) now becomes (omitting minus signs): 
\begin{align}
\label{eq_EnergyRatioShift}
\left(\frac{\omega_{1234}}{\omega_{234}}\right)^{s_2} \left(\frac{\omega_{1234}}{\omega_{34}}\right)^{s_3} \left(\frac{\omega_{1234}}{\omega_{4}}\right)^{s_4}
\to &~
\left(\frac{\omega_{1234}}{\omega_{234}}\right)^{q_{1234}-s_{34}+n_2} \left(\frac{\omega_{1234}}{\omega_{34}}\right)^{s_3} \left(\frac{\omega_{1234}}{\omega_{4}}\right)^{s_4}\n\\
=&~\left(\frac{\omega_{1234}}{\omega_{234}}\right)^{q_{1234}+n_2} \left(\FR{\omega_{234}}{\omega_{34}}\right)^{s_3} \left(\frac{\omega_{234}}{\omega_{4}}\right)^{s_4}.
\end{align}
In the last line, the first factor is in good shape, showing that the series could be convergent when $|\omega_{1234}/\omega_{234}|\ll 1$, which \emph{is} what we require from the very beginning. However, the energy ratios in the second and third factors are new. They show that, to proceed, we have to further specify the sizes of $|\omega_{234}/\omega_{34}|$ and $|\omega_{234}/\omega_4|$. In this work, we choose to stay inside the physical region as much as possible, and thus should have $|\omega_{234}/\omega_{34}|>1$ and $|\omega_{234}/\omega_4|>1$. Consequently, we should pick up \emph{left} poles for $s_3$ and $s_4$. 

At this point, if we are going to integrate out $s_3$, we could pick up the left poles of $s_3$ from $\Gamma(s_3)$, or from $\Gamma(1-s_2)$ after setting $s_2=q_{1234}-s_{34}+n_2$. However, choosing $\Gamma(1-s_2)$ is already considered as one option when we integrate out $s_2$ in the first step, so we should not choose it again. This is an important lesson for performing multi-fold MB integrals: What really matters is to find all distinct combinations of $\Gamma$ factors to collect poles; The order of picking up poles from these $\Gamma$ factors is irrelevant. Practically, if we choose to perform MB integrals layer by layer, we have to choose an order to pick up poles. Then it is important to keep in mind the above points to avoid overcounting.

\begin{figure}[t]
\centering
\includegraphics[width=0.78\textwidth]{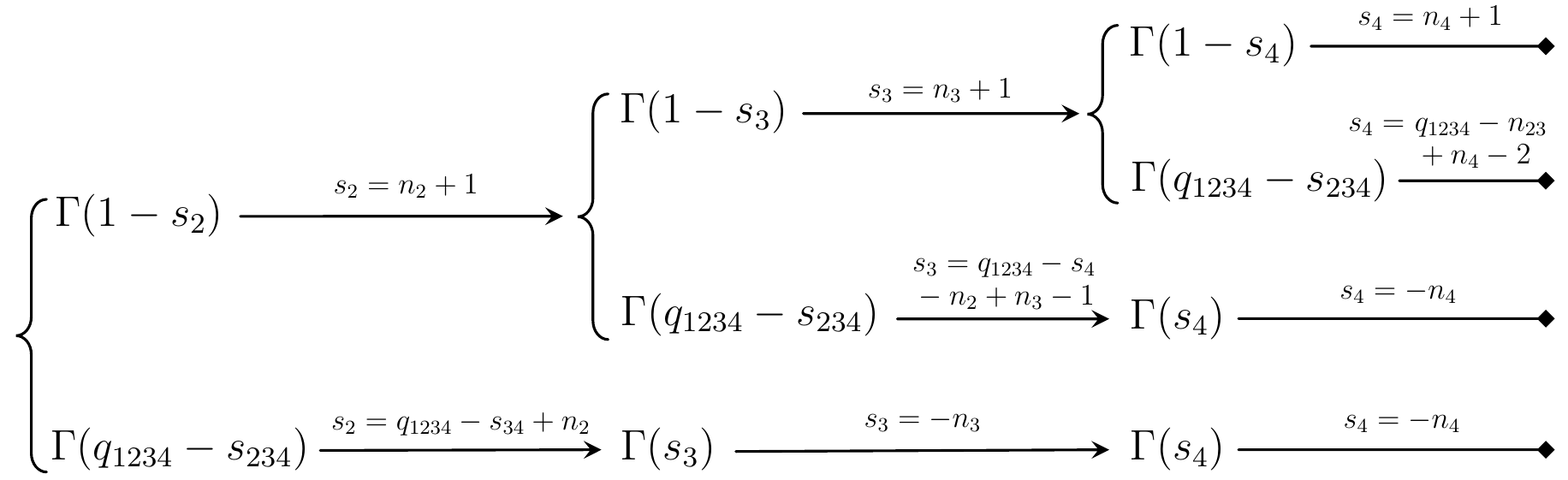}
\caption{The pole collecting algorithm for the MB integral of $\ft{1234}$ in the $\omega_{1234}\to 0$ limit, together with conditions $|\omega_{234}|>|\omega_{34}|$ and $|\omega_{234}|>|\omega_{4}|$. This procedure also applies to obtaining the series of $\ft{1(2)(34)}$ in the region $|\omega_2|>|\omega_{34}|>|\omega_4|>|\omega_{1234}|$.}
\label{fig_ChainPoleCollect}
\end{figure}
With all above points clarified, it is straightforward to see that there are 4 distinct ways of picking up poles, as summarized in Fig.\;\ref{fig_ChainPoleCollect}, and it is also straightforward to write down the result after summing up residues of all these poles: 
\begin{align}
\label{eq_1234TEseries}
    &\ft{1234}=\frac{-1}{(\ii \wt\omega_{1})^{\wt q_{1}}}\sum_{n_2,n_3,n_4=0}^{\infty}  \FR{\Gamma(q_{1234}-n_{234}-3)(\wt\omega_1/\wt\omega_2)^{n_2+1}(\wt\omega_1/\wt\omega_3)^{n_3+1}(\wt\omega_1/\wt\omega_4)^{n_4+1}}{(\wt q_{2}-n_{34}-2)_{-n_2}(\wt q_{3}-n_4-1)_{-n_3}(\wt q_4)_{-n_4}}\notag \\ 
    &+\frac{\pi\csc(\pi\wt{q}_1)}{(-\ii\wt\omega_{4})^{\wt q_1}} \sum_{n_2,n_3,n_4=0}^{\infty}  \Gamma \left[
    \begin{matrix}
         q_4 \\
        3-q_{123}+n_{23\bar 4}
    \end{matrix}
    \right]  \FR{ (\wt\omega_4/\wt\omega_2)^{n_2+1}(\wt\omega_4/\wt\omega_3)^{n_3+1}(-\wt\omega_1/\wt\omega_4)^{n_4}}{n_4!(1-q_1+n_{2\bar 4})_{-n_2}(2-q_{12}+n_{23\bar 4})_{-n_3}}\n\\ 
    & -\frac{\pi\csc(\pi\wt{q}_1)}{(-\ii \wt\omega_{3})^{\wt q_1}} \sum_{n_2,n_3,n_4=0}^{\infty} \Gamma \left[
    \begin{matrix}
       q_{34}+n_4 \\
       2-q_{12}+n_{2\bar 3}
    \end{matrix}
    \right] 
    \FR{(\wt\omega_3/\wt\omega_2)^{n_2+1}(-\wt\omega_1/\wt\omega_3)^{n_3}(\wt\omega_4/\wt\omega_3)^{n_4}}{n_3!(1-q_1+n_{2\bar 3})_{-n_2} (q_4)_{1+n_4}}\n\\
    & +\frac{\pi\csc(\pi\wt{q}_1)}{(-\ii \wt\omega_{2})^{\wt q_{1}}} \sum_{n_2,n_3,n_4=0}^{\infty}  \Gamma \left[
    \begin{matrix}
        q_{234}+n_{34}\\
        1-q_1+n_{\bar 2}
    \end{matrix} 
    \right] \FR{(-\wt\omega_1/\wt\omega_2)^{n_2}(\wt\omega_3/\wt\omega_2)^{n_3}(\wt\omega_4/\wt\omega_2)^{n_4}}{n_2!(q_{34}+n_4)_{1+n_3}(q_4)_{1+n_4}} .
\end{align}
Here $\wt q_1=q_{1234}$, $\wt\omega_1=\omega_{1234}$, $\wt\omega_2=\omega_{234}$, $\wt\omega_3=\omega_{34}$, and $\wt\omega_4=\omega_4$, and the four lines are in one-to-one correspondence with the four lines on the right of Fig.\;\ref{fig_ChainPoleCollect}. 
Unlike the large-energy expansion, here the result is quite complicated in that a single family tree is expressed as the sum of 4 hypergeometric series.\footnote{In a standard (multivariate) hypergeometric series, all $\Gamma$ factors, including those in the Pochhammer symbols $(a)_n\equiv \Gamma(a+n)/\Gamma(n)$, contain summation variables $n_j$ as $\Gamma(\cdots + n_j)$. Here we have some $n$ variables with negative signs in the $\Gamma$ factors, which might look worrisome at first sight. However, one can always use the identities $\Gamma(z)\Gamma(1-z)=\pi \csc(\pi z)$ to reverse the sign of those $n$ variables, provided that none of $q_i$ is nonnegative integers. Family trees with nonnegative integer $q_j$ reduce to polylogarithms ($q_j=0$) or fractions ($q_j\in \mathbb{Z}_+$) and thus are much simpler than the general case. See \cite{Fan:2024iek} for more discussions.\label{fn_negn}} Nevertheless, there are notable patterns and some of them persist in more general cases. One important pattern is the singular part: Only the first series is singular in the zero total-energy limit $\omega_{1234}\to 0$, and  other three series are regular. This point is general and can be understood from the integral (\ref{eq_4chainMBI}) without actually performing it: If we ever pick up poles from $\Gamma(q_{1234}-s_{234})$ in (\ref{eq_4chainMBI}), then the exponent of $\wt\omega_1$ is forced to be nonnegative integers. Note that this $\Gamma$ factor is from the last-layer time integral and thus general for any family tree whatever representation of ${}_1\mathcal{F}_1$ we choose. Thus, for any family tree, its singular part in the zero total-energy limit must not receive contributions from poles of this $\Gamma$ factor.

We end the discussion of this example by a comment that ``choosing $\Gamma(q_{1234}-s_{234})$ at site $j$'' is indeed an abuse of language. What we did is picking a set of $\Gamma$ functions to solve for all $s_j$. This corresponds to a linear system of equations. It is not true in general that, every single equation determine one single $s_j$ alone. However, in our situations, it turns out always sensible to say that a particular $\Gamma$ factor fixes a particular $s_j$. So, we will be content to use this nonrigorous language in this work. 

\paragraph{Tree with branches}
Now, we can move on and consider a slightly more complicated family tree $\ft{1(2)(34)}$, which is still four-site but has two branches. The MB integral for this family tree is:
\begin{align}
    &\ft{1(2)(34)}=\frac{1}{(\ii\omega_{1234})^{q_{1234}}}\int_{s_2,s_3,s_4}\left(-\frac{\omega_{1234}}{\omega_{2}}\right)^{s_2} \left(-\frac{\omega_{1234}}{\omega_{34}}\right)^{s_3} \left(-\frac{\omega_{1234}}{\omega_{4}}\right)^{s_4} \notag \\
    & \times \Gamma[q_{1234}-s_{234}]
    {\color{BrickRed}\Gamma\left[
    \begin{matrix}
        s_2,1-s_2,q_2\\
        q_{2}-s_{2}+1
    \end{matrix}
    \right]
    \Gamma\left[
    \begin{matrix}
        s_3,1-s_3,q_{34}-s_4 \\
        q_{34}-s_{34}+1
    \end{matrix}
    \right]
    \Gamma\left[
    \begin{matrix}
        s_4,1-s_4,q_4 \\
        q_4-s_4+1
    \end{matrix}
    \right]}   \ .
\end{align}
The way to finish the MB integral is similar to the previous case. Once again, if we first pick up the poles from $\Gamma(q_{1234}-s_{234})$ to integrate out $s_2$, we get power functions of new energy ratios $\omega_{2}/\omega_{34}$ and $\omega_2/\omega_4$. Likewise, if we use $\Gamma(1-s_2)$ to integrate out $s_2$ and use $\Gamma(q_{1234}-s_{234})$ to integrate out $s_3$, then we get another new energy ratio $\omega_{34}/\omega_4$. Therefore, we need to compare the sizes of $|\omega_{2}|$, $|\omega_{34}|$, and $|\omega_{4}|$ to convert the MB integral into a (formally) convergent series. In other words, we need to put all $|\wt\omega_j|$ $(j=1,\cdots,4)$ into a total order, with $|\wt\omega_1|$ being the smallest. 

Of course, there are many ways ($(N-1)!$ ways for an $N$-site family tree) to choose a total order for all $\wt\omega_j$ with $\wt\omega_1$ being the smallest. However, if we decide to reach the limit $\omega_{1234}\to 0$ by analytically continuing $\omega_1$ to negative values and keep all other energies staying in the interior of the physical region, i.e., $\omega_{j}>0$ $(j=2,3,4)$, then the possible choices of energy orders are reduced. It is easy to see that, in this case, the total order of $\wt\omega_j$ must be consistent with the partial order of the family tree except for the root site. That is, we pick up a given term in the family chain decomposition of $\ft{1(2)(34)}$ and use its total order to fix the energy order except that the root tilded energy $\wt\omega_1$ always remains the smallest.

For $\ft{1(2)(34)}$, there are 3 possibilities:  $|\wt\omega_{2}|>|\wt\omega_3|>|\wt\omega_4|>|\wt\omega_1|$, $|\wt\omega_{3}|>|\wt\omega_2|>|\wt\omega_4|>|\wt\omega_1|$, and $|\wt\omega_{3}|>|\wt\omega_4|>|\wt\omega_2|>|\wt\omega_1|$. Each choice of them will yield a series convergent in a distinct region, as we shall show below.

Let us consider the first choice: We fix the energy orders to be $|\wt\omega_{2}|>|\wt\omega_3|>|\wt\omega_4|>|\wt\omega_1|$. Then, we can perform the MB integral of $s_2$ first, $s_3$ second, and $s_4$ last.\footnote{Once again, we remind the reader that choosing a particular order to perform MB integrals is only for practical convenience and has no deeper meaning.} Then, the procedure is identical to the one for the family chain $\ft{1234}$ considered previously, and is still given by Fig.\;\ref{fig_ChainPoleCollect}. Collecting the residues of these poles, we get a series as follows:
\begin{align}
\label{eq_4siteBranchSmallTEOrder1}
    &\ft{1(2)(34)}=\frac{-1}{(\ii \wt{\omega}_{1})^{\wt{q}_{1}}}\sum_{n_2,n_3,n_4=0}^{\infty} \frac{\Gamma(q_{1234}-n_{234}-3) (\wt{\omega}_1/\wt{\omega}_2)^{n_2+1} (\wt{\omega}_1/\wt{\omega}_3)^{n_3+1} (\wt{\omega}_1/\wt{\omega}_4)^{n_4+1}}{(\wt{q}_2)_{-n_2} (\wt{q}_3-n_4-1)_{-n_3} (\wt{q}_4)_{-n_4}} \notag \\
    &  +\frac{\pi\csc(\pi\wt{q}_1)}{(-\ii \wt{\omega}_4)^{\wt{q}_1}} \sum_{n_2,n_3,n_4=0}^{\infty} \Gamma \left[
    \begin{matrix}
        q_4 \\
        3-q_{123}+n_{23\bar{4}}
    \end{matrix}
    \right]
    \frac{(\wt{\omega}_4/\wt{\omega}_2)^{n_2+1} (\wt{\omega}_4/\wt{\omega}_3)^{n_3+1} (-\wt{\omega}_1/\wt{\omega}_4)^{n_4}}{n_4! (q_2)_{-n_2} (2-q_{12}+n_{23\bar{4}})_{-n_3}} \notag \\
    &  -\frac{\pi\csc(\pi\wt{q}_1)}{(-\ii \wt{\omega}_3)^{\wt{q}_1}} \sum_{n_2,n_3,n_4=0}^{\infty} \Gamma \left[
    \begin{matrix}
        q_{34}+n_4 \\
        2-q_{12}+n_{2\bar{3}}
    \end{matrix}
    \right]
    \frac{(\wt{\omega}_3/\wt{\omega}_2)^{n_2+1} (-\wt{\omega}_1/\wt{\omega}_3)^{n_3} (\wt{\omega}_4/\wt{\omega}_3)^{n_4}}{n_3! (q_2)_{-n_2} (q_4)_{n_4+1}} \notag \\
    &  +\frac{\pi\csc(\pi\wt{q}_1)}{(-\ii \wt{\omega}_{2})^{\wt{q}_{1}}} \sum_{n_2,n_3,n_4=0}^{\infty} \Gamma \left[
    \begin{matrix}
        q_2 \\
        1-q_{134}-n_{234}
    \end{matrix}
    \right]
    \frac{(-\wt{\omega}_1/\wt{\omega}_2)^{n_2} (\wt{\omega}_3/\wt{\omega}_2)^{n_3} (\wt{\omega}_4/\wt{\omega}_2)^{n_4}}{n_2! (\wt{q}_3+n_4)_{n_3+1} (q_4)_{n_4+1}} \ .
\end{align}
Again, only the first series involves singularity at $\omega_{1234}\to 0$.

\begin{figure}[t]
\centering
\includegraphics[width=0.78\textwidth]{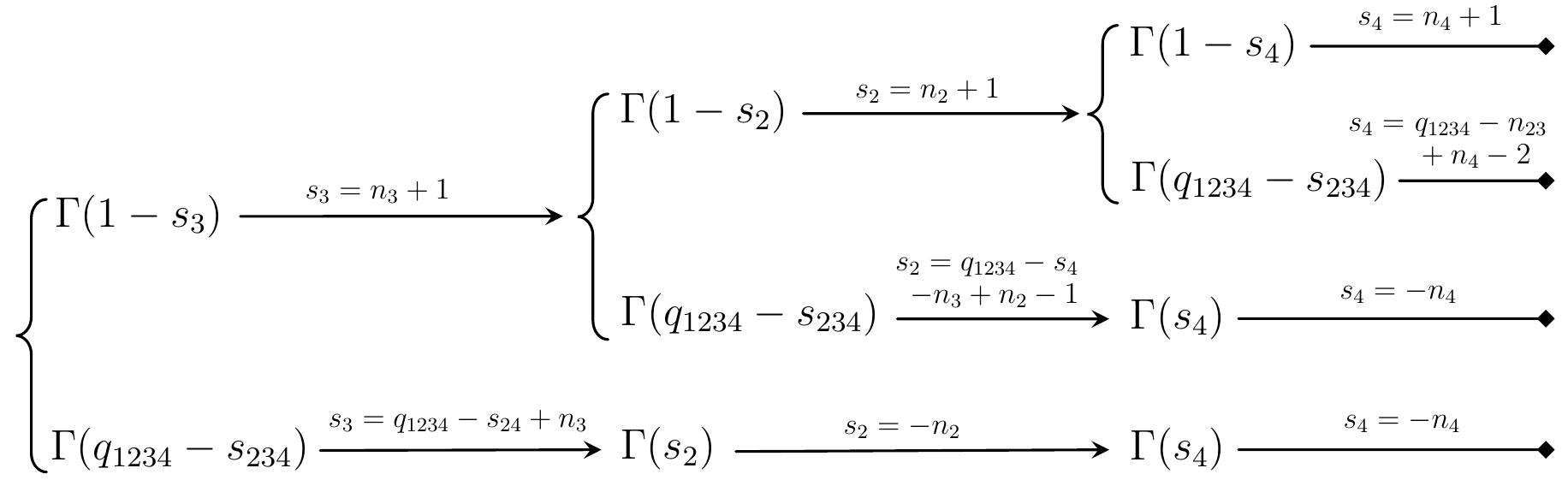}
\caption{The pole collecting algorithm for the MB integral of $\ft{1(2)(34)}$ in the $\omega_{1234}\to 0$ limit, together with conditions $|\omega_2|>|\omega_{34}|>|\omega_4|>|\omega_{1234}|$.}
\label{fig_BranchPoleCollect}
\end{figure}
We can also consider other choices of energy orders, say $|\wt\omega_{3}|>|\wt\omega_2|>|\wt\omega_4|>|\wt\omega_1|$. For this case, it is convenient to carry out the MB integral in the order of $s_3,s_2,s_4$. The corresponding pole collecting algorithm is shown in Fig.\;\ref{fig_BranchPoleCollect} and the result is:
\begin{align}
\label{eq_4siteBranchSmallTEOrder2}
    &\ft{1(2)(34)}=\frac{-1}{(\ii \wt{\omega}_{1})^{\wt{q}_{1}}}\sum_{n_2,n_3,n_4=0}^{\infty} \frac{\Gamma(q_{1234}-n_{234}-3) (\wt{\omega}_1/\wt{\omega}_2)^{n_2+1} (\wt{\omega}_1/\wt{\omega}_3)^{n_3+1} (\wt{\omega}_1/\wt{\omega}_4)^{n_4+1}}{(\wt{q}_2)_{-n_2} (\wt{q}_3-n_4-1)_{-n_3} (\wt{q}_4)_{-n_4}} \notag \\
    &  +\frac{\pi\csc(\pi\wt{q}_1)}{(-\ii \wt{\omega}_4)^{\wt{q}_1}} \sum_{n_2,n_3,n_4=0}^{\infty} \Gamma \left[
    \begin{matrix}
        q_4 \\
        3-q_{123}+n_{23\bar{4}}
    \end{matrix}
    \right]
    \frac{(\wt{\omega}_4/\wt{\omega}_2)^{n_2+1} (\wt{\omega}_4/\wt{\omega}_3)^{n_3+1} (-\wt{\omega}_1/\wt{\omega}_4)^{n_4}}{n_4! (q_2)_{-n_2} (2-q_{12}+n_{23\bar{4}})_{-n_3}}\notag \\
    &  -\frac{\pi\csc(\pi\wt{q}_1)}{(-\ii \wt{\omega}_2)^{\wt{q}_1}} \sum_{n_2,n_3,n_4=0}^{\infty} \Gamma \left[
    \begin{matrix}
        q_2 \\
        2-q_{134}-n_{2\bar{3}4}
    \end{matrix}
    \right]
    \frac{(-\wt{\omega}_1/\wt{\omega}_2)^{n_2} (\wt{\omega}_2/\wt{\omega}_3)^{n_3+1} (\wt{\omega}_4/\wt{\omega}_2)^{n_4}}{n_2! (\wt{q}_3+n_4)_{-n_3} (q_4)_{n_4+1}} \notag \\
    &  +\frac{\pi\csc(\pi\wt{q}_1)}{(-\ii \wt{\omega}_3)^{\wt{q}_1}} \sum_{n_2,n_3,n_4=0}^{\infty} \Gamma \left[
    \begin{matrix}
        q_{34}+n_4 \\
        1-q_{12}-n_{23}
    \end{matrix}
    \right]
    \frac{(\wt{\omega}_2/\wt{\omega}_3)^{n_2} (-\wt{\omega}_1/\wt{\omega}_3)^{n_3} (\wt{\omega}_4/\wt{\omega}_3)^{n_4}}{n_3! (q_2)_{n_2+1} (q_4)_{n_4+1}}  \ .
\end{align}
The first two lines here are identical to the previous case in (\ref{eq_4siteBranchSmallTEOrder1}). In contrast, the last two lines are different, and are formally convergent at small $|\omega_2/\omega_{34}|$.

We can continue this analysis and consider the last choice of total order $|\wt\omega_{3}|>|\wt\omega_4|>|\wt\omega_2|>|\wt\omega_1|$, but the procedure should be clear to readers by now. Without writing down the explicit result, we can anticipate that the result still consists of four series, among which two are identical to the first two lines of (\ref{eq_4siteBranchSmallTEOrder1}) and (\ref{eq_4siteBranchSmallTEOrder2}), and the other two are different. We leave it as an exercise for interested readers.

\paragraph{General tree}
Now we have gained enough insight to move forward to general trees. It is easy to follow and generalize the procedures above to an arbitrary family tree in (\ref{eq_TotalIntMB}).  

The main lesson from the above examples is that, unlike the infinite energy limit, to perform the small total-energy ($\wt\omega_1$) expansion of a family tree, it is not sufficient to require $|\wt\omega_1|<|\wt\omega_j|$ for all $j\neq 1$; We also need to compare tilded energies at all other sites. Of course, the possibilities are many. However, if we require all but the root tilded energies stay in the interior of the physical region, and achieve the small total-energy limit by analytically continuing only $\omega_1$ to negative values, then the possible choices are greatly reduced. In such cases, all but the root sites can be put into a total order consistent with the partial order of the family tree.\footnote{We stress that we choose this particular way of analytical continuation and ordering of energies purely for simplicity. One can certainly consider more general situations where more energies besides the root energy are analytically continued to unphysical region. The procedure described here still applies, but the results typically involve more than $N$ hypergeometric series for an $N$-site family tree. We leave them as exercises for interested readers.}

Since we put all but the root sites into a total order consistent with the partial order of the family tree, to simplify expressions, we can always relabel the sites according to the total order. Then, we look for a small total-energy expansion of the family tree in the region $|\wt{\omega}_1|<|\wt{\omega}_N|<|\wt{\omega}_{N-1}|<\cdots<|\wt{\omega}_2|$. From the previous examples, we know that such an expansion consists of $N$ independent hypergeometric series. (See Footnote\;\ref{fn_negn} for comments of being ``hypergeometric.'')

For convenience, we still adopt the non-rigorous language that we integrate out all $s_j$ following the total order: first $s_2$, then $s_3$, and $s_N$ last.  

Then, there is one hypergeometric series that corresponds to choosing poles from $\Gamma(1-s_j)$ for all $j=2,\cdots,N$. Thus, we collect residues of $\Gamma(1-s_j)$ at these poles, set $s_j=n_j+1$ in all other factors, and get the following result: 
\begin{align}
\label{eq_TESingSeries}
\frac{(-1)^{N-1}}{(\ii \wt{\omega}_1)^{\wt{q}_1}} \sum_{n_2,\cdots,n_N=0}^{\infty}   \Gamma(\wt{q}_1-\wh n_1-d_1)\prod_{j=2}^{N}  \FR{(\wt\omega_1/\wt\omega_j)^{n_j+1}}{(\wt q_j-\wh n_j-d_j)_{-n_j}},
\end{align}
where $d_j$ is the number of descendants of Site $j$ defined below (\ref{eq_wtqjDef}). 

The other $N-1$ series correspond to using the poles of root $\Gamma$ to integrate out $s_k$, with $k=2,\cdots,N$. Now, if we do so for Site $k$, then we should choose the right poles from $\Gamma(1-s_j)$ to integrate out $s_j$ for all $j<k$, and choose the left poles from $\Gamma(s_j)$ to integrate out $s_j$ for all $j>k$. In this case, we should set MB variables to:
\begin{align}
    \label{eq_TotalEnergyPolePosK}
    s_j=\si_j^{(k)}\equiv\begin{cases}
        n_j+1, \quad &(j<k) \\
        \wt{q}_1-n_{2\cdots(k-1)}+n_{k\cdots N}-k+2, \quad &(j=k)\\
        -n_j. \quad &(j>k) \\
    \end{cases}  
\end{align}
We stress that $\si_j^{(k)}$ stand for solutions to the Mellin variables, and are in general linear combinations of $n_j$ and constants. They are \textit{not} integral variables of the MB integral. After picking up these poles to finish the integral, we get the following series: 
\begin{align}
    &\frac{(-1)^{k}\pi}{(-\ii\wt{\omega}_k)^{\wt{q}_1}\sin(\pi\wt q_1) }\sum_{n_2,\cdots,n_N=0}^{\infty}  
    \FR{(-\wt\omega_1/\wt\omega_k)^{n_k}\prod_{i=2}^{k-1}(\wt\omega_k/\wt\omega_i)^{n_i+1}\prod_{j=k+1}^N(\wt\omega_j/\wt\omega_k)^{n_j}}{n_k!\prod _{j=2}^N\big(\wt q_j-\wh \si_j^{(k)}\big)_{1-\si_j^{(k)}}} \ .
\end{align}
Once again, here $k=2,\cdots,N$, and thus we have got $N-1$ series. Combining the previous one in (\ref{eq_TESingSeries}), we arrive at the full series expansion of an arbitrary family tree at small total energy:
\begin{keyeqn}
\begin{align}
\label{eq_TEseries}
  &\ii^N\ft{\mathscr{P}(\wh 12\cdots N)}
  =\frac{(-1)^{N-1}}{(\ii \wt{\omega}_1)^{\wt{q}_1}} \sum_{n_2,\cdots,n_N=0}^{\infty}   \Gamma(\wt{q}_1-\wh n_1-d_1)\prod_{j=2}^{N}  \FR{(\wt\omega_1/\wt\omega_j)^{n_j+1}}{(\wt q_j-\wh n_j-d_j)_{-n_j}}\n\\
  &+\sum_{k=2}^N\frac{(-1)^{k}\pi\csc(\pi\wt q_1)}{(-\ii\wt{\omega}_k)^{\wt{q}_1} }\sum_{n_2,\cdots,n_N=0}^{\infty}  
    \FR{(-\wt\omega_1/\wt\omega_k)^{n_k}\prod_{i=2}^{k-1}(\wt\omega_k/\wt\omega_i)^{n_i+1}\prod_{j=k+1}^N(\wt\omega_j/\wt\omega_k)^{n_j}}{n_k!\prod _{j=2}^N\big(\wt q_j-\wh \si_j^{(k)}\big)_{1-\si_j^{(k)}}}\ ,
\end{align}
\end{keyeqn}
where $\si_j^{(k)}$ are given by (\ref{eq_TotalEnergyPolePosK}). Although more complicated, this expression has a clear pattern: The single hypergeometric function in the first line is singular as $\wt\omega_1\to 0$ due to the overall factor $(\ii\wt\omega_1)^{-\wt q_1}$, while all other $k-1$ series in the second line are regular in the same limit, and the mass dimensions of these regular series are carried by the factor $(-\ii\wt\omega_k)^{\wt q_1}$. Notice that these factors do not imply that the series in the second line are singular in the limit $\wt\omega_k\to 0$, since this limit is outside the convergent region of the above expression. 

One more remark is that some summation variables in (\ref{eq_TEseries}) appear with a negative sign in $\Gamma$ (or Pochhammer) factors like $\Gamma(\cdots-n_j)$. We can always flip these signs using the identity $\Gamma(1-z)\Gamma(z)=\pi\csc(\pi z)$, provided that none of the twists $q_j$ are nonnegative integers. We will come back to the cases of nonnegative integer twists in Sec.\;\ref{sec_twists}.

To conclude this subsection, we note that the singular part of the series (\ref{eq_TEseries}) is considerably simpler than the full result in that it does not care about the energy orderings of $\wt\omega_j$ for all $j\neq 1$. Also, the singular part is a single convergent hypergeometric series multiplied by a singular power factor $1/(\ii\wt\omega_1)^{\wt q_1}$, which is very similar to the structure seen in infinite energy limits. The singular part will be useful in the following studies, so we give it a special notation:
\begin{align}
  \mathop{\text{Sg}}\limits_{\wt\omega_{1}\to 0}\ft{\mathscr{P}(\wh{1}2\cdots N)}&=\frac{-\ii^N}{(\ii \wt{\omega}_1)^{\wt{q}_1}} \sum_{n_2,\cdots,n_N=0}^{\infty}   \Gamma(\wt{q}_1-\wh n_1-d_1)\prod_{j=2}^{N}  \FR{(\wt\omega_1/\wt\omega_j)^{n_j+1}}{(\wt q_j-\wh n_j-d_j)_{-n_j}} ,
 \label{eq_SgZeroTotal}
\end{align} 
where $\mathop{\text{Sg}}\limits_{\wt\omega_{1}\to 0}$ means to extract the singular part of the family tree in the region where $\wt\omega_1$ is small. We emphasize that this result is exact to all orders in $\wt\omega_1$. However, it is interesting to show the leading order term of this singular piece:
\bge
  \mathop{\text{Sg}}\limits_{\wt\omega_{1}\to 0}\ft{\mathscr{P}(\wh 12\cdots N)}=\FR{-\ii^N\Gamma(\wt q_1-N+1)}{(\ii\wt\omega_1)^{\wt q_1-N+1}(\ii\wt\omega_2)\cdots(\ii\wt\omega_N)}\Big[1+\order{\wt\omega_1}\Big].
\ede
The interesting point is that the leading behavior in $\wt\omega_1$ is generally softer than a single power factor $(\ii\wt\omega_1)^{\wt q_1}$ as in the case of large partial-energy series. There are additional factors of $1/(\wt{\omega}_2\cdots\wt{\omega}_N)$ contributing at the leading order.  

\subsection{Small partial-energy series}
\label{sec_SmallPE}

Now we proceed to the series representation around a small partial energy. We only consider small partial energies where the family tree is singular. As discussed in the previous section, these are the root-bearing partial energies. Since the total energy is also a root-bearing partial energy, the small total-energy series obtained in the previous subsection is a special case.

As in Sec.\;\ref{sec_LargePartialSeries}, suppose we choose a root-bearing $M$-site subgraph $\mathcal{C}$ of an $N$-site family tree with $1\leq M\leq N$. We want to study the behavior of the family tree when the total energy of $\mathcal{C}$ becomes small. Again, for convenience, we relabel the family tree such that all $M$ sites of $\mathcal{C}$ acquire the first $M$ labels $(1,\cdots,M)$ and that the root site is fixed to be Site 1. Then, the corresponding MB integral is:
\begin{align}
    \label{eq_PartialIntMB}
    \ii^N\ft{\mathscr{P}(\wh{1}2\cdots N)}=\int_{s_2,\cdots,s_N}\frac{\Gamma(\wt{q}_1-\wh{s}_1)}{(\ii \Omega_1)^{\wt{q}_1}} & \prod_{j=2}^{M} \left(-\frac{\Omega_1}{\Omega_j}\right)^{s_j} 
    {\color{BrickRed}\Gamma \left[
    \begin{matrix}
        s_j, 1-s_j, \wt{q}_j-\wh{s}_j \\
        \wt{q}_j-\wt{s}_j+1
    \end{matrix}
    \right]} \notag \\
    \times & \prod_{j=M+1}^{N} \left(\frac{\Omega_1}{\omega_j}\right)^{s_j} 
    {\color{RoyalBlue}\Gamma \left[
    \begin{matrix}
        \wt{q}_j-\wt{s}_j,s_j \\
        \wt{q}_j-\wt{s}_j+1
    \end{matrix}
    \right] } \ .
\end{align}
The structure of the MB integrand is by now familiar: For $j=2,\cdots,M$ where we choose to collect energies, we can not simultaneously pick the ``poles'' of $\Gamma(1-s_j)$ and $\Gamma(\wt{q}_j-\wh{s}_j)$ due to the zero contributed by the $\Gamma$ factor in the denominator. For $j=M+1,\cdots,N$,  $\Gamma(\wt{q}_j-\wt{s}_j)$ is always a single pole due to the $\Gamma(\wt{q}_j-\wt{s}_j+1)$ in the denominator. Thus only the leading term in the corresponding series survives. We will see that choosing such single pole leads to a factorization in the result.

Once again, we use a few examples to illustrate the basic structure of these series before considering the most general case. 

\paragraph{An illustrating example}
We consider the family tree $\ft{1(2)(34)}$ at small partial energy $\omega_{12}$. The MB integrand is now 
\begin{align}
    \ft{1(2)(34)}=&~\frac{1}{(\ii\omega_{12})^{q_{1234}}}\int_{s_2,s_3,s_4} \left(-\frac{\omega_{12}}{\omega_2}\right)^{s_2} \left(\frac{\omega_{12}}{\omega_3}\right)^{s_3} \left(\frac{\omega_{12}}{\omega_4}\right)^{s_4} \notag \\
    & \times \Gamma(q_{1234}-s_{234}){\color{BrickRed}\Gamma\left[
    \begin{matrix}
        s_2,1-s_2,q_2 \\
        q_2-s_2+1
    \end{matrix}
    \right]}
{\color{RoyalBlue}\Gamma\left[
    \begin{matrix}
       s_3,q_{34}-s_{34}\\
       q_{34}-s_{34}+1
    \end{matrix}
    \right] 
  \Gamma\left[
    \begin{matrix}
        s_4,q_4-s_4 \\
        q_4-s_4+1
    \end{matrix}
    \right] }\ .
\end{align}

Just like before, we need to pick a total order of energies. Once again, for simplicity, we only consider total orders that is consistent with the partial order of the family tree, except that the root capitalized energy ($\omega_{12}$ in this example) remains the smallest. 

Let us first consider the total order $|\omega_{12}|<|\omega_4|<|\omega_3|<|\omega_2|$. 

\begin{figure}[t]
\centering
\includegraphics[width=0.78\textwidth]{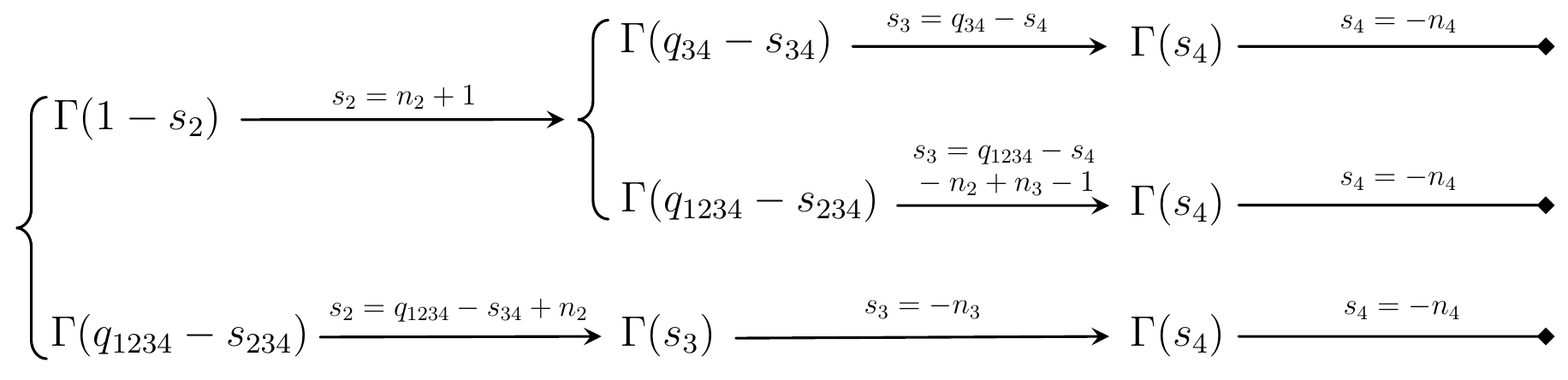}
\caption{The pole collecting algorithm for the MB integral of $\ft{1(2)(34)}$ in the $\omega_{12}\to 0$ limit, in the region with $|\omega_{12}|<|\omega_{4}|<|\omega_3|<|\omega_2|$.}
\label{fig_polecollection_4chainPE}
\end{figure}

Again, there is a possibility that we pick up no poles from the root $\Gamma$ (namely, $\Gamma(q_{1234}-s_{234})$ in this example). In this case, we can only pick up poles from $\Gamma(1-s_2)$ to integrate out $s_2$ since $|\omega_{12}|<|\omega_2|$. This is the same as before. However, a new phenomenon appears when we consider Site $3$. To integrate out $s_3$, we have to pick up right poles, and the only candidate is from $\Gamma(q_{34}-s_{34})$. Picking up poles from this $\Gamma$ factor has two consequences: 1) The pole from $\Gamma(q_{34}-s_{34})$ is really one pole instead of a series of poles, due to the denominator. As a result, there is no summation variable corresponds to $s_3$, or equivalently, the summation variable $n_3$ is forced to take only one value $n_3=0$. 2) Since $\Gamma(q_{34}-s_{34})$ also involves $s_4$, picking up poles from this factor generates a new energy ratio $(\omega_{3}/\omega_4)^{s_4}$ which lets us compare $|\omega_3|$ and $|\omega_4|$. Since we have chosen $|\omega_3|>|\omega_4|$, we need to pick up left poles to integrate out $s_4$, and the only possibilities are from $\Gamma(s_4)$. 

The above choices of poles are summarized in the first line of Fig.\;\ref{fig_polecollection_4chainPE}. With this set of choices of poles, we get a series:
\begin{align}
\label{eq_4site12PEpoleterm}
  &\bigg[\frac{1}{(\ii\omega_{12})^{q_{12}}}
  \sum_{n_2=0}^\infty\FR{\Gamma(q_{12}-n_2-1)(\omega_{12}/\omega_2)^{n_2+1}}{(q_2)_{-n_2}}\bigg]
  \bigg[\frac{1}{(\ii\omega_{3})^{q_{34}}} \sum_{n_4=0}^\infty\FR{\Gamma(q_{34}+n_4)(-\omega_4/\omega_3)^{n_4}}{n_4!(q_4+n_4)}\bigg] \n\\
   = &~\bigg\{-\FR{\omega_{12}/\omega_2}{(\ii\omega_{12})^{q_{12}}}\FR{\pi\csc(\pi q_{12})}{\Gamma(1-q_2)}\,{}_2\mathcal{F}_1\left[\bgm 1, 1-q_2\\2-q_{12}\edm\middle|\FR{\omega_{12}}{\omega_2}\right]\bigg\}\bigg\{\FR{1}{(\ii\omega_{3})^{q_{34}}}\,{}_2\mathcal{F}_1\left[\bgm q_{34}, q_4\\1+q_4\edm\middle|-\FR{\omega_{4}}{\omega_3}\right]\bigg\} .
\end{align}
As discussed around Fig.\;\ref{fig_contour}, the single pole corresponds to a factorized line. Thus it is not surprising that we get a factorized series. Indeed, picking up the single pole is a hallmark of a factorization of the graph at the MB integrand level. We will see this phenomenon again below. 

Furthermore, we recognize that the first factor in (\ref{eq_4site12PEpoleterm}) is the singular part of the family tree $\ft{12}$ in the zero total-energy limit and that the second factor is the family tree $\ft{34}$. We will see below that this is an example of a general factorization theorem in zero partial-energy limits.

Now we consider other possibilities of pole collecting. Just like the small total-energy expansion, we can use the root $\Gamma$ to integrate out various Mellin variables. However, as discussed before, we never have a chance to use it to integrate out $s_4$. Thus we have 2 possibilities remaining, corresponding to using the root $\Gamma$ to integrate out $s_2$ and $s_3$. They correspond to the second and third lines in Fig.\;\ref{fig_polecollection_4chainPE}.
Summing all three sets together, we get a small partial-energy expansion for $\ft{1(2)(34)}$ which is formally convergent in the region $|\omega_{12}|<|\omega_4|<|\omega_3|<|\omega_2|$:
\begin{align}
    \ft{1(2)(34)}=&~\bigg(\mathop{\text{Sg}}_{\omega_{12}\to 0}\ft{12}\bigg) \times \ft{34} \notag \\
    &+\frac{\pi\csc(\pi\wt q_1)}{(-\ii\omega_2)^{\wt q_{1}}} \sum_{n_2,n_3,n_4=0}^{\infty} \Gamma\left[
    \begin{matrix}
        q_2 \\
        1-q_{134}-n_{234}
    \end{matrix}
    \right]
  \FR{(-\omega_{12}/\omega_2)^{n_2}(-\omega_3/\omega_2)^{n_3}(-\omega_4/\omega_2)^{n_4}}{n_2!n_3!n_4!(q_{34}+n_{34})(q_4+n_4)} \notag \\
    &-\frac{1}{(\ii\omega_3)^{\wt q_{1}}} \sum_{n_2,n_3,n_4=0}^{\infty} \FR{\Gamma(\wt q_1+n_{\bar 234}-1)(\omega_3/\omega_2)^{n_2+1}(-\omega_{12}/\omega_3)^{n_3}(-\omega_4/\omega_3)^{n_4}}{n_3!n_4!(q_2)_{-n_2}(1-q_{12}+n_{2\bar 3})(q_4+n_4)} \ .
\end{align}
Here the three lines are in one-to-one correspondence with the three choices of poles in Fig.\;\ref{fig_polecollection_4chainPE}.

Next, we look at a different total order of energies: $|\omega_{12}|<|\omega_{4}|<|\omega_{2}|<|\omega_3|$. Something interesting happens for this ordering. Here we still have three ways of collecting poles:
\begin{align}
\label{eq_GammaFactors1(2)(34)}
  &\Gamma[q_{34}-s_{34},1-s_2,s_4],
  &&\Gamma[q_{34}-s_{34},q_{1234}-s_{234},s_4],
  &&\Gamma[q_{1234}-s_{234},s_2,s_4],
\end{align}
and the three arguments of each $\Gamma$ products are written in an order which corresponds to using them to integrate out $s_3,s_2,s_4$, respectively. By collecting poles accordingly, we get the following series for $\ft{1(2)(34)}$:
\begin{align}
\label{eq_1(2)(34)in1324}
      \ft{1(2)(34)}=&~\bigg(\mathop{\text{Sg}}_{\omega_{12}\to 0}\ft{12}\bigg) \times \ft{34} +\frac{1}{(-\ii \omega_{2})^{q_{12}}} \sum_{n_2=0}^{\infty} \FR{\pi\csc(\pi q_{12})\Gamma(q_2)(-\omega_{12}/\omega_2)^{n_2}}{n_2!\Gamma(1-q_1-n_2)}
    \times \ft{34}\notag  \\
    &-\frac{1}{(\ii\omega_3)^{\wt q_{1}}} \sum_{n_2,n_3,n_4=0}^{\infty} \FR{\Gamma(\wt q_1+n_{234})(\omega_2/\omega_3)^{n_2}(-\omega_{12}/\omega_3)^{n_3}(-\omega_4/\omega_3)^{n_4}}{n_3!n_4!(q_2)_{n_2+1}(q_{12}+n_{23})(q_4+n_4)} \ .
\end{align} 
The interesting thing here is that the single pole from $\Gamma(q_{34}-s_{34})/\Gamma(q_{34}-s_{34}+1)$ appears in the first two sets of $\Gamma$ factors in (\ref{eq_GammaFactors1(2)(34)}) which signal factorizations for not only the singular series (the first one) but also a regular series (the second). Indeed, by comparing with the general expression (\ref{eq_TEseries}), it is not hard to recognize that the second series in (\ref{eq_1(2)(34)in1324}) multiplied by $\ft{34}$ is nothing but the regular part of $\ft{12}$ in the limit $\omega_{12}\to 0$ and can be summed: 
\begin{align}
\label{eq_RegSeriesSum}
    \frac{1}{(-\ii \omega_{2})^{q_{12}}} \sum_{n_2=0}^{\infty} \FR{\pi\csc(\pi q_{12})\Gamma(q_2)(-\omega_{12}/\omega_2)^{n_2}}{n_2!\Gamma(1-q_1-n_2)}
 =\FR{\pi\csc(\pi q_{12})\Gamma(q_2)}{(-\ii\omega_1)^{q_1}(-\ii\omega_2)^{q_2}\Gamma(1-q_1)} .
\end{align}
The simplification here is not entirely accidental; It originates from rewriting the original family tree: $\ft{1(2)(34)}=\ft{12}\ft{34}-\ft{3(12)(4)}$. Indeed, the two series in the first line of (\ref{eq_1(2)(34)in1324}) sum into $\ft{12}\ft{34}$, and the second line is a series representation of $\ft{3(12)(4)}$ in the large $\omega_3$ limit.\footnote{Note that this is an unconventional representation of $\ft{3(12)(4)}$ which includes a non-root-bearing partial energy $\omega_{12}$. We can obtain it by collecting energy $\omega_2$ to Site 1 but not collecting $\omega_{12}$ or $\omega_4$ to Site 3.} As explained in Sec.\;\ref{sec_SingularityProof}, the family tree $\ft{3(12)(4)}$ is completely regular in the limit $\omega_{12}\to 0$ since the partial energy $\omega_{12}$ does not contain the root.

This example is suggestive: Suppose we have a general family tree $\mathcal{F}$ and we want to find its series representation in some small (root-bearing) partial-energy limit $\omega_\mathcal{G}\to 0$, which is always the small total-energy series of a root-bearing subgraph $\mathcal{G}\subset \mathcal{F}$. The above example suggests that we may flip all lines connecting $\mathcal{G}$ and $\mathcal{F}\backslash\mathcal{G}$. This flipping recasts the original family tree $\mathcal{F}$ into a sum of many terms, among which there is one term where $\mathcal{G}$ is totally factorized. The singular series in $\omega_\mathcal{G}\to 0$ is totally from this term (since $\mathcal{G}$ is singular in $\omega_\mathcal{G}\to 0$), and all other terms are regular (since $\omega_\mathcal{G}$ is not root bearing for all these terms).

This pattern is indeed seen in the most general situation to be discussed below, with one more complication: Since the above procedure involves flipping of lines, the resulting terms may not be family trees at all (i.e., not partially ordered), and we cannot directly apply the formulae for family trees to these terms. For this reason, we will still work with the MB representation and go through a careful pole collecting procedure. Still, the flipping described here provides useful intuition for understanding the final results.

\paragraph{General tree}
With all above discussions, we are ready to consider an arbitrary family tree. Suppose we have a family tree $\mathcal{F}$ and we want to find the series representation for any of its root-bearing partial energy. As described above, this amounts to specify a root-bearing subgraph $\mathcal{G}\subset\mathcal{F}$ and consider the small total-energy limit of $\mathcal{G}$. In general, the complement $\mathcal{F}\backslash\mathcal{G}$ is a collection of $P$ disjoint subgraphs of $\mathcal{F}$, which we call $\mathcal{F}_1,\cdots,\mathcal{F}_P$. Clearly, $P\geq 0$ and the equal sign is reached only when $\mathcal{G}=\mathcal{F}$. 

\begin{figure}[t]
\centering
\includegraphics[width=0.3\textwidth]{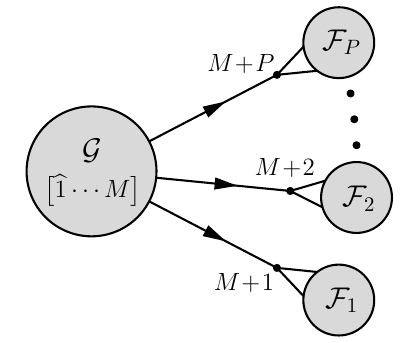}
\caption{The labeling of the family tree $\mathcal{F}=\ft{\mathscr{P}(\wh 12\cdots N)}$ featuring the small partial energy $\omega_{1\cdots M}$. We show explicitly the root site $M+a$ of each subgraph $\mathcal{F}_a$ $(a=1,\cdots,P)$ and all her descendants in a blob, with the understanding that $M+a$ is part of $\mathcal{F}_a$.}
\label{fig_SmallPElabels}
\end{figure}

For future convenience, we can always relabel a family tree without loss of generality. Suppose $\mathcal{F}$ has $N$ sites and $\mathcal{G}$ has $M\leq N$ sites. Then we adopt the following label scheme: 
\begin{enumerate}
  \item We label the $M$ sites of $\mathcal{G}$ by $1,\cdots,M$ with $1$ denoting the root.
  \item Each of $\mathcal{F}_a$ $(a=1,\cdots,P)$ is still a family tree with its own root. We label the root of $\mathcal{F}_a$ by $M+a$. 
  \item The rest of labels, from $M+P+1$ to $N$, are left to all non-root sites of $\mathcal{F}_a$ in a way that is consistent with the partial order of $\mathcal{F}$. That is, if $j$ is a descendant of $i$, then the label of $j$ is strictly larger than the label of $i$.    
\end{enumerate}
We illustrate this labeling in Fig.\;\ref{fig_SmallPElabels}. 

Next, since the resulting series also compare the (capitalized) energies of different sites, we need to choose a total order of (capitalized) energy variables. We will simply choose $|\Omega_1|<|\Omega_N|<|\Omega_{N-1}|<\cdots<|\Omega_2|$. This is not the most general choice but this choice bring certain simplifications to the final results. One can consider more general total orders and the procedure to get the corresponding series is completely the same. 

Now, we are ready to derive the small partial-energy series for $\mathcal{F}$ in the limit $\omega_{1\cdots M}\to 0$, again starting from the MB integral (\ref{eq_PartialIntMB}). The previous discussions on $\ft{1(2)(34)}$ generalizes directly. We would get one series where we do not pick poles of the root $\Gamma$, which is the only singular series, plus another $M+P-1$ (possibly factorized) series, which correspond to using the poles of root $\Gamma$ to integrate out $s_j$ for $j=2,\cdots,M+P$. As before, we never have chances to use the poles of root $\Gamma$ to integrate $s_j$ for any $j>M+P$. 

Below, we look into the details of these series.

First, we have a series which does not pick up poles of the root $\Gamma$. Given the total order of the energy variables, there is only one way to pick up poles in this case. That is, we use (the poles of) $\Gamma(1-s_j)$ to integrate out all $s_j$ for $j=2,\cdots,M$, use $\Gamma(\wt q_j-\wt s_j)$ to integrate out all $s_j$ for $j=M+1,\cdots,M+P$, and use $\Gamma(s_j)$ to integrate out all $s_j$ for $j=M+P+1,\cdots,N$. Since this series is the only singular series in $\omega_{1\cdots M}\to 0$ limit, we can write its explicit expression as:
\begin{align}
  \mathop{\text{Sg}}_{\omega_{1\cdots M}\to 0}\ft{\mathscr{P}(\wh 12\cdots N)}=\bigg(\mathop{\text{Sg}}_{\omega_{1\cdots M}\to 0}\ft{\mathscr{P}(\wh 12\cdots M)}\bigg)\prod_{j=M+1}^{M+P}\ft{\mathscr{P}(\wh j\cdots)},
\end{align}
or, equivalently:
\bge
  \mathop{\text{Sg}}_{\omega_{1\cdots M}\to 0}\mathcal{F}=\bigg(\mathop{\text{Sg}}_{\omega_{1\cdots M}\to 0}\mathcal{G}\bigg)\prod_{a=1}^P\mathcal{F}_a.
\ede
As expected, picking up the single poles from $\Gamma(\wt q_j-\wt s_j)$ for all $j=M+1,\cdots,M+P$ makes the final expression factorized into $P+1$ pieces, which is also consistent with the direction-flip argument given below (\ref{eq_RegSeriesSum}). 

Second, we have $M-1$ series corresponding to using the root $\Gamma$ to integrate out $s_k$ for $k=2,\cdots,M$. This is similar to the regular part of the small total-energy expansion discussed around (\ref{eq_TotalEnergyPolePosK}). Specifically, for a given choice of $k\in\{2,\cdots,M\}$, we should use $\Gamma(1-s_j)$ to integrate out $s_j$ for all $j<k$, $\Gamma(\wt q_1-\wh s_1)$ (root $\Gamma$) for $j=k$, and $\Gamma(s_j)$ for all $j>k$. These choices fix the solutions to all MB variables $s_j$. The solutions are identical to (\ref{eq_TotalEnergyPolePosK}).
With these solutions for $s_j$, we get the following series:
\begin{align}
    \frac{(-1)^k}{(-\ii \Omega_k)^{\wt{q}_1}} \sum_{n_2,\cdots,n_N=0}^{\infty} \FR{\pi(-\Omega_1/\Omega_k)^{n_k}\prod_{i=2}^{k-1}(\Omega_k/\Omega_i)^{n_i+1}\prod_{j={k+1}}^{M}(\Omega_{j}/\Omega_k)^{n_j}}{ n_k!\sin (\pi\wt q_1)\prod_{j=2}^{M} (\wt q_j-\wh \si_{j}^{(k)})_{1-\si_j^{(k)}}} 
    \prod_{j=M+1}^{N} \frac{(-\omega_{j}/\Omega_k)^{n_j}}{n_j!\left(\wt{q}_j+\wt{n}_j\right)} \ ,
\end{align}
where $\si_j^{(k)}$ is given by (\ref{eq_TotalEnergyPolePosK}). Of course, this is the series for one choice of $k\in\{2,\cdots,M\}$, and we have $M-1$ such series and should sum them up in the final result. 

Third, we have $P$ series corresponding to using the root $\Gamma$ to integrate out $s_k$ for $k=M+1,\cdots,M+P$. For a given $k$ in this range, we should use $\Gamma(1-s_j)$ to integrate out all $s_j$ for $j=2,\cdots,M$, use $\Gamma(\wt q_j-\wt s_j)$ to integrate out $s_j$ for $j=M+1,\cdots,k-1$, use $\Gamma(\wt q_1-\wh s_1)$ to integrate out $s_k$, and finally use $\Gamma(s_j)$ to integrate out $s_j$ for all $j>k$. These choices again fix the values of all $s_j$, which we denote by $\rho_j^{(k)}$:
\begin{align}
    \label{eq_PartialEnergyPolePosK2}
    \rho_j^{(k)}=\begin{cases}
        n_j+1,  &(j=2,\cdots,M) \\
        \wt{q}_j+\wt{n}_j, \quad &(j=M+1,\cdots,k-1) \\
        \wt{q}_1+n_{k\cdots N}-n_{2\cdots M}-(M-1)-\sum_{j=M+1}^{k-1}\rho_j^{(k)}, &(j=k) \\
        -n_j. \quad &(j=k+1,\cdots,N)
    \end{cases} 
\end{align}
With this solutions, we get the following series for a given $k\in\{M+1,\cdots,M+P\}$:
\begin{align}
\label{eq_partiallyFactSeries1}
    &\frac{(-1)^{M-1}}{(\ii\omega_k)^{\wt{q}_1}} \sum_{n_2,\cdots,n_M=0}^{\infty} \sum_{n_k,\cdots,n_N=0}^{\infty} 
      \frac{\Gamma (\rho_k^{(k)} )(-\Omega_1/\omega_k)^{n_k}}{n_k! (\wt{q}_k+\wh{n}_k-\rho_k^{(k)} )}\prod_{j=2}^{M}\FR{(\omega_k/\Omega_j)^{n_j+1}}{(\wt q_j-\wh\rho_j^{(k)})_{-n_j}}
    \notag \\
    & \times  \prod_{j=M+1}^{k-1}\Gamma(\wt{q}_j+\wh{n}_j)(\omega_k/\omega_j)^{\wt q_j+\wh n_j} \prod_{j=k+1}^{N} \frac{(-\omega_j/\omega_k)^{n_j}}{n_j! \left(\wt{q}_j+\wt{n}_j\right)} \ .
\end{align}
In the final results, we should also sum over all $k\in\{M+1,\cdots,M+P\}$. However, before taking the summation, we expect further simplifications of this expression, in particular because we have chosen single poles from $\Gamma(\wt q_j+\wt n_j)$ to integrate out $s_j$ for $M<j<k$. As repeatedly seen before, choosing single poles leads to factorizations, which is not obvious in (\ref{eq_partiallyFactSeries1}). To make the factorization manifest, we need a closer look at various parameters in the expression. For instance, immediate after the summation variable, we have a factor of $\Gamma(\rho_k^{(k)})$. According to (\ref{eq_PartialEnergyPolePosK2}), it can be written as:
\begin{align}
  \rho_k^{(k)}=\wt{q}_1+n_{k\cdots N}-n_{2\cdots M}-(M-1)-\sum_{j=M+1}^{k-1}(\wt q_j+\wt n_j).
\end{align}
Superficially, it involves parameters at all descendant sites through $\wt q_1$, $\wt q_j$, and $\wt n_j$. However, we recognize that all parameters at Sites $j\in\{M+1,\cdots,k-1\}$ get canceled in $\rho_k^{(k)}$:
\begin{align}
  &\wt q_1-\sum_{j=M+1}^{k-1} \wt q_j=\big[\wt q_1\big]_\text{C},
  &&n_{k\cdots N}-\sum_{j=M+1}^{k-1}\wt n_j=\bigg[\sum_{j\geq k}n_j\bigg]_\text{C}.
 \label{eq_ExcludeFactorized_C}
\end{align}
Here the notation $[\cdots]_\text{C}$ means that we only include sites from $\mathcal{F}\backslash\big(\mathcal{F}_1\cup\cdots\cup\mathcal{F}_{k-M-1}\big)$. Thus $\rho_k^{(k)}$ does not involve any parameter from the subgraphs $\mathcal{F}_a$ with $a=1,\cdots,k-M-1$, so is the factor in the denominator $(\wt q_k+\wh n_k-\rho_k^{(k)})$ right after the summation in (\ref{eq_partiallyFactSeries1}). 

Similarly, let us look at the denominator in the last term of the first line of (\ref{eq_partiallyFactSeries1}), namely $(\wt q_j-\wh\rho_j^{(k)})_{-n_j}$, where $j\in\{2,\cdots,M\}$. Naively, this factor can depends on parameters from the supposedly factorized graphs $\mathcal{F}_a$ $(a=1,\cdots,k-M-1)$. However, let us take a closer look: Suppose that, say, Site $M+1$ is a descendant of Site $j$, and thus so are all descendants of Site $M+1$. (This amounts to saying that $\mathcal{F}_1$ is a descendant subgraph of Site $j$.) Then, according to (\ref{eq_PartialEnergyPolePosK2}), we can rewrite $ \wt q_j-\wh\rho_j^{(k)} $ as:
\begin{align}
  \wt q_j-\wh\rho_j^{(k)} =&~\wt q_j-\sum_{\ell\in\mathcal{D}(j)\backslash\mathcal{F}_1}\rho_\ell^{(k)}-\rho_{M+1}^{(k)}-\sum_{\ell\in\mathcal{D}(M+1)}\rho_{\ell}^{(k)}\n\\
  =&~\wt q_j-\sum_{\ell\in\mathcal{D}(j)\backslash\mathcal{F}_1}\rho_\ell^{(k)}-\big(\wt q_{M+1}+\wt n_{M+1}\big)+\sum_{\ell\in\mathcal{D}(M+1)}n_\ell\n\\
  =&~q_j+\sum_{\ell\in\mathcal{D}(j)\backslash\mathcal{F}_1}\Big[q_\ell-\rho_\ell^{(k)}\Big]=\Big[\wt q_\ell-\wh\rho_\ell^{(k)}\Big]_{\mathcal{F}\backslash\mathcal{F}_1}.
\end{align} 
Here, in the first equality in the last line, we have used the fact that $n_{M+1}=0$. We can repeat this arguments for all descendants $s_\ell$ of $s_j$ with $\ell\in\{M+1,k-M-1\}$, and prove the following equality:
\bge
  \wt q_j-\wh\rho_j^{(k)}=\Big[\wt q_j-\wh\rho_j^{(k)}\Big]_\text{C}.~~~~(j=2,\cdots,M)
\ede  
With all above points understood, and also with some recombination of energy factors, we can massage the series (\ref{eq_partiallyFactSeries1}) into a factorized form:
\begin{align}
    &\bigg[\frac{(-\ii)^{M-1}}{(\ii\omega_k)^{\wt{q}_1}} \sum_{\{n\}}     \frac{\Gamma (\rho_k^{(k)} )(-\Omega_1/\omega_k)^{n_k}}{n_k! (\wt{q}_k+\wh{n}_k-\rho_k^{(k)} )}
      \prod_{j=2}^{M}\FR{(\omega_k/\Omega_j)^{n_j+1}}{(\wt q_j-\wh\rho_j^{(k)})_{-n_j}}\prod_{j=k+1}^{N} \frac{(-\omega_j/\omega_k)^{n_j}}{n_j! \left(\wt{q}_j+\wt{n}_j\right)}\bigg]_\text{C}
    \notag \\
    & \times \prod_{j=M+1}^{k-1}\FR{1}{(\ii\omega_j)^{\wt q_j}}\sum_{\{n\}}\Gamma(\wt{q}_j+\wh{n}_j) \prod_{\ell\in\mathcal{D}(j)} \frac{(-\omega_\ell/\omega_j)^{n_\ell}}{n_\ell! \left(\wt{q}_\ell+\wt{n}_\ell\right)} \ .
\end{align}
Here the second line is nothing but the products of all $\mathcal{F}_a$ with $a=1,\cdots,k-M-1$. Thus, we can rewrite it as:
\begin{align}
    &\bigg[\frac{(-\ii)^{M-1}}{(\ii\omega_k)^{\wt{q}_1}} \sum_{\{n\}}     \frac{\Gamma (\rho_k^{(k)} )(-\Omega_1/\omega_k)^{n_k}}{n_k! (\wt{q}_k+\wh{n}_k-\rho_k^{(k)} )}
      \prod_{j=2}^{M}\FR{(\omega_k/\Omega_j)^{n_j+1}}{(\wt q_j-\wh\rho_j^{(k)})_{-n_j}}\prod_{j>k}\frac{(-\omega_j/\omega_k)^{n_j}}{n_j! \left(\wt{q}_j+\wt{n}_j\right)}\bigg]_\text{C}\times\!\!\prod_{a=1}^{k-M-1}\!\!\mathcal{F}_a \ .
\end{align}
\begin{figure}[t]
\centering
\includegraphics[width=0.75\textwidth]{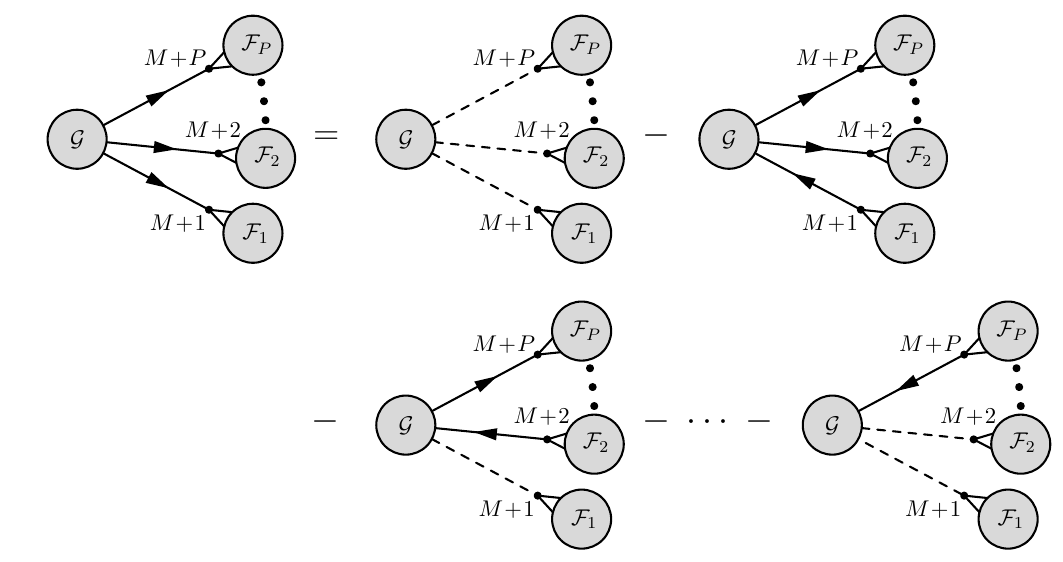}
\caption{The structure of series representation of the family tree $\ft{\mathscr{P}(\wh 12\cdots N)}$ around the zero partial-energy limit $\omega_{1\cdots M}\to 0$. We show explicitly the root site $M+a$ of each subgraph $\mathcal{F}_a$ $(a=1,\cdots,P)$ and all her descendants in a blob, with the understanding that $M+a$ is part of $\mathcal{F}_a$.}
\label{fig_SmallPE}
\end{figure}

Now, we are ready to sum over all series obtained above and write down our final expression for the family tree $\mathcal{F}=\ft{\mathscr{P}(\wh 12\cdots N)}$ expanded in the small $\omega_{1\cdots M}$ limit:

\begin{keyeqn}
\begin{align}
\label{eq_SmallPESeries}
  &\ft{\mathscr{P}(\wh 12\cdots N)}=\mathop{\text{Sg}}_{\omega_{1\cdots M}\to0}[\mathscr{P}(\wh 12\cdots M)]\times \prod_{a=1}^P\mathcal{F}_a\n\\
  &+\sum_{k=2}^M\frac{(-1)^k}{(-\ii \Omega_k)^{\wt{q}_1}} \sum_{\{n\}}^{\infty} \FR{\pi(-\Omega_1/\Omega_k)^{n_k}\prod_{i=2}^{k-1}(\Omega_k/\Omega_i)^{n_i+1}\prod_{j={k+1}}^{M}(\Omega_{j}/\Omega_k)^{n_j}}{ n_k!\sin (\pi\wt q_1)\prod_{j=2}^{M} (\wt q_j-\wh\si_{j}^{(k)})_{1-\si_j^{(k)}}} 
    \prod_{j=M+1}^{N} \frac{(-\omega_{j}/\Omega_k)^{n_j}}{n_j!\left(\wt{q}_j+\wt{n}_j\right)}
\n\\
  &+\sum_{k=M+1}^{M+P}\!\bigg[\frac{(-\ii)^{M-1}}{(\ii\omega_k)^{\wt{q}_1}} \!\sum_{\{n\}}     \frac{\Gamma (\rho_k^{(k)} )(-\Omega_1/\omega_k)^{n_k}}{n_k! (\wt{q}_k+\wh{n}_k-\rho_k^{(k)} )}
      \prod_{j=2}^{M}\FR{(\omega_k/\Omega_j)^{n_j+1}}{(\wt q_j-\wh\rho_j^{(k)})_{-n_j}}\prod_{j>k}\frac{(-\omega_j/\omega_k)^{n_j}}{n_j! \left(\wt{q}_j+\wt{n}_j\right)}\bigg]_\text{C}\!  \times\!\! \prod_{a=1}^{k-M-1}\!\!\mathcal{F}_a.
\end{align}
\end{keyeqn}
Here, $\si_j^{(k)}$ are given by (\ref{eq_TotalEnergyPolePosK}) and $\rho_j^{(k)}$ by (\ref{eq_PartialEnergyPolePosK2}), and $[\cdots]_\text{C}$ means that we exclude all sites from the factorized graphs. For a given $k\in\{M+1,M+P\}$, these are the sites from $\mathcal{F}_a$ with $a\in\{1,\cdots,k-M-1\}$.

Although (\ref{eq_SmallPESeries}) looks quite complicated, it does have the form we expected from the flip-direction arguments. We show this structure in Fig.\;\ref{fig_SmallPE}. On the right hand side of the equality in Fig.\;\ref{fig_SmallPE}, the first graph corresponds to the singular series, which is the first line of the right hand side of (\ref{eq_SmallPESeries}). The second graph corresponds to the totally unfactorized series, and is the second line of (\ref{eq_SmallPESeries}) plus the $k=M+1$ term of the third line of (\ref{eq_SmallPESeries}). The rest of graphs are partially factorized series, and correspond to all but $k=M+1$ series in the third line of (\ref{eq_SmallPESeries}).

\begin{figure}[t]
\centering
\includegraphics[width=0.95\textwidth]{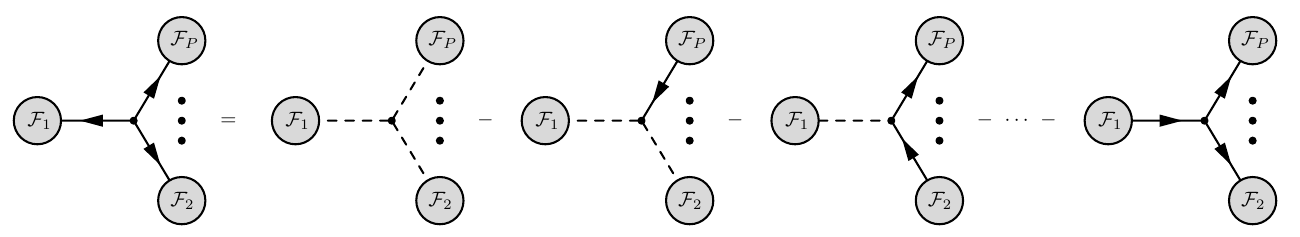}
\caption{The decomposition of a family tree at the zero root-energy limit.}
\label{fig_smallroot}
\end{figure}

\paragraph{Zero root-energy series} 
The series (\ref{eq_SmallPESeries}) is the most general expression for any root-bearing partial energy of a family tree. It is trivial to see that it reduces to the small total-energy expansion (\ref{eq_TEseries}) when taking $M=N$. As another check, it is also interesting to look at the case of $M=1$. That is, we consider the small root-energy limit $\omega_1\to 0$ of an arbitrary family tree $\ft{\mathscr{P}(\wh 12\cdots N)}$. Then, the second line of (\ref{eq_SmallPESeries}) disappears completely and the whole expression collapses to:
\begin{align}
\label{eq_ZeroRootSeries}
  &\ft{\mathscr{P}(\wh 12\cdots N)}=\frac{-\ii\Gamma(q_1)}{(\ii {\omega}_1)^{{q}_1}} \prod_{a=1}^P\mathcal{F}_a 
   +\sum_{k=2}^{P+1}\bigg[\frac{-1}{(\ii\omega_k)^{\wt{q}_1}} \sum_{\{n\}}\Gamma (\wt q_k+\wh n_k) 
\prod_{\substack{j=1\text{\,or\,}>k}}\frac{(-\omega_j/\omega_k)^{n_j}}{n_j! \left(\wt{q}_j+\wt{n}_j\right)}\bigg]_\text{C}\!\! \times  \prod_{a=1}^{k-2}  \mathcal{F}_a.
\end{align}
Note that, for a given $k\in\{2,\cdots,P+1\}$, the series in $[\cdots]_\text{C}$ is in a standard form of the family tree $\mathcal{F}\backslash\{\mathcal{F}_1\cup\cdots\cup\mathcal{F}_{k-2}\}$ with Site $k$ being the root, and is expanded in the large $\omega_k$ limit. This is expected because the flipping-direction argument is precise in this case, as shown in Fig.\;\ref{fig_smallroot}. We could have derived this result directly without any heavy machinery developed in this section, but it is reassuring that our full result (\ref{eq_SmallPESeries}) does reduce to this simple and expected expression in a special case.

\section{Factorization Theorem at Zero Partial Energies}
\label{sec_Factorization}

In the previous section, we obtain the most general series representation of a family tree $\mathcal{F}$ at arbitrary small root-bearing partial energy limits $\omega_\text{G}\to 0$ where $\mathcal{G}$ is the subgraph determined by the partial energy. An interesting corollary of our explicit expression (\ref{eq_SmallPESeries}) is that the singular part of $\mathcal{F}$ at $\omega_{G}\to 0$ is from and only from the part where $\mathcal{G}$ is totally factorized (disconnected) from $\mathcal{F}\backslash\mathcal{G}$. Moreover, the factorization is not just a leading-order result; It holds to all orders in $\omega_\mathcal{G}$. We summarize this result as a theorem:
\begin{keytext}
\textbf{Theorem}~(\emph{Factorization theorem at zero partial energies})~Let $\mathcal{F}$ be an arbitrary family tree and $\mathcal{G}$ be a proper root-bearing sub-family tree of $\mathcal{F}$ (proper means $\mathcal{G}\neq \mathcal{F}$). The complement of $\mathcal{G}$, $\mathcal{F}\backslash\mathcal{G}$, is generally $P\geq 1$ disconnected sub-family trees which we denote by $\mathcal{F}_a$ $(a=1,\cdots,P)$. Let the total energy of $\mathcal{G}$ be $\omega_\mathcal{G}$. Then, the singular part of $\mathcal{F}$ is:
\bge
\label{eq_PEFactThm}
  \mathop{\text{Sg}}_{\omega_\mathcal{G}\to 0}\mathcal{F}=\bigg(\mathop{\text{Sg}}_{\omega_\mathcal{G}\to 0}\mathcal{G}\bigg)\prod_{a=1}^P\mathcal{F}_a.
\ede
\end{keytext}
Although we get this theorem as a corollary of the explicit series representation, it can be more easily proved by Landau analysis, which comes in the following simple steps:
\begin{enumerate}
  \item We use the identity $\theta_{ij}+\theta_{ji}=1$ to flip all lines connecting $\mathcal{G}$ and $\mathcal{F}\backslash\mathcal{G}$, as shown in Fig.\;\ref{fig_SmallPE}.
  \item For all but the first terms on the right hand side of the equality in Fig.\;\ref{fig_SmallPE}, we further flip lines inside $\mathcal{G}$ to make all these terms to be sum of products of family trees. (It is trivial to see that this is always possible.) Importantly, in these terms, $\omega_\mathcal{G}$ never appears as a root-bearing partial energy. 
  \item Then the analysis of Sec.\;\ref{sec_SingularityProof} shows that only the first graph on the right hand side of Fig.\;\ref{fig_SmallPE} could possibly be singular in the $\omega_\mathcal{G}\to 0$ limit, and the singular part of this graph is given by (\ref{eq_PEFactThm}). This completes the proof.
\end{enumerate}

The proof here may look a bit formal, but it is actually quite intuitive from the bulk viewpoint. To see this, we return to the original time integral (\ref{eq_NestedIntegral}). In general, such integrals are convergent in the early time limit thanks to the $\ii\epsilon$-prescription of an exponential function $e^{\pm\ii\omega\tau}$. However, a singularity in the partial energy $\omega_\mathcal{G}\to 0$ may arise whenever there is a factor of $e^{\pm\ii\omega_\mathcal{G}\tau}$ where the time $\tau$ can reach $-\infty$. This is the case for the first (fully factorized) term on the right hand side of Fig.\;\ref{fig_SmallPE}, but not the case for other terms where $\mathcal{G}$ is not factorized. In the latter cases, the time $\tau$ in the factor $e^{\ii\omega_\mathcal{G}\tau}$ is always bounded by some earlier time from $\mathcal{F}_a$ due to the lines going into $\mathcal{G}$. So, the upshot is that only the fully factorized graph on the right hand side of Fig.\;\ref{fig_SmallPE} could be singular.

\section{Analytical Structure of Family Trees in Twist Space}
\label{sec_twists}

The FTD is designed to deal with the nested time integral of powers and exponentials encountered in the cosmological correlators. However, such nested time integrals are not always the final result. For massive particles, we usually dissolve special functions (like Hankel's functions) into powers using PMB representation. Then, some Mellin variables enter the power functions $(-\tau_j)^{q_j-1}$. Since the MB integral from the special functions should be finished in the end, it would be useful to know the analytical property of family trees with respect to the twists $q_j$. 

The analytical property of family trees in twist space is much simpler than in energy space. From the bulk (time integral) viewpoint via (\ref{eq_NestedIntegral}), it is immediately clear that, for generic choices of energy variables where the family trees are away from all energy singularities discussed before, all singularities at finite values of twists $q_j$ arise from divergences in the late-time limit. Furthermore, it is easy to see that all such singularities can be directly isolated as simple poles of $\Gamma$ factors. 

To make it clear, let us look at a familiar example of the two-site chain, which is nothing but a dressed Gauss hypergeometric function ${}_2\mathcal{F}_1$. This function can be defined by a series, and is related to a ``regularized'' hypergeometric function ${}_2\wt{\text{F}}_1$ in the following way:
\begin{align}
    \ _2\mathcal{F}_1 \left[
    \begin{matrix}
        a,b \\
        c
    \end{matrix} \middle| z
    \right]
&=\sum_{n=0}^{\infty} \frac{z^n}{n!} \Gamma \left[
\begin{matrix}
    a+n,b+n \\
    c+n
\end{matrix}
\right]  
=\Gamma\left[a,b\right]\ _2\wt{\text{F}}_1 \left[
    \begin{matrix}
        a,b \\
        c
    \end{matrix} \middle| z
    \right] \ .
\end{align}
From the series above, it is clear that the dressed function ${}_2\mathcal{F}_1$ could be singular whenever $a$ or $b$ takes nonpositive integer values, because when this happens, a finite number of terms in the series hit poles of $\Gamma(a+n,b+n)$. In contrast, the regularized function ${}_2\wt{\text{F}}_1$ is free of such poles, because all poles from $\Gamma[a+n,b+n]$ are killed by zeroes of $1/\Gamma[a,b]$. Therefore, we have identified all poles of the dressed hypergeometric function as $\Gamma$ poles, while the regularized version ${}_2\wt{\text{F}}_1$ is an entire function of $a$, $b$, and $c$, when it is away from singularities of $z$.

It is straightforward to ``regularize'' a general family tree in the same way. That is, for every family tree $\ft{\cdots}$, we define a regularized family tree $\ft{\cdots}_\text{R}$ by: 
\begin{align}
    \ft{\mathscr{P}(\wh{1}2\cdots N)}\equiv\  \ft{\mathscr{P}(\wh{1}2\cdots N)}_\text{R} \times \prod_{j=1}^{N} \Gamma(\wt{q}_j) \ ,
    \label{eq_FamilyTreeRegularized}
\end{align}
Then, using the large root-energy series (\ref{eq_LargeMaximalSeries}), we see that the regularized family tree has the following series representation: 
\begin{align}
    \ii^N  \ft{\mathscr{P}(\wh{1}2\cdots N)}_\text{R}=\frac{1}{(\ii \omega_1)^{\wt{q}_1}} \sum_{n_2,\cdots,n_N=0}^{\infty}\left(\wt{q}_1\right)_{\wh n_1}\prod_{j=2}^{N} \frac{\left(\wt{q}_j\right)_{\wt{n}_j}\left(-\omega_j/\omega_1\right)^{n_j}}{\Gamma(\wt{q}_j+\wt{n}_j+1)n_j!} \ .
    \label{eq_FamilyTreeRegularizedMaximalEnergySeries}
\end{align}
From this series representation, it is clear that the regularized family tree $\ft{\cdots}_\text{R}$ is an entire function of all twists $q_j$ $(j=1,\cdots, N)$ at least in the convergence domain of above series. Also, by analytical continuation in $\omega_j$, the regularized family tree remains an entire function of $q_j$ for all nonsingular values of energy variables. Thus, all singularities in the original family tree $\ft{\cdots}$ in $q_j$ are simple poles and are all from the factor $\Gamma[\wt{q}_1,\cdots,\wt{q}_N]$ in (\ref{eq_FamilyTreeRegularized}). 

\paragraph{Positive integer twists}
In the small partial-energy series (\ref{eq_SmallPESeries}), as well as its special case of small total-energy series (\ref{eq_TEseries}), we have various $\Gamma$ factors whose arguments contain summation variables with negative signs, such as $\Gamma(q_{1234}-n_{234}-3)$ in (\ref{eq_1234TEseries}). It seems that such $\Gamma$ factors lead to poles at nonnegative integer values of twists, contradicting our previous conclusion that all poles of family trees in twists $q_j$ are from $\Gamma[\wt q_1,\cdots,\wt q_N]$. 

The solution to this puzzle is that all poles in twists from factors like $\Gamma(\cdots -n_j)$ must cancel among different series in (\ref{eq_TEseries}) and (\ref{eq_SmallPESeries}). Below we show this cancelation with an example of $\ft{12}$. 

First, the large total-energy series (\ref{eq_LargeTEseries}) of $\ft{12}$ can be written as:
\begin{align}
    \ft{12}&=
    \frac{-1}{(\ii \omega_{12})^{q_{12}}} \sum_{n_2=0}^\infty 
    \FR{\Gamma(q_{12}+n_2)(\omega_2/\omega_{12})^{n_2}}{(q_2)_{n_2+1}}
     =\frac{-\Gamma\left[q_{12},q_2\right]}{(\ii \omega_{12})^{q_{12}}} \ _2 \wt{\text F}_1 \left[
    \begin{matrix}
        1,q_{12}\\
        q_2+1
    \end{matrix}
    \middle| \frac{\omega_2}{\omega_{12}}
    \right] .
\end{align}
Since ${}_2\wt{\text{F}}_1$ is an entire function in $q_1$ and $q_2$, we know that all singularities of $\ft{12}$ in twists are at nonpositive integers $q_{12}$ and $q_2$ through the factor $\Gamma[q_{12},q_2]$. On the other hand, we can write the small total-energy series (\ref{eq_TEseries}) for $\omega_{12}$ as:
\begin{align}
    \ft{12}= &~\frac{1}{(\ii \omega_{12})^{q_{12}}} \sum_{n_2=0}^\infty 
    \FR{\Gamma(q_{12}-n_2-1)(\omega_{12}/\omega_2)^{n_2+1}}{(q_{2})_{-n_2}}
     -\FR{\pi\csc(\pi q_{12})}{(-\ii \omega_2)^{q_{12}}} \sum_{n_2=0}^\infty 
    \FR{\Gamma(q_2)(-\omega_{12}/\omega_2)^{n_2}}{n_2!\Gamma(1-q_1-n_2)}\n\\
    =&-\FR{\pi\csc(\pi q_{12})}{(\ii\omega_{12})^{q_{12}}}\FR{\omega_{12}}{\omega_2}{}_2\wt{\text F}_1\left[\bgm 1,1-q_2\\ 2-q_{12} \edm\middle|\FR{\omega_{12}}{\omega_2}\right]-\FR{\pi\csc(\pi q_{12})\Gamma(q_2)}{\Gamma(1-q_1)(\ii\omega_{1})^{q_1}(-\ii\omega_{2})^{q_{2}}} .
\end{align}
Here the first term (singular in $\omega_{12}\to 0$) contains poles for all $q_{12}\in\mathbb{Z}$ from $\csc(\pi q_{12})$, and the second term (regular in $\omega_{12}\to 0$) contains the same $q_{12}\in\mathbb{Z}$ poles, as well as poles at $q_2\in\mathbb{Z}_-$ from $\Gamma(q_2)$ factors. As mentioned above, we expect that all $q_{12}\in\mathbb{Z}_+$ poles cancel between the two terms. Indeed, when $q_{12}=n\in\mathbb{Z}_+$, ${}_2\wt{\text F}_1$ reduces to an elementary function: 
\bge
  {}_2\wt{\text F}_1\left[\bgm 1,1-q_2 \\ 2-n \edm\middle|z \right]=(1-q_2)_{n-1}\FR{z^{n-1}}{(1-z)^{n-q_2}}.
\ede
Then it is elementary algebra to show that the residues of poles at $q_{12}\in\mathbb{Z}_+$ from the first term equal to the negative of the corresponding residues from the second term, consistent with our expectation that $\ft{12}$ has no poles at these positions.

Notice that when all $q_j$ are positive integers, all family trees simply reduce to rational functions of energy variables \cite{Fan:2024iek}, and all expansions are trivial. However, what we have been considering here are slightly more general: In the $\ft{12}$ example, we can have $q_{12}$ being a positive integer but $q_1$ and $q_2$ can take general real values. In these cases, the family trees are not elementary and our series representations are useful. Also, we don't consider the case when any of $q_j$ or $\tilde q_j$ equal to 0. In these cases, there are infrared divergences in the family trees which must cancel out when combining family trees into correlators, and the resulting correlators are polylogarithmic functions \cite{Hillman:2019wgh,Arkani-Hamed:2023kig,Fan:2024iek}.

\section{Conclusion and Outlook}
\label{sec_Outlook}
The cosmological correlators are the central objects in the study of the primordial universe, both for theoretical study and observational purposes. The analytical approach plays a crucial role in our understanding of cosmological correlators, but is plagued by numerous technical complications, such as the use of special functions and nested time integrals for Feynman diagrams. With the recently developed family-tree decomposition method, the nested time integrals have been systematically reduced into canonical objects called family trees. These objects capture the time-order structure of both the cosmological correlators and the wavefunction coefficients, in the dS spacetime background and a more general power-law FRW background. However, they are represented by MHFs, which are poorly understood. Therefore, it is beneficial to have a comprehensive understanding of the analytical structure of the family trees. 

In this work, we identify all the singularities of family trees with respect to their arguments (energy variables) as well as parameters (twists). The singularities of energy space can be put into two categories: the zero and infinite partial-energy singularities (where the single energy and total energy are recognized as two special cases of partial energies). We derive a series representation of family trees around each of these singular points. All these series are hypergeometric series over energy ratios, and the singular term is always associated with a monodromy factor responsible for the nonanalyticity. As a byproduct, we derive a factorization theorem, showing that the nonanalytical part of a family tree at a zero partial-energy limit is factorized to all orders in the small partial energy. 

On the other hand, the analytical structure of family trees in the twist space is much simpler: They have only simple poles coming from Euler $\Gamma$ factors and thus are meromorphic. After regulating family trees by removing these $\Gamma$ factors, the regularized family trees are entire functions of twists as long as the energy variables are away from their singularities. 

With all results obtained in this work, we have a firm understanding of family trees in the vicinity of every one of their singularities. This knowledge opens up new possibilities and also brings up new questions in this direction, which merit future studies. We finish this work by mentioning several of them.

First, it would be useful to explore the consequences of our results for general massive cosmological correlators. In the scheme of PMB+FTD, the family trees are part of the MB integrand for massive correlators. The singularities of the integrand usually control the  analytical property of the integrated function (as in Landau analysis). In this regard, the singularities of family trees in energy space and twist space are both useful. On the one hand, the energies in family trees correspond to energies of external lines in the full correlators (``vertex energy'' in the terminology of \cite{Liu:2024str}). Thus, the energy singularities of family trees are useful to analyze the nonanalytic behavior of the full correlators in vertex energies, such as the local signals or total-energy singularity. Moreover, we believe that the factorization theorem for partial energies given in this work has a counterpart in general massive correlators, even at the loop level. On the other hand, the twist singularities (simple poles) of family trees translate to poles in Mellin variables for the full correlator. These poles could make contributions to the final result of the massive correlators in certain parameter space. 

It is potentially more straightforward to analyze massive correlators by directly working with massive family trees introduced in \cite{Liu:2024str}. If we can reproduce what we have here for massive family trees, we will have full knowledge about the singularities of massive correlators, bypassing the procedure of PMB.

Second, it would be interesting to investigate the analytical structure of loop correlators using the results here. In particular, the FTD still works for loop graphs simply because the time flow does not form a loop, i.e., $\theta_{12}\cdots\theta_{N-1,N}\theta_{N1}=0$. Thus, we can expect to get similar MHF expressions at least for loop integrands with no essential difference from tree graphs. Then we can imagine studying the analytical properties of loop correlators as momentum integrals of either family trees or massive family trees.

Last but not least, we can use the results of this work to explore the family trees in the entire parameter space: The convergence domain of our series representations has covered some regions around singularities but unfortunately not the entire parameter space. Currently, finding series representations for entire parameter space is a challenging mathematical problem that has not been solved, even for simple bivariate hypergeometric functions such as Appell or Kampé de Fériet functions. However, we know that the values of the family tree outside the convergence domain are uniquely fixed by analytical continuation (at a given sheet when there are branch points). Thus, it is a well-defined mathematical problem to numerically work out the values of these functions for the entire parameter space, using the singular series in this work as boundary data, which may facilitate numerical implementations of massive correlators \cite{Wang:2021qez,Werth:2023pfl,Pinol:2023oux,Werth:2024aui}. We leave all these interesting questions for future works.

\paragraph{Acknowledgments} We thank Jiaju Zang for initial collaboration on this project and Jiayi Wu for useful discussions. Partial results of this work have been presented at the workshop ``Cosmology Beyond the Analytic Lamppost (CoBALt)'' at the Institut Pascal at Université Paris-Saclay, and we thank the organizers and participants of the workshop for useful discussions and feedback. 
This work is supported by NSFC under Grants No.\ 12275146 and No.\ 12247103, the National Key R\&D Program of China (2021YFC2203100), and the Dushi Program of Tsinghua University.

\newpage
\begin{appendix}

\section{Notations}
\label{app_Notations}
For readers' convenience, we compile the notations frequently used in this paper in Table \ref{tab_notations} along with the equation numbers where they are defined or first appear.
\begin{table}[htbp]
    \centering
    \caption{List of Symbolic Notations}
    \vspace{2mm}
    \begin{tabular}{lll}
        \toprule[1.5pt]
        Notation & \multicolumn{1}{c}{Description} & Equation \\
        \hline
        $\tau_j$ & Time variable at vertex $j$ & (\ref{eq_GeneralCorrelator}) \\
        $\aa_j$ & SK index at vertex $j$ & (\ref{eq_GeneralCorrelator}) \\
        $E_j$ & Vertex energy & (\ref{eq_GeneralCorrelator}) \\
        $K_i$ & Line energy & (\ref{eq_GeneralCorrelator}) \\
        $\wt\nu$ & Mass parameter, $\sqrt{m^2-9/4}$ & Above (\ref{eq_SK_Propagators}) \\
        $p_j$ & Time dependence exponent & (\ref{eq_GeneralCorrelator}) \\
        $q_j$ & Twist parameter at site $j$ & (\ref{eq_NestedIntegral}) \\
        $\mb{T}^{(q_1\cdots q_N)}_{\mathscr{N}}$ & General time integral & (\ref{eq_NestedIntegral}) \\
        $\omega_j$ & Energy variable & (\ref{eq_NestedIntegral}) \\
        $\mathscr{N}$ & Set of ordered pairs & (\ref{eq_NestedIntegral}) \\
        $\theta_{ij}\equiv\theta(\tau_i-\tau_j)$ & Time-ordering Heaviside $\theta$ function & (\ref{eq_FamilyTreeExamples}) \\
        $\ft{\mathscr{P}(\wh{1}2\cdots N)}$ & Family tree with partial order $\mathcal{P}$ & (\ref{eq_LargeMaximal_1})  \\
        $\mathcal{R}\ft{\mathscr{P}(\wh{1}2\cdots N)}$ & Reversed family tree & Sec.\;\ref{sec_MBRofTI} \\
        $\mathcal{D}(j)$ & Descendants of $j$ & (\ref{eq_wtqjDef}) \\
        $d_j$ & Number of descendants of $j$ & Below (\ref{eq_wtqjDef}) \\
        $\wt q_j$ & Sum of $q$'s of $j$ and her descendants & (\ref{eq_LargeMaximal_1}) \\
        $\wh n_j$ & Sum of $n$'s of descendants & (\ref{eq_LargeMaximal_1}) \\
        $\wt\omega_j$ & Sum of $\omega$'s of $j$ and her descendants (``tilde energy'') & (\ref{eq_LargeTotal_1}) \\
        $\Omega_j$ & Collected energy (``capitalized energy'') & (\ref{eq_RevTreeMB}) \\
        $\mathcal{I}_j(s,q)$ & Integrand factor in MB integrals & (\ref{eq_RevGammaProducts},\ref{eq_GammaProducts}) \\
        $s_j$ & Mellin variable & (\ref{eq_RevTreeMB},\ref{eq_FTMB}) \\
        $\mathcal{C}$ & Root-bearing subgraph & Sec.\;\ref{sec_LargePartialSeries} \\
        $\si_j^{(k)}$ & Pole solutions in total-energy series & (\ref{eq_TotalEnergyPolePosK}) \\
        $\rho_j^{(k)}$ & Pole solutions in partial-energy series & (\ref{eq_PartialEnergyPolePosK2}) \\
        $\mathcal{F}$ & General family tree & (\ref{eq_PEFactThm}) \\
        $\mathcal{G}$ & Root-bearing subtree & (\ref{eq_PEFactThm}) \\
        $\mathcal{F}_a$ & Disconnected subtree & (\ref{eq_PEFactThm}) \\
        $[\cdots]_\text{C}$ & Exclude factorized subtrees & After (\ref{eq_ExcludeFactorized_C}) \\
        $\mathop{\text{Sg}}\limits_{\omega\to 0}$ & Singular part at $\omega\to 0$ & (\ref{eq_SgZeroTotal}) \\
        $\bigshuffle$ & Shuffle product & Sec.\;\ref{sec_SingularityProof} \\
        $\epsilon_i$ & Dual energy variable & (\ref{eq_ChainEnergyInt}) \\
        \bottomrule[1.5pt]
    \end{tabular}
    \label{tab_notations}
\end{table}

\section{Special Functions and Useful Formulae }
\label{app_Functions}
In this appendix we collect the definitions and useful properties of the special functions used in this paper. The Hankel's functions will not be discussed here, as they are not related to the main topic of this paper.

\paragraph{Gamma functions}
The $\Gamma$ function is defined by the Euler's integral 
\begin{align}
    \Gamma(q)\equiv\int_0^{+\infty} \di t\, e^{-t} t^{q-1} \ .    
\end{align}
$\Gamma(q)$ is known to have first order poles with residue $(-1)^n/n!$ at $q=-n$ for $n=0,1,2,\cdots$. This would be our main ingredient when finishing the MB integrals by residue theorem.

In this and previous related works, we use the following compact notations of products and fractions of $\Gamma$ functions:
\begin{align}
    \Gamma\left[q_1,\cdots,q_a\right]&\equiv \Gamma(q_1)\cdots\Gamma(q_a) \ , \\
    \Gamma\left[
    \begin{matrix}
      q_1,\cdots,q_a \\
      p_1,\cdots,p_b
    \end{matrix}
    \right]
    &\equiv \frac{\Gamma(q_1)\cdots\Gamma(q_a)}{\Gamma(p_1)\cdots\Gamma(p_b)} \ .
\end{align}

The $\Gamma$ function has a reflection identity 
\begin{align}
    \Gamma\left[z,1-z\right] = \frac{\pi}{\sin (\pi z)} \ .
\end{align}
Together with the periodicity of the sine function, we have 
\begin{align}
    \Gamma\left[z,1-z\right]=(-1)^n \Gamma\left[z+n,1-z-n\right] \quad (n \in \mathbb{Z}) \ .
\end{align}
This formula is usually used to adjust the $\Gamma$ factors throughout our calculation.

We use a Pochhammer symbol, defined as:
\begin{align}
    (q)_n\equiv \Gamma \left[
    \begin{matrix}
        q+n \\
        q
    \end{matrix}
    \right] \ .
\end{align}

The incomplete $\Gamma$ function is defined as 
\begin{align}
    \Gamma(q,z)\equiv\int_z^{+\infty} \di t\, e^{-t} t^{q-1} \ .    
\end{align}
This should not be confused with $\Gamma[a,b]=\Gamma(a)\Gamma(b)$.

\paragraph{Exponential integral}
We use the exponential integral, defined as 
\begin{align}
    \text{E}_p (z)\equiv z^{p-1} \int_z^{+\infty} \di t\, \frac{e^{-t}}{t^p} \ .
\end{align}
The following relation between incomplete $\Gamma$ function and exponential integral  follows directly from their definitions: 
\begin{align}
    \text{E}_p (z)=z^{p-1} \Gamma(1-p,z) \ .
\end{align}

\paragraph{Hypergeometric functions}
We will encounter the Gauss hypergeometric $_2 \text{F}_1$ function and the Kummer confluent hypergeometric $_1 \text{F}_1$ function. They are defined by hypergeometric series 
\begin{align}
\label{eq_HGdef}
    _2 \text{F}_1 \left[
    \begin{matrix}
        a,b \\
        c
    \end{matrix}
    \middle| z
    \right] &\equiv \sum_{n=0}^{\infty} \frac{(a)_n (b)_n}{(c)_n} \frac{z^n}{n!} \ , \\
    _1 \text{F}_1 \left[
    \begin{matrix}
        a \\
        b
    \end{matrix}
    \middle| z
    \right] &\equiv \sum_{n=0}^{\infty} \frac{(a)_n}{(b)_n} \frac{z^n}{n!} \ .
\end{align}
The series of $_2\text{F}_1$ converges for $|z|<1$ while that of $_1\text{F}_1$ converges for all $z$. Outside the unit disk, $_2\text{F}_1$ is defined through analytical continuation.

Alongside these standard versions, we also have the \textit{regularized} and the \textit{dressed} version of these hypergeometric functions. The regularized versions are defined as 
\begin{align}
    _2 \wt{\text{F}}_1 \left[
    \begin{matrix}
        a,b \\
        c
    \end{matrix}
    \middle| z
    \right] \equiv \frac{1}{\Gamma(c)} \ _2 \text{F}_1 \left[
    \begin{matrix}
        a,b \\
        c
    \end{matrix}
    \middle| z
    \right]  \ , \quad
    _1 \wt{\text{F}}_1 \left[
    \begin{matrix}
        a \\
        b
    \end{matrix}
    \middle| z
    \right] &\equiv \frac{1}{\Gamma(b)} \ _1 \text{F}_1 \left[
    \begin{matrix}
        a \\
        b
    \end{matrix}
    \middle| z
    \right] \ .
\end{align}
These regularized hypergeometric functions are entire functions with respect to their parameters $a,b,c$.

The dressed version of the hypergeometric functions is usually more convenient for the calculation related to the family trees:
\begin{align}
        _2 \mathcal{F}_1 \left[
    \begin{matrix}
        a,b \\
        c
    \end{matrix}
    \middle| z
    \right] \equiv \frac{\Gamma[a,b]}{\Gamma(c)} \ _2 \text{F}_1 \left[
    \begin{matrix}
        a,b \\
        c
    \end{matrix}
    \middle| z
    \right]  \ , \quad
    _1 \mathcal{F}_1 \left[
    \begin{matrix}
        a \\
        b
    \end{matrix}
    \middle| z
    \right] &\equiv \frac{\Gamma(a)}{\Gamma(b)} \ _1 \text{F}_1 \left[
    \begin{matrix}
        a \\
        b
    \end{matrix}
    \middle| z
    \right] \ .
\end{align}
Alternatively, one could define the dressed version from series (\ref{eq_HGdef}) by the replacement $(a)_n\to \Gamma(a+n)$, $(b)_n\to \Gamma(b+n)$, etc.

The asymptotic behavior of the confluent hypergeometric function at $z\to\infty$ is 
\begin{align}
    _1 \mathcal{F}_1 \left[
    \begin{matrix}
        a \\
        b
    \end{matrix}
    \middle| z
    \right]\sim e^z z^{a-b} \ .
\end{align}
This formula is needed when discussing the origin of the zero-energy singularity from the time-integral perspective.

The confluent hypergeometric function has a relation to the incomplete $\Gamma$ function as 
\begin{align}
    \Gamma(q)-\Gamma(q,z)= z^q \ _1\mathcal{F}_1 \left[
    \begin{matrix}
        q \\
        q+1
    \end{matrix}
    \middle| -z
    \right] \ .
\end{align}

\paragraph{Mellin-Barnes representations}
Now we collect the MB representations of special functions used in this work.

The representations of the confluent hypergeometric function are our main tools for performing nested time integral. We can represent the function as
\begin{align}
    _1 \mathcal{F}_1 \left[
    \begin{matrix}
        a \\
        b
    \end{matrix}
    \middle| z
    \right]
    =\int_{-\ii\infty}^{+\ii\infty}\frac{\di s}{2\pi\ii} z^{-s} \Gamma \left[
    \begin{matrix}
        a-s,s \\
        b-s
    \end{matrix}
    \right] \ .
\end{align}
We have kept the parameters general here. When we go to a special case like $a=q,b=q+1$ (the one we encounter in this work), the ratio $\Gamma(a-s)/\Gamma(b-s)$ becomes $1/(q-s)$ with a single pole at $s=q$. The general representation tells us to treat this pole as a right pole. That is, the choice of integral path (which is part of the \textit{definition} of the integral) is such that this pole lies on the right side of the path.

We can also leave an exponential function out and represent the remaining part as 
\begin{align}
    _1 \mathcal{F}_1 \left[
    \begin{matrix}
        a \\
        b
    \end{matrix}
    \middle| z
    \right]
    =e^z \int_{-\ii\infty}^{+\ii\infty}\frac{\di s}{2\pi\ii} z^{-s} \Gamma \left[
    \begin{matrix}
        s,b-a-s,a \\
        b-s,b-a
    \end{matrix}
    \right] \ .
\end{align}
When converting reversed family trees into MB integrals, we use the representations for the exponential integrals. Similarly we have two choices, either completely or partially resolved. The completely resolved version is:
\begin{align}
    \text{E}_p(z)=\int_{-\ii\infty}^{+\ii\infty}\frac{\di s}{2\pi\ii} z^{-s} \Gamma \left[
    \begin{matrix}
        s,s+p-1 \\
        s+p
    \end{matrix}
    \right] \ .
\end{align}
Like before, we have a single pole at $s=1-p$, but this time we should treat it as a left pole. Again, we emphasize that treating the pole as being left or right is a part of the definition of the integral. We make this definition manifest by writing the single-pole fraction as a $\Gamma$ function ratio.

Similar to before, we have the partially resolved representation with an exponential function left:
\begin{align}
    \text{E}_p(z)=e^{-z} \int_{-\ii\infty}^{+\ii\infty}\frac{\di s}{2\pi\ii} z^{-s} \Gamma \left[
    \begin{matrix}
        s,1-s,s+p-1 \\
        p
    \end{matrix}
    \right] \ .
\end{align}

\section{Infinite-Energy Singularities and the Projective Kinematic Space}
\label{app_CPN}

In this work, we have frequently discussed limits where one or more energy variables become infinite. A proper analysis of such limits requires a consistent way to handle the ``infinity'' of the kinematic space $\mathbb{C}^N$ through compactification. This appendix clarifies why the complex projective space $\mathbb{CP}^N$ is an appropriate way to compactify $\mathbb{C}^N$ and how our series expressions provide a local description of family trees around different points on its singularities at infinity.

When $N>1$, ``infinity'' is not a single point, and the behavior of a function depends on the direction of approaching infinity. A naive compactification using reciprocal coordinates ($\si_j=1/\omega_j$) fails to capture the correct physics. To see this, we compare two types of limits in the $N=2$ case.
\begin{enumerate}
    \item Limit A: $|\omega_1| \to \infty$, while $\omega_2$ is held at a finite constant $C$.
    \item Limit B: Both $|\omega_1|, |\omega_2| \to \infty$, while their ratio $\omega_2/\omega_1$ is held at a fixed constant $R$.
\end{enumerate}
Our results in Sec.\;\ref{sec_LargeEnergy} show that family trees are primarily functions of energy ratios. This means the function's behavior should be insensitive to the specific value of $C$ in Limit A (since $\omega_2/\omega_1 \to 0$ in all cases), but sensitive to the value of the ratio $R$ in Limit B. However, reciprocal coordinates do the opposite: they distinguish among different limits in A by mapping them to distinct points $(0, 1/C)$, but conflate all limits in B by mapping them to the single point $(0,0)$ approached from different directions. This mismatch shows that reciprocal coordinates are not entirely satisfactory.

For our purpose, a proper framework is the complex projective space $\mathbb{CP}^N$. A point in $\mathbb{CP}^N$ is described by an equivalence class of homogeneous coordinates $[Z_0 : Z_1 : \dots : Z_N]$, where these are complex numbers not all zero, and two points are equivalent if they differ by an overall nonzero complex scaling factor $c$, $[Z_0 : \dots : Z_N] \sim [cZ_0 : \dots : cZ_N]$.

The connection to our physical kinematic space is made by identifying a point $(\omega_1, \dots, \omega_N) \in \mathbb{C}^N$ with the point $[1 : \omega_1 : \dots : \omega_N]$ in the projective space. This mapping describes the part of $\mathbb{CP}^N$ where $Z_0 \neq 0$. The remaining points, where $Z_0=0$, constitute the boundary at ``infinity'' of $\mathbb{C}^N$. The structure of this boundary is itself a projective space. Any point at infinity has the form $[0 : Z_1 : Z_2 : \dots : Z_N]$, where the $Z_1, \dots, Z_N$ are not all zero. The equivalence relation acts only on these $N$ coordinates: $[0 : Z_1 : \dots : Z_N] \sim [0 : cZ_1 : \dots : cZ_N]$. This is precisely the definition of the complex projective space $\mathbb{CP}^{N-1}$ for the coordinates $[Z_1: \dots: Z_N]$. Thus, in this procedure, we are compactifying $\mathbb{C}^N$ into $\mathbb{CP}^{N}$ by adding $\mathbb{CP}^{N-1}$ at infinity.

Let's re-examine our test limits using this formalism:
\begin{enumerate}
    \item Limit A ($|\omega_1|\to\infty, \omega_2=C$): the point $[1:\omega_1:\omega_2]$ is equivalent to $[1/\omega_1:1:\omega_2/\omega_1]$. As $|\omega_1|\to\infty$, this approaches $[0:1:0]$ regardless of the finite value of $C$. All such limits converge to a \textit{single point} at infinity.
    \item Limit B ($|\omega_1|,|\omega_2|\to\infty, \omega_2/\omega_1=R$): the point is equivalent to $[1/\omega_1:1:\omega_2/\omega_1]$, which approaches $[0:1:R]$. Different ratios $R$ correspond to \textit{distinct points} in the $\mathbb{CP}^{N-1}$ infinity boundary.
\end{enumerate}
The projective space correctly groups and distinguishes limits based on ratios, which is precisely the structure exhibited by family trees.

In this picture, the entire hyperplane of complex codimension 1 at infinity acts as a singularity of the family tree. Our large partial-energy series in Sec.\,\ref{sec_LargeEnergy} should be understood as series representations valid around specific points on this singular boundary. For a limit where a partial energy $\Omega_1$ becomes large, the series is expressed in powers of ratios like $\omega_j/\Omega_1$. These ratios are the natural coordinates on the projective space near that limit. The overall monodromy factor, $(\ii\Omega_1)^{-\wt{q}_1}$, captures the nonanalytical behavior of the family tree as it moves around the infinity boundary, while the remaining hypergeometric series describes the regular variations of the function along the singular hyperplane.

The methods developed in this paper are flexible enough to derive local series representations in the vicinity of many other points on the infinity boundary. The series presented in Sec.\;\ref{sec_LargeEnergy} cover the most physically relevant cases (e.g., large single, partial, or total energies), providing a powerful toolkit for understanding the family trees near the singularity at infinity.

\section{Multi-Fold Mellin-Barnes Integrals and Convergence}
\label{app_MB}

In this paper, we have converted nested time integrals into multifold MB integrals, which we then evaluate by summing residues. For an integral involving $N-1$ Mellin variables $\bm{s}\equiv(s_2, \dots, s_N)$, there are often multiple families of poles for each variable. Choosing a combination of poles --- one family for each variable --- yields a particular multivariate hypergeometric series. A crucial task is to identify the correct combinations of poles whose corresponding series converge in a given kinematic region. For integrals with more than one variable, caution should be exercised to avoid overcounting or omitting series, and a systematic procedure is desirable.

In the literature, a well-developed approach for this problem is the conic hull or triangulation method \cite{Ananthanarayan:2020fhl,Banik:2023rrz}. However, we adopted an algebraic procedure in the main text which is convenient enough for our purpose. In this appendix, we systematize our algebraic procedure and use the four-site chain from Sec.\;\ref{sec_SmallTE} as a working example to illustrate the method. Then, we give a simple proof of the equivalence between our algebraic method and the conic hull method.

\paragraph{The algebraic procedure}
A general MB integral encountered in our work takes the form
\begin{align}
    \mathcal{I}=\int \frac{\di^{N-1}\bm{s}}{(2\pi\ii)^{N-1}} \, \mathcal{K}(\bm{s}) \prod_{j=2}^{N} z_j^{s_j} \ ,
\end{align}
where $\mathcal{K}(\bm{s})$ is a product of $\Gamma$ functions of $\bm s$ and $z_j$ are ratios of energy variables. After finishing this integral, the resulting series is expected to converge only in a particular region in the complex space of $\{z_j\}$. Our procedure to find such a series representation has been demonstrated through many examples in the main text, and can be summarized as follows:

\begin{enumerate}
    \item \textbf{Specifying a kinematic region~} We specify a region in the kinematic space where the series expression is formally convergent (see below). For the integrands in this work, it is sufficient to fix the relative magnitudes of the energy variables. For more general cases, one may need to specify the magnitudes of power combinations like $|z_2^3|<|z_3^2 z_4|$ but this never occurs for our representations of family trees.

    \item \textbf{Identifying pole-generating factors~} We list all $\Gamma$ functions in the numerator of the integrand $\mathcal{K}(\bm{s})$ whose arguments contain Mellin variables $\bm s$. In our cases, the arguments of a $\Gamma$ factor is always at most a linear function of (some components of) $\bm s$ in the form of $\Gamma(a+\bm b\cdot\bm s)$ where the vector $\bm b$'s components take values from $\{0,\pm 1\}$. Thus, picking up poles from such a $\Gamma$ factor gives rise to a linear equation in $\bm s_j$, i.e., $a+\bm b\cdot\bm s=-n$ where $n\in\mathbb{N}$.

    \item \textbf{Selecting a set of $\bm\Gamma$ factors~} According to above item, selecting a particular set of poles to finish the MB integral is equivalent to solving for $\bm s$ from the set of linear equations from the selected $\Gamma$ factors. For the solution of $\bm s$ to be uniquely determined, we must select exactly $N-1$ $\Gamma$ factors whose arguments, viewed as linear functions of $\bm{s}$, are linearly independent. There may be many different choices of $\Gamma$ factors and each choice corresponds to a potential series contribution.

    \item \textbf{Checking formal convergence~} For each choice of combination of $\Gamma$ factors as above, we solve for $\bm{s}$ in terms of summation indices $n_j \in \mathbb{N}$ ($j=2,\cdots,N$). We then substitute this solution into the energy-ratio product $\prod z_j^{s_j}$, which becomes a new product $\prod (z'_k)^{n_k}$ up to an overall factor. The series is deemed \emph{formally convergent} and thus makes a valid contribution to the final result if and only if $|z'_k|<1$ for all $k$ in the chosen kinematic region. The final result is simply the sum of all formally convergent series in the chosen kinematic region. 
\end{enumerate}

\paragraph{Example: the four-site chain}
Although we have used the above method in many examples in the main text, here, we use an example to clarify the meaning of each step above. Our example is $\ft{1234}$ in its small total-energy limit. The MB integral, from (\ref{eq_4chainMBI}), is:
\begin{align}
    &\ft{1234}=\frac{1}{(\ii\omega_{1234})^{q_{1234}}}\int_{s_2,s_3,s_4}\left(-\frac{\omega_{1234}}{\omega_{234}}\right)^{s_2} \left(-\frac{\omega_{1234}}{\omega_{34}}\right)^{s_3} \left(-\frac{\omega_{1234}}{\omega_{4}}\right)^{s_4} \notag \\
    &\times \Gamma(q_{1234}- s_{234})
    \Gamma\left[
    \begin{matrix}
        s_2,1-s_2,q_{234}-s_{34} \\
        q_{234}-s_{234}+1
    \end{matrix}
    \right]
    \Gamma\left[
    \begin{matrix}
        s_3,1-s_3,q_{34}-s_4 \\
        q_{34}-s_{34}+1
    \end{matrix}
    \right]
    \Gamma\left[
    \begin{matrix}
        s_4,1-s_4,q_4 \\
        q_4-s_4+1
    \end{matrix}
    \right] \ .
\end{align}
We choose the kinematic region defined by $|\wt\omega_1|<|\wt\omega_4|<|\wt\omega_3|<|\wt\omega_2|$. The initial energy ratios are $z_2 = -\wt\omega_1/\wt\omega_2$, $z_3 = -\wt\omega_1/\wt\omega_3$, and $z_4 = -\wt\omega_1/\wt\omega_4$, all of which have magnitudes less than 1. The pole collecting algorithm has been given in Fig.\;\ref{fig_ChainPoleCollect} and the final result in (\ref{eq_1234TEseries}). Here we comment on a few choices of valid and invalid sets of $\Gamma$ factors:
\begin{enumerate}
  \item \emph{ The simplest choice (no mixing):} We can choose poles from $\{ \Gamma(1-s_2), \Gamma(1-s_3), \Gamma(1-s_4) \}$. These poles do not mix the variables. The pole equations are $s_2=n_2+1$, $s_3=n_3+1$, and $s_4=n_4+1$. The energy product is simply $(z_2)^{n_2+1} (z_3)^{n_3+1} (z_4)^{n_4+1}$. Since $|z_j|<1$ in our chosen region, the series is formally convergent. This basis is valid and gives the first term (the singular term) in Eq.\,(\ref{eq_1234TEseries}).
  
  \item \emph{A valid choice with mixing:} Consider the choice $\{ \Gamma(q_{1234}-s_{234}), \Gamma(s_3), \Gamma(s_4) \}$. Here, the first factor mixes all three variables. The pole equations are:
\begin{align}
    s_2+s_3+s_4 = q_{1234}+n_2, \quad s_3 = -n_3, \quad s_4 = -n_4 \ .
\end{align}
The solution is $s_3=-n_3$, $s_4=-n_4$, and $s_2 = q_{1234}+n_2+n_3+n_4$. Substituting this into the energy product yields:
\begin{align}
    \prod_{j=2}^4 z_j^{s_j} \propto z_2^{n_2+n_3+n_4} z_3^{-n_3} z_4^{-n_4} = (z_2)^{n_2} \left(\frac{z_2}{z_3}\right)^{n_3} \left(\frac{z_2}{z_4}\right)^{n_4} \ .
\end{align}
The new energy ratios for the summation variables are $z'_2 = z_2 = -\wt\omega_1/\wt\omega_2$, $z'_3 = z_2/z_3 = \wt\omega_3/\wt\omega_2$, and $z'_4 = z_2/z_4 = \wt\omega_4/\wt\omega_2$. In our kinematic region, we have $|z'_2|<1$, $|z'_3|<1$, and $|z'_4|<1$. Thus, this basis is also valid and contributes to the final sum. This corresponds to the fourth term in Eq.\,(\ref{eq_1234TEseries}). The other two valid bases can be analyzed similarly.

  \item \emph{An invalid choice:} We show a choice of set of $\Gamma$ factors that does not work. Consider choosing $\{ \Gamma(q_{1234}-s_{234}), \Gamma(1-s_3), \Gamma(1-s_4) \}$. This choice seems plausible, but it is inconsistent with the energy hierarchy. The pole equations are:
\begin{align}
    s_2+s_3+s_4 = q_{1234}+n_2, \quad s_3 = 1+n_3, \quad s_4 = 1+n_4 \ .
\end{align}
The solution is $s_3=1+n_3$, $s_4=1+n_4$, and $s_2 = q_{1234}-2+n_2-n_3-n_4$. The energy product becomes:
\begin{align}
    \prod_{j=2}^4 z_j^{s_j} \propto z_2^{n_2-n_3-n_4} z_3^{n_3} z_4^{n_4} = (z_2)^{n_2} \left(\frac{z_3}{z_2}\right)^{n_3} \left(\frac{z_4}{z_2}\right)^{n_4} \ .
\end{align}
The new ratios are $z'_2 = z_2 = -\wt\omega_1/\wt\omega_2$, $z'_3 = z_3/z_2 = \wt\omega_2/\wt\omega_3$, and $z'_4 = z_4/z_2 = \wt\omega_2/\wt\omega_4$. In our region $|\wt\omega_1|<|\wt\omega_4|<|\wt\omega_3|<|\wt\omega_2|$, we find that $|z'_3| = |\wt\omega_2/\wt\omega_3| > 1$. Thus, the series over $n_3$ would diverge. Therefore, this choice is invalid and must be discarded. The algebraic procedure automatically filters out inconsistent choices such as this one.
\end{enumerate}

\paragraph{Equivalence between the algebraic and geometric procedures}
The algebraic procedure outlined above is equivalent to the geometric conic hull method. For readers familiar with the latter, we sketch the proof here. A basis of $N-1$ pole equations can be written in matrix form as:
\begin{align}
    \bm{E} \bm{s} + \bm{c} = -\bm{n} \quad \implies \quad \bm{s} = -\bm{E}^{-1}(\bm{c}+\bm{n}) \ ,
\end{align}
where $\bm{E}$ is a square matrix whose rows are the coefficient vectors of $\bm{s}$ in the arguments of the chosen $\Gamma$ functions. Let the vector of logarithms of the magnitudes of energy ratios be $\bm{y}$ with components $y_j = \log|z_j|$. The exponent of the energy-ratio product becomes:
\begin{align}
    \bm{y}^\text{T} \bm{s} = -\bm{y}^\text{T}\bm{E}^{-1}(\bm{c}+\bm{n}) \ .
\end{align}
For the resulting series to converge, this term must go to $-\infty$ as each $n_j \to \infty$. This requires all components of the row vector $\bm{y}^\text{T} \bm{E}^{-1}$ to be positive. This gives our algebraic condition:
\begin{align}
    \text{Algebraic Condition:}\quad (\bm{E}^{-1})^\text{T} \bm{y} > 0 \quad (\text{component-wise}) \ .
\end{align}
In the conic hull method, a basis (defined by the rows of $\bm{E}$, i.e., column vectors of $\bm{E}^\text{T}$) contributes if the vector $\bm{y}$ lies in its positive conic hull. This is expressed as $\bm{y} = \bm{E}^\text{T} \bm{c}_{\text{coeffs}}$ for a vector of coefficients $\bm{c}_{\text{coeffs}}$ with all positive components. This gives the geometric condition:
\begin{align}
    \text{Geometric Condition:}\quad \bm{c}_{\text{coeffs}} = (\bm{E}^\text{T})^{-1} \bm{y} > 0 \quad (\text{component-wise}) \ .
\end{align}
Since $(\bm{E}^\text{T})^{-1} = (\bm{E}^{-1})^\text{T}$, the two conditions are identical, proving the equivalence of the two methods.

\section{More Examples}
\label{app_Examples}
In this appendix, we collect the series expressions of several simple family trees at their singular limits. Some of the series can be summed into familiar special functions. For each tree, we list all the possible root-bearing subtrees and write down the corresponding MB integrals. We then present their series expressions at large/small partial energies.

\subsection{Two-site chain}
For the two-site chain $\ft{12}$, the possible root-bearing subtrees are $\ft{1}$ and $\ft{12}$.

\paragraph{Subtree $\ft{1}$}
We don't collect the energy on Site 2. The MB form of the tree is 
\begin{align}
    \ft{12}= \frac{(-\ii)^2}{(\ii\omega_1)^{q_{12}}} \int _{s_2} \left( \frac{\omega_1}{\omega_2} \right)^{s_2} \Gamma(q_{12}-s_2) {\color{RoyalBlue}\Gamma\left[
    \begin{matrix} 
        q_2-s_2, s_2 \\ 
        q_2-s_2+1 
    \end{matrix}\right]}  \ .
\end{align}
For large $\omega_1$, the series expression is 
\begin{align}
    \ft{12}&=\frac{-1}{(\ii \omega_1)^{q_{12}}} \sum_{n_2=0}^\infty \Gamma(q_{12}+n_2) \frac{(-\omega_2/\omega_1)^{n_2}}{(q_2+n_2)n_2!} \notag \\
    &= -\frac{1}{(\ii \omega_1)^{q_{12}}} \ _2\mathcal{F}_1 \left[
    \begin{matrix}
        q_2,q_{12} \\
        q_2+1
    \end{matrix}
    \middle| -\frac{\omega_2}{\omega_1}
    \right] \ .
\label{eq_2siteLarge1}
\end{align}
For small $\omega_1$, we have $|\omega_1|<|\omega_2|$ and the series expression is 
\begin{align}
    \ft{12}&=-\frac{\Gamma(q_1)}{(\ii\omega_1)^{q_1}}\frac{\Gamma(q_2)}{(\ii\omega_2)^{q_2}} + \frac{1}{(\ii\omega_2)^{q_{12}}} \sum_{n_2=0}^\infty \Gamma(q_{12}+n_2)\frac{(-\omega_1/\omega_2)^{n_2}}{(q_1+n_2)n_2!} \notag \\
    &=-\frac{\Gamma(q_1)}{(\ii\omega_1)^{q_1}}\frac{\Gamma(q_2)}{(\ii\omega_2)^{q_2}} + \frac{1}{(\ii \omega_2)^{q_{12}}} \ _2\mathcal{F}_1 \left[
    \begin{matrix}
        q_1,q_{12} \\
        q_1+1
    \end{matrix}
    \middle| -\frac{\omega_1}{\omega_2}
    \right] \ .
\label{eq_2siteSmall1}
\end{align}
We can directly see that the first term is just the factorized $\ft{1}\ft{2}$ while the second term is the large $\omega_{2}$ expansion of $-\ft{21}$.

\paragraph{Subtree $\ft{12}$}
We collect the energy on Site 2. The MB form of the tree is 
\begin{align}
    \ft{12}= \frac{(-\ii)^2}{(\ii\omega_{12})^{q_{12}}} \int _{s_2} \left( -\frac{\omega_{12}}{\omega_2} \right)^{s_2} \Gamma(q_{12}-s_2) {\color{BrickRed}\Gamma\left[
    \begin{matrix} 
        s_2,1-s_2,q_2 \\ 
        q_2-s_2+1 
    \end{matrix}\right]}  \ .
\end{align}
At large $\omega_{12}$, the series expression is 
\begin{align}
    \ft{12}&=\frac{-1}{(\ii \omega_{12})^{q_{12}}} \sum_{n_2=0}^\infty \Gamma(q_{12}+n_2) \frac{(\omega_2/\omega_{12})^{n_2}}{(q_2)_{n_2+1}} \notag \\
    &= -\frac{\Gamma(q_2)}{(\ii \omega_{12})^{q_{12}}} \ _2\mathcal{F}_1 \left[
    \begin{matrix}
        1,q_{12} \\
        q_2+1
    \end{matrix}
    \middle| \frac{\omega_2}{\omega_{12}}
    \right] \ .
\label{eq_2siteLarge12}
\end{align}
At small $\omega_{12}$, we have $|\omega_{12}|<|\omega_2|$ and the series expression is 
\begin{align}
    \ft{12}&= \frac{1}{(\ii\omega_{12})^{q_{12}}} \sum_{n_2=0}^\infty \Gamma(q_{12}-n_2-1) \frac{(\omega_{12}/\omega_2)^{n_2+1}}{(q_2)_{-n_2}} - \frac{\pi\csc(\pi q_{12})}{(-\ii \omega_2)^{q_{12}}}\sum_{n_2=0}^\infty \frac{\Gamma(q_2)(-\omega_{12}/\omega_2)^{{n_2}}}{n_2! (1-q_1-n_2)} \notag \\
    &= -\frac{\pi \csc(\pi q_{12})}{(\ii \omega_{12})^{q_{12}}} \frac{\omega_{12}}{\omega_2} \ _2\tilde{\text{F}}_1 \left[
    \begin{matrix}
        1,1-q_2 \\
        2-q_{12}
    \end{matrix}
    \middle| \frac{\omega_{12}}{\omega_2}
    \right] -\frac{\pi \csc(\pi q_{12}) \Gamma(q_2)}{\Gamma(1-q_1) (\ii \omega_1)^{q_1} (-\ii \omega_2)^{q_2}}  \ .
\label{eq_2siteSmall12}
\end{align}
We can equate all 4 representations (\ref{eq_2siteLarge1}), (\ref{eq_2siteSmall1}), (\ref{eq_2siteLarge12}), and (\ref{eq_2siteSmall12}) to get a few transformation-of-variable identities of ${}_2\text{F}_1$ function which are well known. The same comment apply to expressions below, which yield more complicated and most likely new identities of MHFs. 

\subsection{Three-site chain}
For the three-site chain $\ft{123}$, the possible root-bearing subtrees are $\ft{1}$, $\ft{12}$, and $\ft{123}$.

\paragraph{Subtree $\ft{1}$}
We collect the energy on neither Site 2 or 3. The MB form of the tree is 
\begin{align}
    \ft{123}= \frac{(-\ii)^3}{(\ii\omega_{1})^{q_{123}}} \int _{s_2,s_3} \left( \frac{\omega_1}{\omega_2} \right)^{s_2} \left( \frac{\omega_1}{\omega_3} \right)^{s_3} \Gamma(q_{123}-s_{23}) {\color{RoyalBlue}\Gamma\left[
    \begin{matrix} 
        q_{23}-s_{23},s_2 \\ 
        q_{23}-s_{23}+1 
    \end{matrix}\right]} {\color{RoyalBlue}\Gamma\left[
    \begin{matrix} 
        q_3-s_3,s_3 \\ 
        q_3-s_3+1 
    \end{matrix}\right]}  \ .
\end{align}
At large $\omega_{1}$, the series expression is 
\begin{align}
    \ft{123}&=\frac{\ii}{(\ii \omega_{1})^{q_{123}}} \sum_{n_2,n_3=0}^\infty \Gamma(q_{123}+n_{23}) \frac{(-\omega_2/\omega_{1})^{n_2}}{(q_{23}+n_{23})n_2!} \frac{(-\omega_3/\omega_{1})^{n_3}}{(q_3+n_3)n_3!}   \ .
\end{align}
At small $\omega_{1}$, we choose $|\omega_1|<|\omega_3|<|\omega_2|$ and the series expression is 
\begin{align}
    \ft{123}&=  \frac{\ii\Gamma(q_1)}{(\ii \omega_1)^{q_{1}}} \frac{1}{(\ii \omega_{2})^{q_{23}}} \sum_{n_3=0}^\infty \Gamma(q_{23}+n_3) \frac{(-\omega_3/\omega_2)^{n_3}}{(q_3+n_3)n_3!} \notag \\
    &\quad -\frac{\ii}{(\ii \omega_2)^{q_{123}}} \sum_{n_2,n_3=0}^\infty \Gamma(q_{123}+n_{23}) \frac{(-\omega_1/\omega_2)^{n_2}}{(q_1+n_2)n_2!} \frac{(-\omega_3/\omega_2)^{n_3}}{(q_3+n_3)n_3!} \notag \\
    &=  \frac{\ii\Gamma(q_1)}{(\ii \omega_1)^{q_{1}}} \frac{1}{(\ii \omega_{2})^{q_{23}}} \ _2\mathcal{F}_1 \left[
    \begin{matrix}
        q_3,q_{23} \\
        q_3+1
    \end{matrix}
    \middle| -\frac{\omega_3}{\omega_2}
    \right] \notag \\
    &\quad -\frac{\ii}{(\ii \omega_2)^{q_{123}}} \sum_{n_2,n_3=0}^\infty \Gamma(q_{123}+n_{23}) \frac{(-\omega_1/\omega_2)^{n_2}}{(q_1+n_2)n_2!} \frac{(-\omega_3/\omega_2)^{n_3}}{(q_3+n_3)n_3!}   \ .
\end{align}
The first singular term is exactly $\ft{1}\ft{23}$, since the singular part of $\ft{1}$ is just itself. Thus the second term must be $\ft{2(1)(3)}$, which we can check using the large root-energy series of three-site star presented later.

\paragraph{Subtree $\ft{12}$}
We collect the energy on Site 2 but not on Site 3. The MB form of the tree is 
\begin{align}
    \ft{123}= \frac{(-\ii)^3}{(\ii\omega_{12})^{q_{123}}} \int _{s_2,s_3} \left( -\frac{\omega_{12}}{\omega_2} \right)^{s_2} \left( \frac{\omega_{12}}{\omega_3} \right)^{s_3} \Gamma(q_{123}-s_{23}) {\color{BrickRed}\Gamma\left[
    \begin{matrix} 
        s_2,1-s_2,q_{23}-s_3 \\ 
        q_{23}-s_{23}+1 
    \end{matrix}\right]} {\color{RoyalBlue}\Gamma\left[
    \begin{matrix} 
        q_3-s_3,s_3 \\ 
        q_3-s_3+1 
    \end{matrix}\right]}  \ .
\end{align}
At large $\omega_{12}$, the series expression is 
\begin{align}
    \ft{123}&=\frac{\ii}{(\ii \omega_{12})^{q_{123}}} \sum_{n_2,n_3=0}^\infty \Gamma(q_{123}+n_{23}) \frac{(\omega_2/\omega_{12})^{n_2}}{(q_{23}+n_{3})_{n_2+1}} \frac{(-\omega_3/\omega_{12})^{n_3}}{(q_3+n_3)n_3!}   \ .
\end{align}
At small $\omega_{12}$, we choose $|\omega_{12}|<|\omega_3|<|\omega_2|$ and the series expression is 
\begin{align}
    \ft{123}&=  \frac{\ii\pi \csc(\pi q_{12})}{(\ii \omega_{12})^{q_{12}}} \frac{\omega_{12}}{\omega_2} \ _2\tilde{\text{F}}_1 \left[
    \begin{matrix}
        1,1-q_2 \\
        2-q_{12}
    \end{matrix}
    \middle| \frac{\omega_{12}}{\omega_2}
    \right] \frac{\Gamma(q_3)}{(\ii \omega_3)^{q_3}} \notag \\
    &\quad + \frac{\ii\pi \csc(\pi q_{123})}{(\ii \omega_2)^{q_{123}}} \sum_{n_2,n_3=0}^\infty \Gamma \left[
    \begin{matrix}
        q_{23}+n_3 \\
        1-q_1-n_2
    \end{matrix}
    \right] \frac{(-\omega_{12}/\omega_2)^{n_2} (-\omega_3/\omega_2)^{n_3}}{(q_3+n_3)n_2! n_3!} \notag \\
    &\quad + \frac{\ii}{(\ii \omega_3)^{q_{123}}} \sum_{n_2,n_3=0}^\infty \Gamma (q_{123}-1+n_{\bar{2}3}) \frac{(\omega_3/\omega_2)^{n_2+1} (-\omega_{12}/\omega_2)^{n_3}}{(q_{12}-1+n_{\bar{2}3}) (1-q_1+n_{2\bar{3}})_{-n_2} n_3!} \ .
\end{align}
We can see that the singular term (the first term) is factorized explicitly, here and in the remaining examples in this appendix.

\paragraph{Subtree $\ft{123}$}
We collect the energy on both Sites 2 and 3. The MB form of the tree is 
\begin{align}
    \ft{123}= &~\frac{(-\ii)^3}{(\ii\omega_{123})^{q_{123}}} \int _{s_2,s_3} \left( -\frac{\omega_{123}}{\omega_{23}} \right)^{s_2} \left( -\frac{\omega_{123}}{\omega_3} \right)^{s_3} \notag \\
    &~ \times \Gamma(q_{123}-s_{23}) {\color{BrickRed}\Gamma\left[
    \begin{matrix} 
        s_2,1-s_2,q_{23}-s_3 \\ 
        q_{23}-s_{23}+1 
    \end{matrix}\right]} {\color{BrickRed}\Gamma\left[
    \begin{matrix} 
        s_3,1-s_3,q_3 \\ 
        q_3-s_3+1 
    \end{matrix}\right]}  \ .
\end{align}
At large $\omega_{123}$, the series expression is 
\begin{align}
    \ft{123}&=\frac{\ii}{(\ii \omega_{123})^{q_{123}}} \sum_{n_2,n_3=0}^\infty \Gamma(q_{123}+n_{23}) \frac{(\omega_{23}/\omega_{123})^{n_2}}{(q_{23}+n_{3})_{n_2+1}} \frac{(\omega_3/\omega_{123})^{n_3}}{(q_3)_{n_3}}   \ .
\end{align}
At small $\omega_{123}$, we choose $|\omega_{123}|<|\omega_3|<|\omega_{23}|$ and the series expression is 
\begin{align}
    \ft{123}&= \frac{\ii}{(\ii \omega_{123})^{q_{123}}} \sum_{n_2,n_3=0}^\infty \Gamma(q_{123}-2-n_{23}) \frac{(\omega_{123}/\omega_{23})^{n_2+1} (\omega_{123}/\omega_3)^{n_3+1}}{(q_{23}-1-n_3)_{-n_2} (q_3)_{-n_3}} \notag \\
    &\quad + \frac{\ii\pi \csc(\pi q_{123})}{(-\ii\omega_{23})^{q_{123}}} \sum_{n_2,n_3=0}^\infty \Gamma \left[
    \begin{matrix}
        q_{23}+n_3 \\
        1-q_1-n_2
    \end{matrix}
    \right] \frac{(-\omega_{123}/\omega_{23})^{n_2} (\omega_3/\omega_{23})^{n_3}}{(q_3)_{n_3+1} n_2!}   \notag \\
    &\quad -  \frac{\ii\pi \csc(q_{123})}{(-\ii \omega_3)^{q_{123}}} \sum_{n_2,n_3=0}^\infty \Gamma\left[
    \begin{matrix}
        q_3 \\
        2-q_{12}+n_{2\bar{3}}
    \end{matrix}
    \right] \frac{(\omega_3/\omega_{23})^{n_2+1} (-\omega_{123}/\omega_3)^{n_3}}{(1-q_1+n_{2\bar{3}})_{-n_2} n_3!}     \ .
\end{align}

\subsection{Three-site star}
For the three-site star $\ft{1(2)(3)}$, the possible root-bearing subtrees are $\ft{1}$, $\ft{12}$, $\ft{13}$, and $\ft{1(2)(3)}$. We only present the series expression for one valid total order of energies, since the result for the other can be obtained by simply switching the label of Sites 2 and 3.

\paragraph{Subtree $\ft{1}$}
We collect the energy on neither Site 2 nor 3. The MB form of the tree is 
\begin{align}
    \ft{1(2)(3)}= \frac{(-\ii)^3}{(\ii\omega_{1})^{q_{123}}} \int _{s_2,s_3} \left( \frac{\omega_1}{\omega_2} \right)^{s_2} \left( \frac{\omega_1}{\omega_3} \right)^{s_3} \Gamma(q_{123}-s_{23}) {\color{RoyalBlue}\Gamma\left[
    \begin{matrix} 
        q_2-s_2,s_2 \\ 
        q_2-s_2+1 
    \end{matrix}\right]} {\color{RoyalBlue}\Gamma\left[
    \begin{matrix} 
        q_3-s_3,s_3 \\ 
        q_3-s_3+1 
    \end{matrix}\right]}  \ .
\end{align}
At large $\omega_{1}$, the series expression is 
\begin{align}
    \ft{1(2)(3)}&=\frac{\ii}{(\ii \omega_{1})^{q_{123}}} \sum_{n_2,n_3=0}^\infty \Gamma(q_{123}+n_{23}) \frac{(-\omega_2/\omega_{1})^{n_2}}{(q_{2}+n_{2})n_2!} \frac{(-\omega_3/\omega_{1})^{n_3}}{(q_3+n_3)n_3!}   \ .
\end{align}
At small $\omega_{1}$, we choose $|\omega_1|<|\omega_3|<|\omega_2|$ and the series expression is 
\begin{align}
    \ft{1(2)(3)}&= \frac{\ii\Gamma(q_1)}{(\ii \omega_1)^{q_1}} \frac{\Gamma(q_2)}{(\ii \omega_2)^{q_2}} \frac{\Gamma(q_3)}{(\ii \omega_3)^{q_3}} - \frac{\ii}{(\omega_2)^{q_{123}}} \sum_{n_2,n_3=0}^\infty \Gamma(q_{123}+n_{23}) \frac{(-\omega_1/\omega_2)^{n_2} (-\omega_3/\omega_2)^{n_3}}{(q_{13}+n_{23})(q_3+n_3)n_2! n_3!} \notag \\
    &\quad -  \frac{\ii\Gamma(q_2)}{(\ii\omega_2)^{q_2}} \frac{1}{(\ii \omega_3)^{q_{13}}} \ _2\mathcal{F}_1 \left[
    \begin{matrix}
        q_1,q_{13} \\
        q_1+1
    \end{matrix}
    \middle| -\frac{\omega_1}{\omega_3}
    \right]  \ .
\end{align}
We recognize the equation simply as $\ft{1(2)(3)}=\ft{1}\ft{2}\ft{3}-\ft{213}-\ft{2}\ft{31}$. This explicitly shows our flipping-direction arguments about zero root-energy series.

\paragraph{Subtree $\ft{12}$}
We collect the energy on Site 2 but not on Site 3. The MB form of the tree is 
\begin{align}
    \ft{1(2)(3)}= \frac{(-\ii)^3}{(\ii\omega_{12})^{q_{123}}} \int _{s_2,s_3} \left( -\frac{\omega_{12}}{\omega_2} \right)^{s_2} \left( \frac{\omega_{12}}{\omega_3} \right)^{s_3} \Gamma(q_{123}-s_{23}) {\color{BrickRed}\Gamma\left[
    \begin{matrix} 
        s_2,1-s_2,q_2 \\ 
        q_2-s_2+1 
    \end{matrix}\right]} {\color{RoyalBlue}\Gamma\left[
    \begin{matrix} 
        q_3-s_3,s_3 \\ 
        q_3-s_3+1 
    \end{matrix}\right]}  \ .
\end{align}
At large $\omega_{12}$, the series expression is 
\begin{align}
    \ft{1(2)(3)}&=\frac{\ii}{(\ii \omega_{12})^{q_{123}}} \sum_{n_2,n_3=0}^\infty \Gamma(q_{123}+n_{23}) \frac{(\omega_2/\omega_{12})^{n_2}}{(q_{2})_{n_2}} \frac{(-\omega_3/\omega_{12})^{n_3}}{(q_3+n_3)n_3!}   \ .
\end{align}
At small $\omega_{12}$, we choose $|\omega_{12}|<|\omega_3|<|\omega_2|$ and the series expression is 
\begin{align}
    \ft{1(2)(3)}&=  \frac{\ii\pi \csc(\pi q_{12})}{(\ii \omega_{12})^{q_{12}}} \frac{\omega_{12}}{\omega_2} \ _2\tilde{\text{F}}_1 \left[
    \begin{matrix}
        1,1-q_2 \\
        2-q_{12}
    \end{matrix}
    \middle| \frac{\omega_{12}}{\omega_2}
    \right] \frac{\Gamma(q_3)}{(\ii \omega_3)^{q_3}} \notag \\
    &\quad +  \frac{\ii\pi \csc(\pi q_{123})}{(-\ii \omega_2)^{q_{123}}} \sum_{n_2,n_3=0}^\infty \Gamma\left[
    \begin{matrix}
        q_2 \\
        1-q_{13}-n_{23}
    \end{matrix}
    \right] \frac{(-\omega_{12}/\omega_2)^{n_2} (-\omega_3/\omega_2)^{n_3}}{(q_3+n_3)n_2! n_3!} \notag \\
    &\quad + \frac{\ii}{(\ii \omega_3)^{q_{123}}} \sum_{n_2,n_3=0}^\infty \Gamma(q_{123}-1+n_{\bar{2}3}) \frac{(\omega_3/\omega_2)^{n_2+1} (-\omega_{12}/\omega_{3})^{n_3}}{(q_{12}-1+n_{\bar{2}3}) (q_2)_{-n_2} n_3!}  \ .
\end{align}

\paragraph{Subtree $\ft{13}$}
We collect the energy on Site 3 but not on Site 2. The MB form of the tree is 
\begin{align}
    \ft{1(2)(3)}= \frac{(-\ii)^3}{(\ii\omega_{13})^{q_{123}}} \int _{s_2,s_3} \left( \frac{\omega_{13}}{\omega_2} \right)^{s_2} \left( -\frac{\omega_{13}}{\omega_3} \right)^{s_3} \Gamma(q_{123}-s_{23}) {\color{RoyalBlue}\Gamma\left[
    \begin{matrix} 
        q_2-s_2,s_2 \\ 
        q_2-s_2+1 
    \end{matrix}\right]} {\color{BrickRed}\Gamma\left[
    \begin{matrix} 
        s_3,1-s_3,q_3 \\ 
        q_3-s_3+1 
    \end{matrix}\right]}  \ .
\end{align}
At large $\omega_{13}$, the series expression is 
\begin{align}
    \ft{1(2)(3)}&=\frac{\ii}{(\ii \omega_{13})^{q_{123}}} \sum_{n_2,n_3=0}^\infty \Gamma(q_{123}+n_{23}) \frac{(-\omega_2/\omega_{12})^{n_2}}{(q_{2}+n_2){n_2}!} \frac{(\omega_3/\omega_{12})^{n_3}}{(q_3)_{n_3}}   \ .
\end{align}
Obviously, this is simply the series express at $\omega_{12}\to \infty$ with Sites 2 and 3 exchanged.

At small $\omega_{13}$, we choose $|\omega_{13}|<|\omega_3|<|\omega_2|$ and the series expression is 
\begin{align}
    \ft{1(2)(3)}&=  \frac{\ii\pi \csc(\pi q_{13})}{(\ii \omega_{13})^{q_{13}}} \frac{\omega_{13}}{\omega_3} \ _2\tilde{\text{F}}_1 \left[
    \begin{matrix}
        1,1-q_3 \\
        2-q_{13}
    \end{matrix}
    \middle| \frac{\omega_{13}}{\omega_3}
    \right] \frac{\Gamma(q_2)}{(\ii \omega_2)^{q_2}} \notag \\
    &\quad -  \frac{\ii}{(\ii \omega_2)^{q_{123}}} \sum_{n_2,n_3=0}^\infty \Gamma(q_{123}+n_{23}) \frac{(-\omega_{13}/\omega_2)^{n_2} (\omega_3/\omega_2)^{n_3}}{(q_{13}+n_{23})(q_3)_{n_3+1} n_2!} \notag \\
    &\quad +  \frac{\ii \pi\csc(\pi q_{13})\Gamma(q_3)}{\Gamma(1-q_1) (\ii \omega_1)^{q_1} (-\ii \omega_3)^{q_3}} \frac{\Gamma(q_2)}{(\ii \omega_2)^{q_2}} \ .
    \label{eq_3Star_Small_13}
\end{align}
We can see that the first and third terms combine into $\ft{13}\ft{2}$ (using small total-energy series of $\ft{13}$). Thus the second term must be $-\ft{213}$. This can actually be obtained from an unused version of the MB integral
\begin{align}
    \ft{213}=\frac{(-\ii)^3}{(\ii \omega_2)^{q_{123}}} \int_{s_1,s_3} \left(\frac{\omega_2}{\omega_{13}}\right)^{s_1} \left(-\frac{\omega_2}{\omega_{3}}\right)^{s_3} \Gamma(q_{123}+n_{13}) {\color{RoyalBlue}\Gamma \left[
    \begin{matrix}
        q_{13}-s_{13},s_1 \\
        q_{13}-s_{13}+1
    \end{matrix}
    \right]} {\color{BrickRed}\Gamma \left[
    \begin{matrix}
        s_3,1-s_3,q_3 \\
        q_3-s_3+1
    \end{matrix}
    \right]} \ .
\end{align}
Here we collect energy $\omega_3$ to Site 1, but do not continue collecting $\omega_{13}$ to the root (Site 2). We choose $\omega_2$ to be large and the only valid choice of $\Gamma$ functions is $\Gamma(s_1),\Gamma(s_3)$. We can then obtain the series identical (up to a minus sign) to the second term in (\ref{eq_3Star_Small_13}).

\paragraph{Subtree $\ft{1(2)(3)}$}
We collect the energy on both Sites 2 and 3. The MB form of the tree is 
\begin{align}
    \ft{1(2)(3)}&= \frac{(-\ii)^3}{(\ii\omega_{123})^{q_{123}}} \int _{s_2,s_3} \left( -\frac{\omega_{123}}{\omega_2} \right)^{s_2} \left( -\frac{\omega_{123}}{\omega_3} \right)^{s_3}  \notag \\
    &\qquad\qquad\qquad \times  \Gamma(q_{123}-s_{23}) {\color{BrickRed}\Gamma\left[
    \begin{matrix} 
        s_2,1-s_2,q_2 \\ 
        q_2-s_2+1 
    \end{matrix}\right]} {\color{BrickRed}\Gamma\left[
    \begin{matrix} 
        s_3,1-s_3,q_3 \\ 
        q_3-s_3+1 
    \end{matrix}\right]}  \ .
\end{align}
At large $\omega_{123}$, the series expression is 
\begin{align}
    \ft{1(2)(3)}&=\frac{\ii}{(\ii \omega_{123})^{q_{123}}} \sum_{n_2,n_3=0}^\infty \Gamma(q_{123}+n_{23}) \frac{(\omega_2/\omega_{123})^{n_2}}{(q_{2})_{n_2}} \frac{(\omega_3/\omega_{123})^{n_3}}{(q_3)_{n_3}}   \ .
\end{align}
At small $\omega_{123}$, we choose $|\omega_{123}|<|\omega_3|<|\omega_2|$ and the series expression is 
\begin{align}
    \ft{1(2)(3)}&=  \frac{\ii}{(\ii \omega_{123})^{q_{123}}} \sum_{n_2,n_3=0}^\infty \Gamma(q_{123}-n_{23}-2) \frac{(\omega_{123}/\omega_2)^{n_2+1} (\omega_{123}/\omega_3)^{n_3+1}}{(q_2)_{-n_2} (q_3)_{-n_3}} \notag \\
    &\quad + \frac{\ii\pi \csc(\pi q_{123})}{(-\ii\omega_2)^{q_{123}}} \sum_{n_2,n_3=0}^\infty \Gamma\left[
    \begin{matrix}
        q_2 \\
        1-q_{13}-n_{23}
    \end{matrix}
    \right] \frac{(-\omega_{123}/\omega_2)^{n_2} (\omega_3/\omega_2)^{n_3}}{(q_3)_{n_3+1} n_2!} \notag \\
    &\quad -\frac{\ii \pi \csc(\pi q_{123})}{(-\ii\omega_3)^{q_{123}}} \sum_{n_2,n_3=0}^\infty \Gamma\left[
    \begin{matrix}
        q_3 \\
        2-q_{12}+n_{2\bar{3}}
    \end{matrix}
    \right] \frac{(\omega_3/\omega_2)^{n_2+1} (-\omega_{123}/\omega_3)^{n_3}}{(q_2)_{-n_2} n_3!}      \ .
\end{align}

\end{appendix}

\newpage
\bibliography{CosmoCollider} 
\bibliographystyle{utphys}

\end{document}